\documentclass[final,3p,review]{elsarticle}

\usepackage{amssymb}
\usepackage{color}
\usepackage{tikz}
\usepackage{algpseudocode}
\usetikzlibrary {plotmarks}
\usetikzlibrary{shapes,arrows}
\usepackage{booktabs}
\usepackage{color}
\usepackage{amsmath}
\usepackage{amsthm}
\usepackage[small,bf]{caption}
\usepackage{wrapft}
\usepackage{fancyhdr}
\usepackage{verbatim}
\usepackage{fancybox}
\usepackage{float}
\usepackage{listings}
\usepackage{setspace}
\usepackage{subfig}
\usepackage{caption}
\usepackage{algorithm}
\usepackage{graphicx}
\usepackage{wasysym}
\usepackage{textcomp}
\usepackage{endnotes}
\usepackage{textcomp}
\usepackage{soul}
\usepackage{lscape}
\usepackage{longtable}

\usepackage[
pdfauthor={Fragkiskos D. Malliaros and Michalis Vazirgiannis},
pdftitle={Clustering and Community Detection in Directed Networks: A Survey},
pdfsubject={A Review Article on Clustering and Community Detection in Directed
Networks},
pdfkeywords={Directed Graph Clustering, Community Detection, Graph Mining},
bookmarks=false,
colorlinks=true,
citecolor=blue,
]{hyperref}
\newtheorem{mydef}{Definition}
\journal{Physics Reports}

\begin{document}

\begin{frontmatter}


\title{Clustering and Community Detection in Directed Networks: \\ A Survey}

\author[ep]{Fragkiskos D. Malliaros\corref{cor1}}
\ead{fmalliaros@lix.polytechnique.fr}
\author[ep,aueb]{Michalis Vazirgiannis}
\ead{mvazirg@lix.polytechnique.fr}

\cortext[cor1]{Corresponding author. Full postal address: Laboratoire
d'Informatique (LIX), B\^{a}timent Alan Turing, 1 rue Honor\'{e} d'Estienne
d'Orves, Campus de l'\'{E}cole Polytechnique, 91120 Palaiseau, France. Tel: +33
01 7757 8045.}

\address[ep]{Computer Science Laboratory, \'{E}cole Polytechnique, 91120
Palaiseau, France}
\address[aueb]{Department of Informatics, Athens University of Economics and
Business, Patision 76, 10434 Athens, Greece}

\begin{abstract}
Networks (or graphs) appear as dominant structures in diverse domains,
including sociology, biology, neuroscience and computer science.  In most of the
aforementioned cases graphs are directed -- in the sense that there is
directionality on the edges, making the semantics of the edges non symmetric as
the source node transmits some property to the target one but not vice versa.  An
interesting feature that real networks present is the clustering or community
structure property, under which the graph topology is organized into modules
commonly called communities or clusters. The essence here is that nodes of
the same community are highly similar while on the contrary, nodes across
communities present low similarity. Revealing the underlying community structure
of directed complex networks has become a crucial and interdisciplinary topic
with a plethora of relevant application domains. Therefore, naturally there is
a recent wealth of research production in the area of mining directed graphs --
with clustering being the primary method sought and the primary tool for
community detection and evaluation.  The goal of this paper is to offer an
in-depth comparative review of the methods presented so far for clustering
directed networks along with the relevant necessary methodological background
and also related applications.  The survey commences by offering a concise
review of the fundamental concepts and methodological base on which graph
clustering algorithms capitalize on. Then we present the relevant work along two
orthogonal classifications. The first one is mostly concerned with the
methodological principles of the clustering algorithms, while the second one
approaches the methods from the viewpoint regarding the properties of a good
cluster in a directed network. Further, we present methods and metrics for
evaluating graph clustering results, demonstrate interesting  application
domains and provide promising future research directions.
\end{abstract}

\begin{keyword}
Community detection \sep graph clustering \sep directed networks \sep complex
networks \sep graph mining
\end{keyword}

\end{frontmatter}

% \linenumbers

\tableofcontents

%====================== Sections ===============================================
\section{Introduction} \label{sec:introduction}
Networks have become ubiquitous as data from many different disciplines can be 
naturally mapped to graph structures \cite{newman}. Technological networks,
including the Internet, electrical grids, telephone networks and road networks
are an important part of everyday life. Information networks, such as the
hyperlink structure of the Web and citation networks, offer an effective way to
represent content and information and navigate through it. Biological networks,
including protein-protein interaction networks, neural networks, gene regulatory
networks and food webs, can be used to model the function and interaction of
natural entities. Social networks, such as collaboration networks,
sexual networks and interaction networks over online social networking
applications are used to represent and model the social ties among individuals. 
Due to the extent and the diversity of contexts in which graphs appear, 
the area of network analysis has become both crucial and interdisciplinary, in
order to understand the features, the structure and the dynamics of these
complex systems.

\par Real-world networks, as the ones presented above, are not classified as 
random networks (e.g., the Erd\"{o}s-R\'{e}nyi random graph model
\cite{erdos}); that is, they present fascinating patterns and properties 
conveying that their inherent structure is not governed by randomness. The
degree distribution is skewed, following a power-law distribution
\cite{barabasi-albert-science,fff}, the average distance between  
nodes in the network is short (the so-called small-world phenomenon
\cite{milgram, web-diameter, leskovec-msn}), the ties between 
entities do not always represent reciprocal relations forming directed
networks with non symmetric links \cite{newman}, while edge distribution 
is inhomogeneous resulting in node groups with high internal edges' density
and low density between them \cite{newman, girvan-newman}. 
The last property is referred to \textit{clustering} or
\textit{community structure} and is of great interest in various fields and real-world 
applications. Detecting clusters in graphs with directed edges among nodes, is
the focus of this survey paper.

\par Informally, a \textit{cluster} or \textit{community}  can be considered
as a set of entities that are closer each other, compared to the rest of the
entities in the dataset. The notion of closeness is based on a similarity
measure, which is usually defined over the set of entities. In the areas of
machine learning and data mining, the task of clustering is also referred as
``unsupervised learning'' where the aim is to group (cluster) together similar
objects without any prior knowledge about the clusters (e.g., see Ref.
\cite{jain-clustering}). 

\par In the case of networks, the clustering (or community detection) problem
refers to grouping nodes into clusters according to their similarity, which 
usually considers either topological features (e.g., features extracted from the
graph), or other characteristics related to the nodes and edges of the
graph (e.g., additional information that may be associated with the nodes and
edges), or both of them. In other words, the clusters typically correspond to
groups of nodes sharing common properties and characteristics. Although there
are several definitions for the graph clustering problem, the most common one
states that a cluster corresponds to a set of nodes with more edges inside the
set than to the rest of the graph. 
 
\par It is important to stress out here that the task of graph clustering can be
distinguished into two different problems. The first and most studied one --
which is the focus of this paper -- aims to group the nodes of a single
graph according to some clustering definition (e.g., density). On the
other hand, the second problem refers to the task where the goal is to cluster
a set of graphs -- treating them as individual objects -- based on their
similarity (e.g., see Ref. \cite{aggarwal-10-clustering}).

\par Finding clusters in \textit{directed networks} is a challenging task with
several important applications in a wide range of domains. However, the problem of graph
clustering has mainly been considered and studied for the case of undirected
networks. A plethora of diverse algorithms have been proposed for the undirected
settings, involving contributions from the fields of computer science,
statistical physics and biology (e.g., see Ref. \cite{fortunato}). Nevertheless,
numerous graph data in several applications are by nature
directed and thus it is meaningful to incorporate all the available information
during the clustering process (i.e., the directionality of the edges). Some
illustrative examples include (see Section \ref{sec:applications} for a detailed
list of applications related to clustering directed networks):
 
\begin{itemize}
 \item \textit{Social and information networks:} Clusters in the directed
hyperlink structure of the Web correspond to sets of web pages that
share some common topics. Similarly, communities in a social network with
non-symmetric links (e.g., twitter) correspond to individuals with common
interests or friendship relationships.

\item \textit{Biology:} In prokaryote genome sequence data, the donor-recipient
relations among genomes are modeled by directed networks (called Lateral Gene
Transfer networks - LGT). Applying graph clustering methods to these
directed networks enables testing hypotheses relevant to LGT patterns and
mechanisms operating in nature \cite{genome-research}.

\item \textit{Neuroscience:} Analyzing directed brain networks produced by
neuron interactions, neuroscientists are able to comprehend the
functional architecture of the brain \cite{neuro}.

\item \textit{Clustering non-graph data:} Except from the cases where the data
naturally can be  modeled as graphs, graph clustering algorithms can be also
applied on data with no inherent graph structure, operating thus as general
purpose algorithms. In such cases, the data (e.g., points in a $d$-dimensional
Euclidean space) is represented in terms of a similarity graph corresponding 
to topological relationships and distance among them. Hence, the problem of
clustering a set of data points is transformed to a graph clustering
problem (e.g., see  tutorial by von Luxburg \cite{von-luxburg}). Depending
on the way the similarity graph is constructed, the final graph can
contain directed edges (e.g., using $k$-nearest neighbor graphs or based on
probabilistic dependence relations between data points
\cite{clustering-aaai08}).
\end{itemize}

\par It is clearly evident that the clustering problem in directed networks is
particularly significant with many important applications in several areas.
Nevertheless, despite its importance, the problem has not received significant
attention from the research community. Even though a plethora of directed graph
data exist, the most common way to dealing with edge directionality during the
clustering task, is simply to ignore it. In other words, the directed network is
converted into an undirected one (by assuming edge symmetry), and then
algorithms for the undirected graph clustering problem can be applied. However,
in many cases, this simplistic technique would not be satisfactory, since some
of the underlying semantics are not retained (e.g., in a citation network
between scientific publications or in the hyperlinked structure of the Web). 

\par The goal of this survey paper is to review the methods and
algorithms proposed by the wider research community to deal with the clustering
in directed networks. Some of them include extensions of approaches that have
been previously applied in undirected networks while others propose novel ways
as to how edge directionality can be utilized in the clustering task.

\subsection{Challenges in Clustering Directed Networks}
The problem of clustering in directed networks is considered to be a more
challenging task as compared to the undirected case. Highlighting the
difficulties  of the problem, in his resent work Santo Fortunato stated that
``\textit{Developing methods of community detection for
directed graphs is a hard task. For instance, a directed graph is characterized
by asymmetrical matrices (adjacency matrix, Laplacian, etc.), so spectral
analysis is much more complex. Only a few methods can be easily extended 
from the undirected to the directed case. Otherwise, the problem must be
formulated from scratch}'' \cite{fortunato}. This paragraph summarized the 
basic challenges for the directed graph clustering problem. Moreover, the nature
of relationships captured by the edges are fundamentally different from the ones
in the undirected settings \cite{pnas-directed-reciprocity}. To this direction,
we briefly discuss on some of the main challenges, which actually can be helpful
to understand how the directed graph clustering is differentiated from the
undirected version.

\par It is clear that ignoring edge directionality and considering the
graph as undirected is not a meaningful way to cluster directed networks as it 
fails to capture the asymmetric relationships implied by the edges of a directed
network. Therefore, the main challenge is to propose meaningful ways to
incorporate edge directionality in the clustering process. However, even this is
not sufficient. In the broader literature in graph theory and graph algorithms,
the main focus is on undirected graphs. Therefore, two additional points that
strengthen the challenging nature of the problem are:

\begin{itemize}
\item[(i)] While several graph concepts (e.g., density) are theoretically well
founded for undirected graphs, not enough effort has been put on how to extend
these concepts on directed graphs.

\item[(ii)] Similarly, extending to the directed case the available theoretical
tools that have been already applied to define and propose solutions for the
undirected versions of the problem (including graph theoretic and linear
algebraic tools), is not straightforward (e.g., the tool of spectral clustering
based on the Laplacian matrix \cite{chung-directed-laplacian05}).
\end{itemize}
 
\par In addition to the above points,  a precise and common definition for
the clustering problem in directed networks does not yet exist. The intuition
based on the intra-cluster and inter-cluster edge density cannot be easily
extended to the directed case, due to the absence of link symmetry. Moreover,
the presence of directed edges implies more sophisticated types of clusters that
do not exist in undirected networks and cannot be captured using only density
and edge concentration characteristics (e.g., clusters that represent patterns
of movement or flow circulation among nodes -- see Section
\ref{sec:problem-stm}). As we will see at the rest of the paper, in the very
recent literature several clustering definitions have been proposed  and
various algorithms have been designed  to reveal different ``types'' of graph
clusters (e.g., \cite{rosvall-pnas08}).

\subsection{Goals of the Survey and Contributions}
The main goal of this survey paper is to organize, analyze and present in a
unified and comparative manner the methods and algorithms 
proposed so far for the problem of clustering and community detection in
directed networks. While a large amount of research works and related surveys
have been devoted to the
undirected version of the problem (see Section \ref{sec:rel_surveys} for more
details), our focus is on the clustering problem in the directed settings,
where very recently many diverse methods and algorithms have been proposed.

\par Our survey adopts the following methodology: 

\begin{itemize}
 \item[(i)] As a first resort, we present a broad categorization of the
efforts that have been proposed so far. This classification scheme is mostly
concerned with the methodological principles and the algorithmic approaches for
the graph clustering problem in directed networks and is mostly built upon the
work for the undirected case of the problem. Hence, whenever possible we
organize and review the related work describing how existing methods for
undirected networks are extended in order to deal with edge directionality.
Additionally, the related work is organized according to common methodological
features that the approaches may share. We consider that such a classification
and presentation scheme is a natural way to explore and study the relevant
literature, since a large portion of the proposed approaches constitute
extensions from the undirected case of the problem.

\item[(ii)] At a second step, we present two major categories of clusters 
identified from the already proposed clustering methods, and then, the related
work is classified to these categories. The first category corresponds to
methods which adopt the more traditional density-based definition of clusters,
while the second one includes methods where the extracted clusters present 
interesting patterns, beyond simple edge density, reflecting the existence
of directed edges (e.g., flow-based patterns -- Section \ref{sec:problem-stm}).
\end{itemize}

\par To the best of our knowledge, this is the first comprehensive and extensive
survey fully devoted to the clustering problem in directed networks. We consider
that the two aforementioned axes on which the survey will
align, can be helpful both for researchers in the area and for 
practitioners that are interested in graph clustering algorithms for directed
networks (e.g., see Section \ref{sec:applications} for some important
applications where directed graph clustering methods can be applied). For the
latter case, our ultimate goal is that this survey can be used as
a practitioner's guide.

\subsection{Related Surveys} \label{sec:rel_surveys}
There are many previous related works and surveys that refer to graph clustering
and community detection in undirected networks. Fortunato \cite{fortunato}
presents a comprehensive review in the area of community detection for
undirected networks from a statistical physics perspective, while  Schaeffer
\cite{schaeffer-review} mainly focuses on the graph clustering problem as an
unsupervised learning task. Both surveys briefly discuss the case of directed
networks, however their focus is on the undirected case of the problem. A
similar but more compact description of clustering approaches in undirected
networks is presented in Refs. \cite{communities-ams, danon-jstat05,
arenas-survey}. Coscia et al. \cite{coscia-review} present a categorization for
community discovery methods according to the definition of community they adopt
(e.g., communities based on internal density or bridge detection). Our work
shares some common features with the one of Coscia et al. since part of our
presentation follows a similar categorization scheme. Parthasarathy et al.
\cite{parthasarathy-book-community-detection} present the principal methods for
the undirected community discovery problem, as well as research trends and
emerging tasks in the area. One of them is community discovery in directed
networks. Aggarwal and Wang \cite{aggarwal-10-clustering} elaborate on the basic
principles for finding communities in undirected networks, presenting some
well-established approaches (e.g., spectral clustering, minimum cut problem).
Moreover, a large part of their work is devoted to the problem of clustering
a set of graphs, treating them as individual objects (in contrast to the node
clustering problem of a single graph which is the focus of this paper). Finally,
Papadopoulos et al. \cite{papadopoulos-survey-12} discuss the topic of community
detection in the context of Social Media.

\subsection{Structure of the Survey} 
The rest of the paper is organized as follows. In Section \ref{sec:background}
we commence by providing the background and the basic terminology used
throughout this survey. In Section \ref{sec:problem-stm} we elaborate on the 
problem of clustering in directed networks, providing the basic definitions
proposed in the relevant works. These clustering definitions-notions will later
be used to classify the proposed methods based on the type of clusters they aim
to identify. Then, in Section \ref{sec:edge-dir} we present the first and main
methodological classification based on the algorithmic approaches they adopt to
deal with edge directionality. Whenever the methods are built upon approaches
for undirected networks, an incremental description is followed with respect to
the undirected case. In Section \ref{sec:definition-based} we present a second
classification scheme of the clustering approaches according to the
notion-definition of clusters in directed networks, and we also present an
empirical comparison of the main methods that have been reviewed throughout this
paper. In Section \ref{sec:evaluation} we present the evaluation metrics for
assessing the clustering results and discuss on proposed benchmarking
techniques for the graph clustering task. Section \ref{sec:applications}
presents the main applications of directed graph clustering in
different application domains, while in Section \ref{sec:future} we discuss
future research directions. Finally, in Section \ref{sec:conclusions} we
conclude the survey by summarizing and providing remarks on the problem.
\section{Basic Terminology and Background} \label{sec:background}
In this section we provide the basic terminology and background that will be
used throughout the paper. We give the definitions for basic  graph
theoretic and linear algebraic concepts, and then we describe the main aspects
of random walks which play crucial role in the design of clustering and
community detection algorithms. Finally, we make a brief presentation of the
major metrics used to quantify the quality of a community/cluster in
undirected networks. Table \ref{tbl:symbols} gives a list of used symbols along
with their definition. For a general introduction to the field of complex
networks, the reader may refer to Refs. \cite{phys-reports-complex-networks,
newman, chakrabarti-faloutsos-acm-comput-surv06}.

\begin{table}[t]
\centering
\begin{tabular}{ll}
\toprule
Symbol &   Definition \\
\midrule
$G$  & Directed network \\
$G_U$  & Undirected network  \\
$G_B=(V_h, V_a, E_b)$ & Bipartite network \\
$V, ~ E$  &  Set of nodes and edges for network $G$ \\
$|V| = n, ~~ |E| = m$ &   Number of nodes and edges in the network \\
$e = (u, v)$  & Edge $e \in E$ from node $u$ to node $v$ \\
$\mathbf{A}_U, ~~~~ \mathbf{A}$ &  Adjacency matrix of an undirected and
directed network respectively \\
$k^{in}_u, ~~~~ k^{out}_u$ & In-degree and
Out-degree of
node $u$ \\
$\mathbf{D}_{in}, ~~~~ \mathbf{D}_{out}$ &  Diagonal In- and Out- degree
matrices \\
$A_{ij}$ &  Entry of matrix $\mathbf{A}$ \\
$\mathbf{A}^T$ & The transpose of matrix $\mathbf{A}$ \\
$\lambda_i$ & $i$-th largest eigenvalue of a matrix  \\
$\mathbf{u}_i$ & Eigenvector corresponds to $i$-th  eigenvalue \\
$u_{ij}$  &  $i$-th component of $j$-th eigenvector \\
\bottomrule
\end{tabular}
\caption{Symbols and definitions. \label{tbl:symbols}}
\end{table}

\subsection{Graph Theory}
\begin{figure}[t]
\centering
 \begin{tabular}{cccc}
 \includegraphics[width=.2\textwidth]{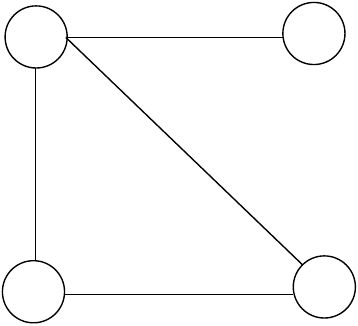} & 
 \includegraphics[width=.2\textwidth]{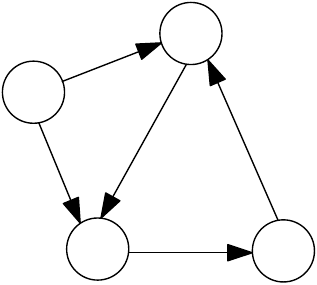} & 
 \includegraphics[width=.2\textwidth]{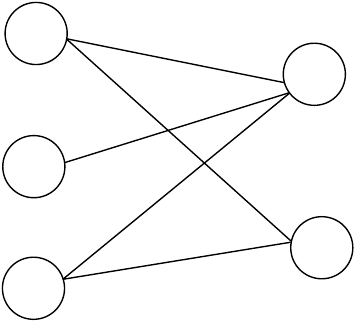} & 
 \includegraphics[width=.2\textwidth]{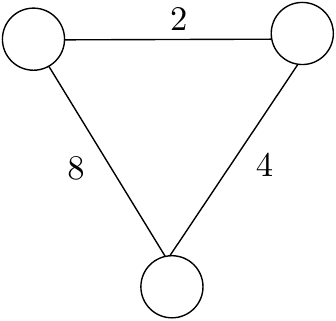} \\
\small{(a) Undirected graph}  & \small{(b) Directed graph} & \small{(c)
Bipartite graph} & \small{(d) Weighted graph} 
\end{tabular}
\caption{Examples of different types of graphs. In the case of directed
graph (b), the arrows indicate the directionality of each edge. In the weighted
graph (d) the values associated with each edge represent the weights (a
weighted graphs can be directed or undirected). \label{fig:graphs}}
\end{figure}

A \textit{network} is usually represented by a \textit{graph} (throughout the
paper we use the terms network and graph interchangeably). A graph $G=(V,E)$
consists
of a set of nodes $V$ and a set of edges $E \subseteq V \times V$ which
connect pairs of nodes (sometimes the nodes and edges of a graph are also
called vertices and links respectively). The number of nodes in the graph is
equal to $n=|V|$ and the number of edges $m=|E|$. A graph may be
\textit{directed} or \textit{undirected}, \textit{unipartite} or
\textit{bipartite} and the edges may contain \textit{weights} or not. Figure
\ref{fig:graphs} depicts some examples of different types of graphs.

\begin{mydef}{\normalfont \bf (Directed and Undirected Graph)}. 
In a directed graph $G=(V,E)$, every edge $(i,j) \in E$ links node $i$ to node
$j$ (ordered pair of nodes). An undirected graph $G_U = (V,E)$ is a directed
one  where if edge $(i,j) \in E$, then edge $(j,i) \in E$ as well.
\end{mydef}

\begin{mydef}{\normalfont \bf (Bipartite Graph)}.
A graph $G_B=(V_h, V_a, E_b)$ is called bipartite if the node set $V$ can
be partitioned into two disjoint sets $V_h$ and $V_a$, where $V = V_h \cup V_a$,
such that every edge $e \in E_b$ connects a node of $V_h$ to a node of $V_a$,
i.e., $e = (i, j) \in E \Rightarrow i \in V_h$ and $j \in V_a$. In other words,
there are no edges between nodes of the same partition.
\end{mydef}

\par Every graph $G=(V,E)$ (directed or undirected, weighted or unweighted) 
can be represented by its \textit{adjacency matrix} $\mathbf{A}$. Matrix
$\mathbf{A}$ has size $|V| \times |V|$ (or $n \times n$), where the rows
and columns represent the nodes of the graph and the entries indicate the
existence of edges.

\begin{mydef}{\normalfont \bf (Adjacency Matrix)}.
The adjacency matrix $\mathbf{A}$ of a graph $G=(V,E)$ is an $|V| \times|V|$
matrix, such that

\[
  A_{ij} = \left\{ 
  \begin{array}{l l}
    w_{ij}, & \quad \text{if $(i,j) \in E$}, ~~\forall ~ i,j \in 1, \ldots, |V|
\\
    0, & \quad \text{otherwise.}\\
  \end{array} \right.
\]
\end{mydef}

\noindent The above definition is rather general and is suitable both for
weighted and unweighted graphs. For the former case, each value $w_{ij}$
represents the weight associated with the edge $(i,j)$, while for the latter
case of unweighted graphs the
weight of each edge is equal to one (i.e., $w_{ij}=1, \forall (i,j) \in E$). If
the graph is undirected, the adjacency matrix $\mathbf{A}$ is symmetric, i.e.,
$\mathbf{A} = \mathbf{A}^T$, while for directed graphs the adjacency matrix is
nonsymmetric.

\par A basic property of the nodes in a graph is their \textit{degree}. In an
undirected graph $G_U$, a node has degree $k$ if it has $k$ incident edges. In
the case of directed graphs, every node is associated with an \textit{in-degree}
and an \textit{out-degree}. The in-degree $k^{in}_i$ of node $i \in V$ is equal
to the number of incoming edges, i.e., $k^{in}_i = \Arrowvert j|(j,i) \in E
\Arrowvert$, while the out-degree $k^{out}_i$ of node $i \in V$ equals to
the number of outgoing edges, i.e., $k^{out}_i = \Arrowvert j|(i,j) \in E
\Arrowvert$. In undirected graphs, the in-degree is
equal to the out-degree, i.e., $k_i = k^{in}_i = k^{out}_i, ~ \forall i \in V$.
The \textit{degree matrix} is defined as the diagonal $n \times n$ matrix
$\mathbf{D}$, with the degree of each node in the main diagonal (zero entries
outside main diagonal). Similarly, in directed graphs we can define the
in-degree matrix $\mathbf{D}_{in}$ and out-degree matrix $\mathbf{D}_{out}$ for
the in- and out- degrees respectively.

\par Let $G_U=(V,E)$ be an undirected graph. A \textit{path}  is
defined as a sequence of nodes $v_1,v_2,\ldots,v_{k-1},v_k$, with the property
that every consecutive pair of nodes $v_i,v_{i+1}$ in the sequence is connected
by an edge. Two nodes $i,j \in V$ are called \textit{connected} if there is
a path in $G_U$ from node $i$ to node $j$. The above definitions can be
extended to directed networks, where in a \textit{directed path}, a directed
edge should exist from each node of the sequence to the next node.

\par An undirected graph $G_U=(V,E)$ is called \textit{connected}, if for every
pair of nodes $i,j \in V$ a path exists from node $i$ to node $j$. In the
case of directed networks, three different notions of connectivity can be
defined. A directed graph  is called   \textit{strongly connected} if for every
pair of nodes $i,j \in V$, there is a directed path from $i$ to $j$ and a
directed path from $j$ to $i$. A directed graph is \textit{connected} if for
every pair of nodes $i,j \in V$, it contains a directed path from $i$ to $j$ or
from $j$ to $i$. Lastly, a directed graph is called \textit{weakly connected}
if ignoring the directionality of the edges (i.e., replacing the directed
edges with undirected), a connected graph is produced.

\par A \textit{connected component} in an undirected graph is a maximal
subgraph where every pair of nodes is connected by a path. For directed graphs,
the notions of \textit{strongly connected component} and
\textit{weakly connected component} can be defined. In the former case, similar
to the definition of strong connectivity  that we described earlier, the
edge directionality is taken into consideration, while a weakly
connected component requires the existence of a path between every pair of
nodes in the maximal subgraph without considering edge directionality.

\subsection{Linear Algebra and Spectral Graph Theory}
As we discussed earlier, every graph can be represented by a matrix, the
so-called \textit{adjacency matrix}. The adjacency matrix $\mathbf{A}$ of a
graph $G=(V,E)$ is the $|V| \times |V|$  matrix  with elements $A_{ij}=1$ if 
there exist an edge between nodes $i,j$ in the graph. In the general case where
the edges of the graph contain weights, the entries of the weighted adjacency
matrix correspond to edge weights. For undirected graphs, the adjacency matrix
$\mathbf{A}$ is symmetric (i.e., $\mathbf{A} = \mathbf{A}^T$), while for
directed graphs the matrix is nonsymmetric.

\par Let $\mathbf{A} \in \mathbb{R}^{n \times n}$ be a symmetric matrix. Then,
$\mathbf{A}$ can be written as $\mathbf{A} = \mathbf{U} \mathbf{\Lambda}
\mathbf{U}^T$, where the orthogonal matrix $\mathbf{U}$ contains as columns the
\textit{eigenvectors} $u_1, u_2, \ldots, u_n$ of $\mathbf{A}$, correspond to
real \textit{eigenvalues} $\lambda_1 \ge \lambda_2 \ge \ldots \ge \lambda_n$
and $\mathbf{\Lambda} = \text{\texttt{diag}}(\lambda_1, \lambda_2, \ldots,
\lambda_n)$ the diagonal matrix with the eigenvalues as entries
\cite{golub1996, strang, graph-spectra-book-2011}. The eigenvalues of the
adjacency matrix define the \textit{spectrum} of a graph and have close
connections with several important graph properties. As we stated above, in the
case of directed graphs the corresponding adjacency matrix is nonsymmetric
and therefore the eigenvalues can be complex. Thus, it is preferable to
work with the singular values of the matrix which can be extracted by the
\textit{singular
value decomposition} (SVD). That is, the SVD of a real  matrix
$\mathbf{A} \in \mathbb{R}^{m \times n}$ is defined as $\mathbf{A} = \mathbf{U}
\mathbf{\Sigma} \mathbf{V}^T$, where $\mathbf{U} \in \mathbb{R}^{m \times m}$
and $\mathbf{V} \in \mathbb{R}^{n \times n}$ contain the left-singular and
right-singular vectors respectively and $\mathbf{\Sigma} =
\texttt{diag}(\sigma_1, \sigma_2, \ldots, \sigma_p) \in \mathbb{R}^{m \times
n}, ~ p = \min\{m,n\}$, the diagonal matrix comprised of singular values
$\sigma_i$ (note that, for symmetric matrices, the singular values correspond to
the absolute values of the eigenvalues).

\par Another matrix commonly used to represent a graph is the \textit{Laplacian}
matrix. 

\begin{mydef}{\normalfont \bf (Laplacian Matrix)}.
In the case of undirected graphs, the Laplacian matrix is defined as 

\begin{equation}
 L_{ij} =
  \begin{cases}
   k_i, & \text{if } i = j, \\
   - 1,       & \text{if } i \text{ and } j \text{ are adjacent}, \\
   0, & \text{otherwise},
  \end{cases}
\end{equation}

\noindent where $k_i$ is the degree of node $i$. In a more compact
form, the Laplacian matrix can be written as $\mathbf{L} = \mathbf{D} -
\mathbf{A}$, where $\mathbf{A}$ is the adjacency matrix of the graph and
$\mathbf{D} = \text{\normalfont\texttt{diag}}(k_1, k_2, \ldots, k_n)$ the
diagonal degree matrix \cite{chung}. 
\end{mydef}

\noindent The \textit{normalized Laplacian matrix} $\mathbf{L_n}$ is symmetric
and defined as $\mathbf{L_n} = \mathbf{D}^{-1/2} \mathbf{L}
\mathbf{D}^{-1/2}$. In other words,
if two nodes $i,j$ are adjacent, the entry $L_{n_{ij}}$ is equal to
$-\dfrac{1}{\sqrt{k_ik_j}}$. The spectrum of the normalized Laplacian matrix 
$\lambda_0 \le \lambda_1 \le \ldots \le \lambda_{n-1}$, presents some
interesting properties:

\begin{itemize}
 \item All eigenvalues are non-negative. Moreover, $0$ is an eigenvalue of
$\mathbf{L_n}$.
 \item The number of eigenvalues with value $0$ corresponds to the number of
connected components in the graph.
 \item The smallest non-zero eigenvalue is called \textit{spectral gap} and the
corresponding eigenvector is used for the task of spectral clustering (e.g.,
see Ref. \cite{von-luxburg}).
\end{itemize}

\noindent A basic difference between the spectrum of the Laplacian matrix and
the one of normalized Laplacian is that in the former case the eigenvalues 
belong to the range $0 = \lambda_0 \le  \lambda_i \le 2k^{max}$, where
$k^{max}$ is  the maximum degree in the graph, while in latter case the
eigenvalues always lying in the range $0 = \lambda_0 \le \lambda_i \le 2$. For a
detailed presentation of the Laplacian matrix and its properties, one can refer
to the Spectral Graph Theory textbook by Chung \cite{chung}.

\par Chung \cite{chung-directed-laplacian05} defined the Laplacian matrix for
directed graphs, showing interesting connections of its spectrum with the mixing
rate of random walks. Very recently, Li and Zhang \cite{diplacian-waw10,
diplacian} proposed another generalization of
the Laplacian matrix, establishing novel perspectives and results for the case
of directed graphs. In Section \ref{sec:spectral} we discuss in detail about the
definition of the Laplacian matrix in directed graphs and how can be used to
solve the clustering problem.

\subsection{Random Walks on Graphs}
Generally, a random walk is a mathematical concept formalizing a procedure
consisting of a sequence of random steps. In the case of graphs, given a node
that corresponds to a starting point, a \textit{random walk} is defined as  the
sequence of nodes formed by a repeating process starting from the
initial node and  randomly moving to neighborhood nodes. In other words, at
each step the random walker is situated on a node of the graph and jumps to a
new node selected randomly and uniformly among its neighbors.

\par More precisely, let $G_U=(V,E)$ be an undirected graph and $v_0$ be the
starting node of the random walk. Let us suppose that at the $t$-th step, the
random walk is situated at node $i$. At $t+1$ step, the random walk is
moving from node $i$ to node $j$ (neighbor of $i$) with transition probability
$\dfrac{1}{k_i}$. This defines the \textit{transition matrix} $\mathbf{P}$ of
the random walk as

\begin{equation}
 P_{ij} =
  \begin{cases}
   \dfrac{A_{ij}}{k_i}, & \text{if } (i,j) \in E, \\
   0, & \text{otherwise}.
  \end{cases}
\end{equation}

\noindent In a compact form, this matrix can be written as $\mathbf{P} =
\mathbf{D}^{-1} \mathbf{A}$, where $\mathbf{D}^{-1}$ is the inverse of the
diagonal degree matrix $\mathbf{D}$. This matrix can also be considered as a
degree normalized version of the adjacency matrix. In the general case, random
walks are considered to be \textit{Markov chains}\footnote{Wikipedia's lemma for
\textit{Markov chain:} \url{http://en.wikipedia.org/wiki/Markov_chain}.}, where
the set of possible states corresponds to the vertex set of the graph.

\par Any distribution on a graph $G$ can be represented by a row vector
$\boldsymbol{\pi}=[\pi_1, \cdots, \pi_n]^T$, where the $i$-th entry captures the
amount of the distribution resides at node $i$. In the case of random walks, the
probability distribution over the graph $G$ for each  node $i \in V$ at any time
step, gives the probability of the random walk of being at node $i$. Thus, if
$\boldsymbol{\pi}$ is the initial distribution, then
$\boldsymbol{\pi}_1 = \boldsymbol{\pi} \mathbf{P}$ is the distribution after one
step and $\boldsymbol{\pi}_t = \boldsymbol{\pi} \mathbf{P}^t$ is the
distribution after $t$ steps. Based on the above idea, we can define a
\textit{stationary distribution} $\boldsymbol{\pi}_s$, as the distribution
where $\boldsymbol{\pi}_s = \boldsymbol{\pi}_s \mathbf{P}^t, \forall t$. In
other words, the stationary distribution corresponds to a distribution that does
not change over time and describes the probability that the walk is being at a
specific node after a sufficiently long time. The \textit{mixing time} is
the time needed by the random walk to reach its stationary
distribution. The spectrum of the transition matrix $\mathbf{P}$ can be used to
bound the mixing time of a random walk in a graph, and specifically the second
largest eigenvalue \cite{sinclair-mixing-time}. In a similar manner, random
walks can be defined over directed graphs. However, in this case, two main
difficulties can occur: (i) at some time point, the random walker can be
situated in a node with no outgoing edges, and (ii) nodes with no incoming edges
will never be reached. As we will present later, the PageRank algorithm is a
random walk process on directed graphs that overcomes the above problems
\cite{pagerank}.

\par The theoretical tool of random walks is closely related to the problem of
clustering and community detection in graphs (e.g., Ref.
\cite{pons-latapy-2006}). For example, it is known that matrix $\mathbf{P}$ has
always the largest eigenvalue equal to one. In the case of networks with very
clear community structure, matrix $\mathbf{P}$ will also have $c-1$ eigenvalues
close to one, where $c$ is the number of well-defined modules (clusters) in the
network; the rest of the eigenvalues will be relatively away from one. The
eigenvectors correspond to the first eigenvalues can be used to extract
the clustering structure: for nodes that belong on the same clusters, their
components in the eigenvectors will have similar values, following a step-wise
form. The number of steps, corresponds to the number of clusters $c$. In Section
\ref{sec:random-walk} we provide more
details on this issue and we present random walk based techniques
for the case of directed networks.
 For a more detailed discussion on various
aspects of random walks, the reader can refer to Refs. \cite{lovasz1993random,
chung}.

\subsection{Quality Measures}
As we have already mentioned, a cluster or community in a network is typically
considered as a group of nodes with better connectivity (and/or stronger
interactions) among its members than with the nodes of different communities.
Usually, the process of detecting communities in networks follows a two step
approach: 

\begin{itemize}
 \item[(i)] First, a quality measure (or objective function) needs to be
specified, that captures the notion of community structure as groups of nodes
with better internal connectivity than external (or more generally, an objective
criterion which quantifies the desired properties of a community).
 \item[(ii)] Then, using algorithmic techniques, the nodes of the network
are assigned to specific communities, optimizing the objective
function. Since the optimization process of the objective functions 
typically leads to computational difficult problems (e.g., see Ref.
\cite{schaeffer-review}), a common approach is to employ heuristics
or other approximation techniques.
\end{itemize}

\noindent In the literature  several measures have been proposed for
quantifying the quality of communities in networks (most of them have been
introduced for the case of undirected networks, but some of them have been also
extended to directed ones; in Section \ref{sec:extending-techniques} we
describe some of these measures in detail and we also present their extensions
to directed networks). Typically, some of the quality measures focus on both
the intra-cluster and inter-cluster edge density (\textit{multi-criterion}
scores), such as normalized cuts \cite{shi-malik-pami00}, conductance and
expansion. Other measures focus only in one of them (\textit{single-criterion}
scores) and a well-known representative of this category is modularity (e.g.,
see Refs. \cite{fortunato, schaeffer-review} and Ref. \cite{leskovec-www10} for
a recent comparative study of quality measures in undirected large scale
networks).

\par \textit{Modularity} \cite{newman-newman-phys-rev-e-04, newman-modularity}
is one of the most popular and widely used metrics to evaluate the quality of
network's partition into communities. Considering a specific partition of the
network into clusters, modularity measures the number of edges that lie within
a cluster compared to the expected number of edges of a null graph (or
configuration model), i.e., a random graph with the same degree distribution.
In other words, the measure of modularity is built upon the idea that random
graphs are not expected to present inherent community structure; thus,
comparing the observed density of a subgraph with the expected density of the
same subgraph in case where edges are placed randomly, leads to a method
for identifying clusters. More precisely, the
modularity value $Q_u$ of a specific partition of an undirected network into
communities is defined as follows

\begin{equation}
 Q_u = \dfrac{1}{2m} \sum_{i,j} \bigg[A_{ij} - \dfrac{k_i k_j}{2m}
\bigg] \delta(c_i, c_j),
\end{equation}

\noindent where $\mathbf{A}$ is the adjacency matrix, $c_i, \forall i \in V$
is the community membership of node $i$ and $\delta(c_i, c_j)=1$ if
$c_i=c_j$ (i.e., if nodes $i,j$ belong to the same community) and $0$
otherwise. The modularity value can be either positive or negative. Higher
positive values indicate better community structure properties and therefore,
finding the partition that maximizes the modularity provides a method for
extracting the underlying community structure. More details on modularity as
well as its extension to directed networks are presented in Section
\ref{sec:modularity}. 

\par Optimizing the modularity function is a computational difficult
task \cite{modularity-completeness}; however several heuristics and
approximation techniques have been proposed. Newman \cite{newman03fast} proposed
a greedy search algorithm for the problem of modularity maximization. Initially,
every node of the graph belongs on its own community; then, iteratively, pairs
of communities are joined  on the same group if they achieve the highest
increase of the modularity value. Thus, the algorithm can be considered as an
agglomerative hierarchical clustering method (e.g., see Ref.
\cite{jain-clustering}) and the whole procedure can be represented by a
dendrogram.  Clauset et al.
\cite{clauset-fast-modularity} presented a faster greedy algorithm (almost
linear time in sparse graphs), based again on an hierarchical clustering
approach. Other well known methods for modularity optimization are the ones
rely on spectral techniques. For example, Newman \cite{newman-modularity} showed
that the measure of modularity can be expressed in terms of the spectrum of a
specific matrix associated to the network (called \textit{modularity matrix}),
and therefore spectral techniques can be applied in the optimization process. In
Section \ref{sec:modularity} we present a similar approach for the case of
directed networks. Other methods
apply simulated annealing techniques \cite{simulated-annealing} (e.g., Refs.
\cite{guimera-simulated-annealing-2004, guimera-nature-2005}) and extremal
optimization \cite{duch-arenas-phys-rev-e-2005}. For a more detailed
presentation of modularity optimization techniques, the reader can refer to the
survey paper of Fortunato \cite{fortunato}. Similar optimization approaches
can be also applied to directed networks.

\par However, as noted by Fortunato and Barth\'{e}lemy
\cite{modularity-resolution},
modularity suffers from the so-called \textit{resolution limit}. That is,
modularity optimization may fail to detect communities with size smaller
than a scale which mainly depends on the size of the network. This point is
particularly significant since typically real world networks contain communities
of various sizes.

\section{Clusters in Directed Networks -- Intuition and Discussion}
\label{sec:problem-stm}
In this section we introduce the notion of clusters or communities in directed
networks and we discuss about their structural properties. We provide different
intuitive definitions regarding the properties of clusters, that will enable the
reader to better comprehend the notion of clustering in directed networks and
subsequently to classify the clustering methods according to the definition
given. It is important to stress out here that there is no well defined
definition for the graph clustering problem, both in the directed and 
undirected cases. Actually the formulation depends either on the application
domain or generally on the type of clusters we are interesting in. Nevertheless,
regardless of the problem definition, the ultimate goal of the clustering task
remains the same: the graph nodes should be assigned to clusters, with
``similar'' nodes belonging to the same cluster.

\par Let us now present a high level definition of a cluster or community in
networks which can be considered as a generic definition for the clustering
task. Later, we capitalize on this to capture and describe different possible
clustering structures for  directed networks, as they have been proposed in the
literature.

\begin{mydef}[\textbf{High Level Definition of a Cluster}] \label{def:clusters}
 A cluster or community in a network can be considered as a set of nodes that
share common or similar features (characteristics).
\end{mydef}

\par In this generic definition, there are two things that need to be specified:
(a) the notion of similarity among the nodes of a directed network and (b) the
features/characteristics we are interesting in. We consider that
specifying these two elements, we are able to capture and describe all 
possible clustering notions in directed networks. In order to become more
precise, let us consider as example the traditional definition of clusters as
modules with dense connections between the nodes of the same cluster but sparser
connections between different clusters \cite{newman-modularity,
leicht-newman-2008}. According to the above definition, the features correspond
to graph's edges, while the number of edges between a set of nodes (density) can
be considered as a similarity node indicator in the graph.

\par Having defined the desired properties of the clustering structure, then
the process of detecting communities is a two step approach:
first, we should specify an objective function that captures the notion of
clustering structure, according to the chosen definition. Then,
using algorithmic techniques, the nodes of the network are assigned to specific
clusters, optimizing the objective function. In Section \ref{sec:background}, we
gave a brief introduction about objective functions for undirected networks.
Later at this paper, we will describe such objective criteria for directed
networks, as well as approaches for detecting the community structure.

\par Next, we present the two main notions-definitions (or categories) for
clusters in directed networks: 

\begin{itemize}
 \item[(a)] \textit{Density-based} clusters, i.e., groups of nodes that follow
the traditional clustering definition based on edge density characteristics.

 \item[(b)] \textit{Pattern-based} clusters, i.e., groups of nodes that go
beyond edge density patterns. As we will describe shortly, an example of
this category is the case of flow circulation, where a pattern of movement
induced within the nodes of the cluster.
\end{itemize}

\noindent This classification can also be extended to undirected networks, as
some research works propose \cite{coscia-review, ljubljana-communities-ejpb12,
communities-icdm07}). Then the question rising is: which clustering definition
for directed networks should one adopt? The answer highly depends on
the application domain and on the features of the network dataset under
consideration. Such features may include the nature of interactions among
graph's nodes (as captured by the edges) and  prior knowledge of the underlying
structure. In Section \ref{sec:applications} we present a list with real
applications of directed graph clustering, along with the most suitable
clustering definition.

\subsection{Density-based Clusters}  \label{sec:density-based}
\begin{figure}[t]
\centering
 \begin{tabular}{ccc}
  \includegraphics[width=.4\textwidth]{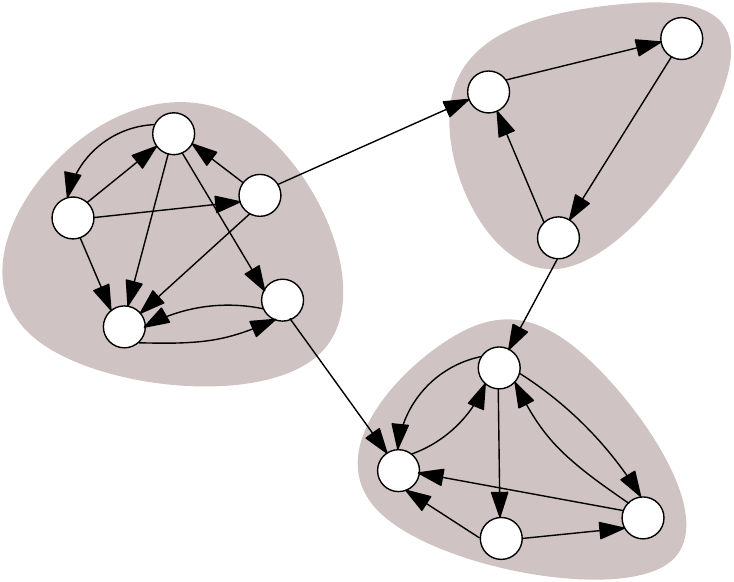} & ~~ &
  \includegraphics[width=.4\textwidth]{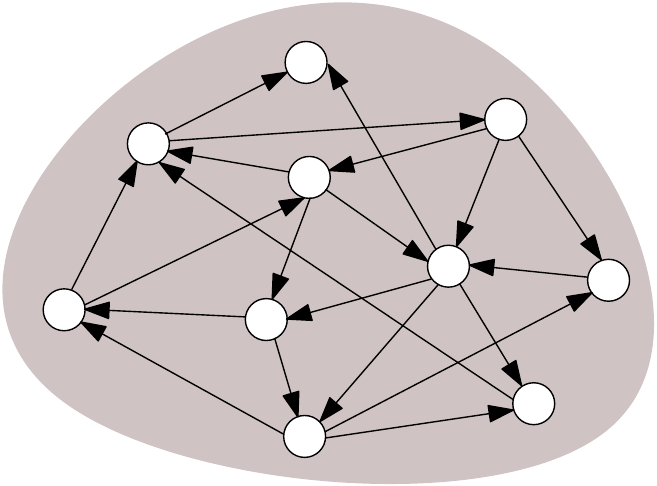} \\
   (a) Graph with density-based clusters & ~~ &  (b) Graph with uniform
structure
 \end{tabular}
 \caption{Two directed graph examples. The left one (a) consists of three
density-based clusters, while the right one shows a homogeneous link density
with the absence of obvious community structure. \label{cluster}}
\end{figure}

We consider \textit{density-based clusters}, that can be regarded as the
more traditional definition of communities/clusters in both directed and
undirected networks, and also the most well studied in the research
community (e.g., Refs. \cite{newman-newman-phys-rev-e-04, girvan-newman,
newman-modularity, leicht-newman-2008}). The notion of density-based clusters
is entirely based on the distribution and topology of the edges inside the
network. As we mentioned earlier, according to this definition, a
cluster in  directed networks is defined as a group of nodes with more
intra-cluster than inter-cluster edges. Figure \ref{cluster} (a) depicts
an example of a directed network which contains three well defined density-based
clusters. It is clear that the edge density within each cluster (shadowed
regions) is much larger than the density between different clusters. On the
other hand, Fig. \ref{cluster} (b) presents a homogeneous directed network with
uniform edge distribution among nodes. The network lacks  a modular
organization and thus there is no obvious density-based community structure.
Based on this definition, the goal of a graph clustering algorithm is to assign
the nodes into clusters, maximizing the number of edges within clusters, while
minimizing the inter-cluster edges. As we will see later, there are several
popular density-based graph clustering techniques, that either trying to
maximize the internal cluster density, either minimize the number of
extra-cluster edges or both of them.

\par The above notion of clustering in directed networks can be
considered as a natural extension from the graph clustering problem in 
undirected networks (e.g., Refs. \cite{newman-newman-phys-rev-e-04,
fortunato, schaeffer-review, coscia-review}). In the next sections of the paper
we will see that some  techniques, initially introduced for undirected networks,
form the basis for dealing with the directed graph clustering problem.

\par Moreover, for the undirected case, the density-based definition has close
connections with the well-known \textit{graph partitioning} problem in the
field of computer science (e.g., Ref. \cite{metis}). However, there are two main
differences between them: (a) in the graph partitioning problem, the
desired number of partitions (or clusters) $k$ is a parameter of the problem and
needs to be specified a priori, while in the case of graph clustering and
community detection problems this is not always prerequisite, and (b) the goal
of the partitioning problem is to equally assign nodes in the different
partitions, where the size of each cluster will be approximately equal to
$\dfrac{n}{k}$ \cite{fortunato}. On the other hand, in the clustering problem,
the distribution of the clusters' sizes may not be uniform.

\par It is important to note here that extending the notion of density-based
clusters to directed networks is not always a trivial procedure. While some of
the proposed objective measures for the undirected case can be easily extended
to directed graphs by considering in a meaningful way the directionality of the
edges (e.g., the criterion of modularity \cite{leicht-newman-2008}), due to the
existence of directed edges, some of the desired cluster properties
may not hold. Even worse, some graph-theoretic measures and concepts that
help us to evaluate the quality of density-based clusters cannot be easily
extended and defined in the directed case. For example, as pointed out by
Schaeffer \cite{schaeffer-review}, each cluster in a graph should be
\textit{connected} (i.e., there should be at least one path between every pair
of nodes in the graph). As mentioned in Section \ref{sec:background},
in directed graphs, the connectivity property can be expressed in three forms:
weak connectivity, connectivity and strong connectivity. Depending on the
required cluster properties, any of the above criteria can be adopted in the
definition of density-based clusters for directed networks. This is just an
indication where simple graph concepts, such as connectivity,  become complex
when edge directionality is taken into consideration.

\subsection{Pattern-based Clusters} \label{sec:pattern-based}
Previously we presented the notion of 
density-based clusters that constitute the major type of clustering structure in
directed networks. Although this represents the most common and well-studied
clustering definition
in both directed and undirected networks, it cannot capture more sophisticated
clustering structures, than the classical well-cohesive groups, where edge
density may not represent the major clustering criterion. More precisely, the
nodes of a directed network can be naturally clustered together according to
similar connectivity patterns that may exist and are not captured completely
applying only density criteria. Actually, in some cases, two or more nodes can
belong to the same cluster even though they are not directly connected by common
edges. We refer to this category of clusters as \textit{pattern-based clusters},
since they represent structures with interesting connectivity  properties in
directed networks\footnote{In the case of undirected networks, similar terms and
definitions have been proposed for characterizing clusters with
interesting connectedness patterns (e.g.,
Refs. \cite{ljubljana-communities-ejpb12, communities-icdm07, coscia-review}).
Here, we extend this notion by considering clustering types that inherently
arise in several real-world directed networks due to the existence of
non-reciprocal relationships.}. 

\par Examples of patterns that are interesting for clustering directed graphs
are the cases of co-citation and flow. Co-citation implies that a set of
nodes $A$ links to a set of nodes $B$ and this structure implies a similarity
among the members of each group -- i.e., the members of $A$ are similar among
them as they all point to the same nodes, group $B$. Another interesting pattern
has to do with the network flow at the cluster level, i.e., the linking
structure within a cluster forces that the flow through the links predominantly
stays within the cluster instead of pouring out of it.

\begin{figure}[t]
\centering
 \begin{tabular}{ccc}
  \includegraphics[width=.2\textwidth]{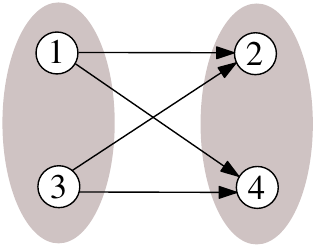} & 
  \includegraphics[width=.25\textwidth]{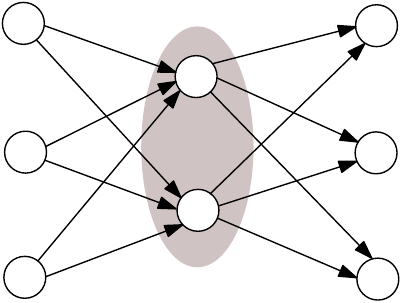} & 
  \includegraphics[width=.3\textwidth]{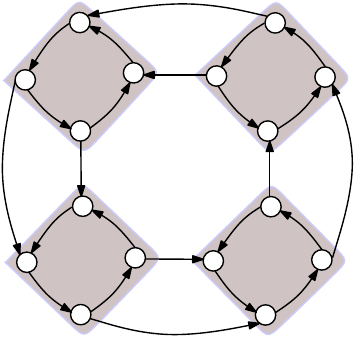} \\
   (a) Citation-based cluster &  (b) Citation-based cluster & 
   (c) Flow-based cluster \\
 \end{tabular}
 \caption{Different examples of pattern-based clusters in directed networks. The
leftmost network (a) and the one in the center (b) represent citation-based
clusters. The graph on the right (c) depicts a graph with four flow-based
clusters. Figures redesigned from Refs. \cite{srini-edbd11, rosvall-pnas08}.
\label{fig:pattern-based-figure}}
\end{figure}

Figure \ref{fig:pattern-based-figure} depicts three cases of graphs that
contain different types of pattern-based clusters (as shown in the shadowed regions).
The first graph (a) forms two clusters -- we will refer on them
as citation-based clusters. The most interesting point in this case is that the
nodes of the graph that are clustered together do not have an edge between
them. Their similarity emanates from the co citation event -- i.e., the
nodes of the leftmost cluster point to the nodes of the right cluster.
Respectively the two nodes of the right cluster are pointed by the same group of
nodes. This is actually a bipartite graph where the partitions represent two
different communities. For example, let us consider the case of a citation
network where nodes correspond to scientific papers and a directed edge from
paper $1$ to paper $2$ implies that the first paper cites the latter. 
Although papers $1$ and $3$ do not share an edge, they form a natural cluster since they
both cite papers $2$ and $4$ and it is probable that they belong on the same
scientific topic. 

\par A similar example of  pattern-based clusters, appears in Fig.
\ref{fig:pattern-based-figure} (b). In this case the two nodes in the shadowed
region form a cluster, since they have out-links to the same nodes, while at
the same time having in-links from the same group of nodes. This structure
constitutes a common situation in the context of directed graphs.
For example, these two nodes may correspond to the websites of two competing
companies of the same market sector; they both link and are linked by a common
group of webpages, but actually they do not have links among them due to
competition \cite{srini-edbd11}. According to our high level definition of clusters
in directed networks (Definition \ref{def:clusters}), in this case the
clustering features correspond to the common neighbors in the graph and thus the
nodes are clustered together if the share common neighbors.

\par A different case of pattern-based clusters is the one presented in Fig.
\ref{fig:pattern-based-figure} (c). The main characteristic of this network is
that the edges form patterns of flow among nodes. In other words, the local
interactions in the network combined with the edge directionality, induce  a
flow of information among the entities and therefore the clustering structure
depends on how information flows (see Ref. \cite{rosvall-pnas08} by Rosvall and
Bergstrom). Then, a cluster or community in the network corresponds to a group
of nodes where the flow is larger (more persistent) as compared to the flow
outside the group. Assuming a user that conducts random walk on the graph, a
flow-based community is a group of nodes where a random surfer is more likely 
to be trapped inside instead of moving out of the group
\cite{linkrank-phys-rev10, lai-physica10}.

\paragraph{Remark} We should note here that both types of clusters --
co-citation and flow based ones -- may co-exist in a directed network. For
example, as we can
see from the related literature, most of the techniques that adopt the
citation-based clustering rule, are also able to identify density-based
clusters. However, the novelty of these techniques resides exactly on this point;
through appropriate transformations, a density-based technique can be enhanced
with pattern-based clustering features.

\section{Dealing with Edge Directionality: Approaches for Identifying
Clusters-Communities in Directed Networks} \label{sec:edge-dir}

There have been different directed network clustering approaches depending on 
the way directed edges are treated. In this section we review the clustering
methods based on the methodological principles and algorithmic approaches
followed. To the best of our knowledge this is the first proposal for a
classification scheme of graph clustering methods for directed networks.

\par Since a large amount of work for the graph clustering problem in directed
networks is built upon clustering approaches for undirected ones, whenever
necessary we review  the basic concepts to make our presentation self contained.
We also consider that giving meaningful connections to the undirected case  will
be helpful to the reader. For a detailed description of the undirected graph
clustering methods, the reader may refer to previous interesting surveys in the
field (see Section \ref{sec:rel_surveys}). 

\par The proposed classification follows:

\begin{itemize}
 \item \textit{Naive graph transformation approach:} in this class we classify 
algorithms that ignore edge directionality and treat graph as undirected. Thus,
clustering algorithms that have been proposed for undirected networks can be
also applied to reveal the
underlying community structure of directed ones (e.g., see Section
\ref{sec:rel_surveys}). However, due to naive graph conversion, the
underlying graph semantics are not retained and useful information is not taken
into consideration during the clustering task. For instance, consider a citation
network, where papers are represented by nodes and edges are the citations.
Then assume a 
paper $i$ cites paper $j$ but not vice versa. Using the naive graph
transformation each directed edge is replaced by an undirected one; thus a
reciprocal relationship is introduced among papers $i$ and $j$ which misses to
represent the  endorsement of paper $i$ to paper $j$.

 \item \textit{Transformations maintaining directionality:} In this class we
have methods where the directed graph is converted into 
an undirected one, either unipartite or bipartite, and
edge direction is meaningfully maintained in the produced
network. For example, in some approaches the directed network is converted
into an undirected and weighted one, where information about directionality is
introduced via weights on the edges of the graph (e.g., Ref.
\cite{srini-edbd11}). Then, algorithms and tools for
clustering undirected weighted graphs can be applied. In other 
approaches, the directed network is converted into a bipartite one and then
appropriate clustering algorithms are applied to the bipartite graph (e.g., Ref.
\cite{zhou-nips05}).

 \item \textit{Extending clustering objective functions and methodologies to 
 directed networks:} this category includes approaches that constitute
extensions of
methodologies from the undirected case. Thus objective criteria are extended to
meet the requirements of the
problem. The graph clustering problem is typically
expressed as an optimization problem, where an objective criterion, capturing
the desired cluster properties, is optimized by re-assigning nodes into
clusters. The algorithm iterates usually until a local min/max value is
reached. 
Since most of the research literature has focused on the undirected version of
problem, a large bulk of interesting approaches regarding the properties and
functionality of the (undirected) objective criteria have been presented (e.g.,
modularity \cite{newman-newman-phys-rev-e-04, newman-modularity} and normalized
cut \cite{shi-malik-pami00}). Hence, a natural way to deal with the directed
version of the graph clustering problem is to extend these measures for directed
networks, where edge directionality is considered as an inherent network
characteristic. Some prominent representatives of this category are the directed
versions of modularity \cite{leicht-newman-2008,
linkrank-phys-rev10, arenas-modularity07} and the objective function of weighted
cuts in directed graphs \cite{meila-sdm07}. 
Similarly, another well established 
approach in this direction is the extension of algorithmic tools that have been
introduced for undirected graphs. One of the most well-known such approaches is
the case of spectral graph clustering based on the Laplacian matrix
\cite{von-luxburg, spectral-clustering-survey-ejor11}. While Laplacian based
spectral clustering
methods initially applied on undirected networks, recent advances
in the field make them also applicable to directed ones (e.g.,
Refs. \cite{chung-directed-laplacian05, zhou-icml05}).

 \item \textit{Alternative approaches:} this category includes approaches that
follow diverse methodologies,  mainly different from the ones described in the
previous three categories. We identify three major types of methods,
namely (i) information-theoretic, (ii)  methods based on probabilistic models
and statistical inference, and (iii) stochastic blockmodeling methods. Even
though the last two methodologies are closely related and both refer
to probabilistic models, we review them independently since they rely on
different statistical inference concepts. Furthermore, in this category we also
review several additional approaches that mostly concern some variations of the
community detection problem in directed networks. Some of these  pose
interesting features and may constitute interesting extensions for future
research work in the field (e.g., community detection in dynamic directed
networks).

\end{itemize}

\par
\begin{figure}
\centering
  \includegraphics[width=.95\textwidth]{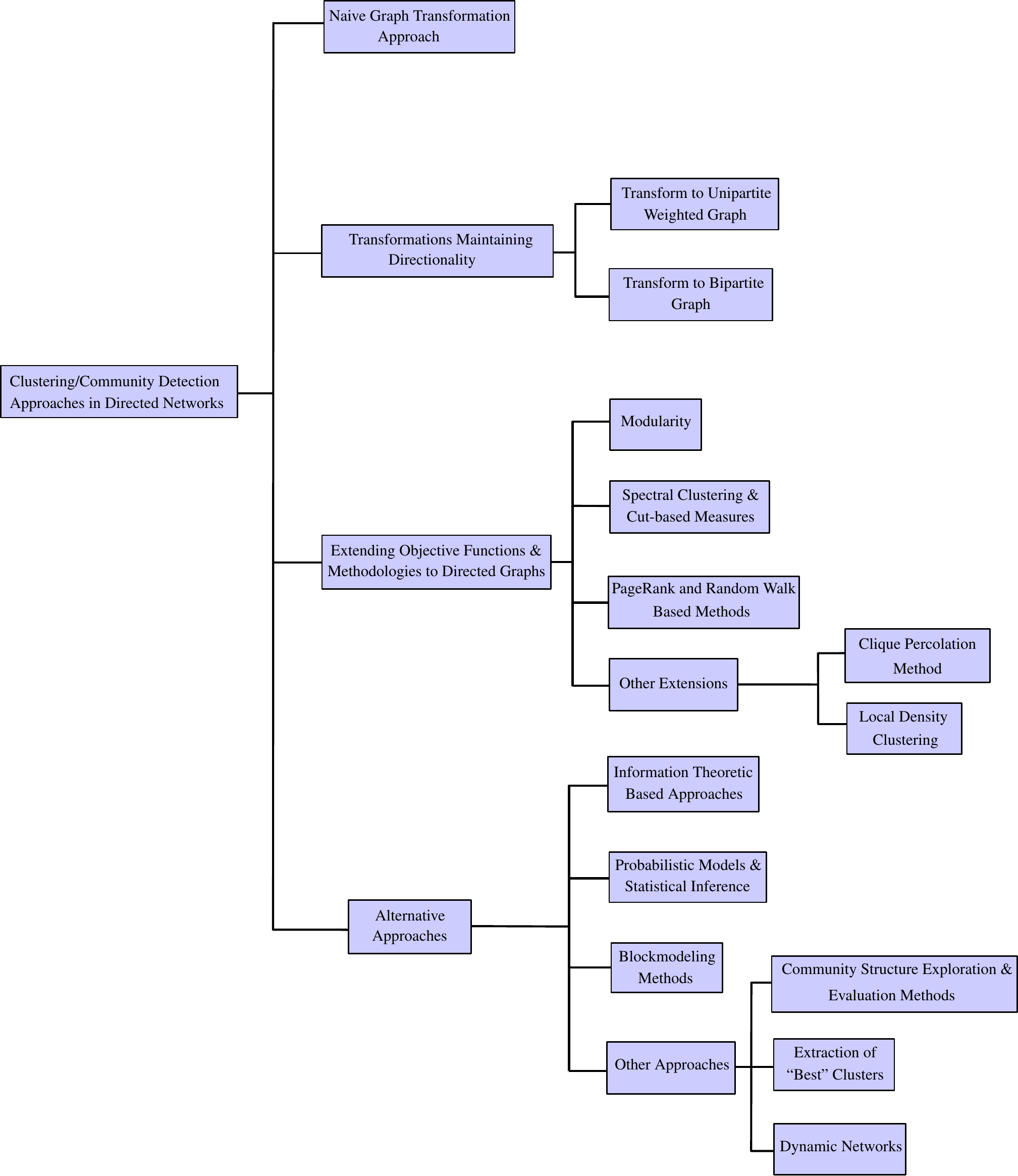}
 \caption{The proposed taxonomy of clustering/community detection
approaches in directed networks. \label{fig:taxonomy}}
\end{figure}

 Next we will elaborate on each of the above categories with more details,
presenting their basic points and classifying the related works. This
is the main proposed  classification scheme for the clustering problem in
directed networks. Figure \ref{fig:taxonomy} depicts schematically the 
proposed taxonomy of the different approaches for the problem.  A large amount of
research work  has been devoted to methods that belong on the third 
category (extending methodologies to directed networks). We also note here
that, some of the approaches  share diverse methodological
features (e.g., methods that transform the graph to undirected but at the same
time propose extensions for the modularity objective criterion that is in
accordance with their framework). Then, we do not assign them crisply in
only one category; whenever necessary, we follow an overlapping classification
trying to capture and present the features of each approach from all
possible viewpoints. Lastly, we will shortly
discuss why the naive approach is not appropriate for dealing with directed
networks.

\subsection{Naive Graph Transformation Approach}
The first and simplest approach for the clustering problem in directed networks,
is to discard edge directionality and treat graphs as undirected. After this
simple transformation step, a large bulk of methods that have been proposed for
undirected networks can be applied to extract the community structure (e.g.,
see Section \ref{sec:rel_surveys} for details).  Even though this is a common
way in the related literature to deal with directed networks, this approach has
several drawbacks that mainly derive from the fact that the information
represented by edges' direction is ignored and not utilized during
the clustering process. In other words, directed edges typically indicate the
existence of non reciprocal relationships between entities represented by the
nodes of the graph, and thus a naive transformation to an undirected graph with
symmetric relationships does not retain the underlying semantics. The two main
drawbacks of this approach can be summarized as follows:

\begin{itemize}
 \item[(a)] \textit{Data ambiguities:} the naive graph transformation
introduce ambiguities and to some degree incorrect
information in the network. For example, let us consider the case of a
citation network where a directed edge $(i, j)$  represents a citation from
paper $i$ to $j$. Converting this edge into an undirected one implies a mutual
relation, i.e. an edge $(j,i)$  that does not exist in the graph. But more
generally, even if someone  argue that the new undirected edge represents 
similarity among papers $i$ and $j$ (since the first one cites the second), 
this does not always hold for both directions (i.e., paper $j$ may be an
important paper, but in a different area; thus mutual relationship and
similarity may not exist). Similar concerns could be used to justify possible
ambiguities introduced by this transformation approach in other domains.

\begin{figure}[t]
\centering
 \begin{tabular}{ccc}
  \includegraphics[width=.3\textwidth]{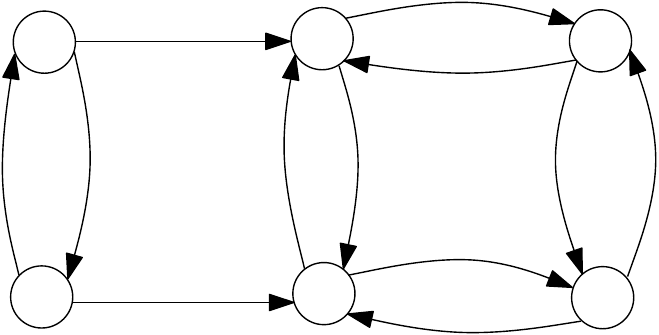} & ~~ &
  \includegraphics[width=.3\textwidth]{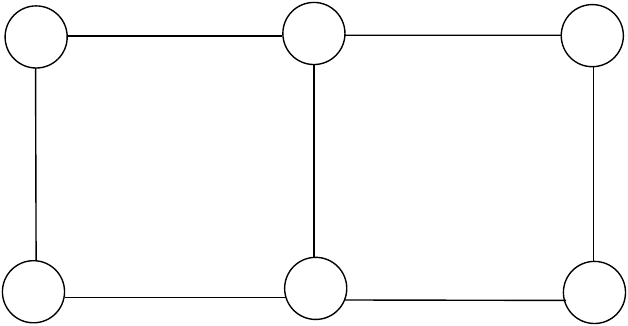} \\
   (a) Directed network & ~~ &  (b) Undirected network after transformation
 \end{tabular}
 \caption{Example of a naive graph transformation. The  directed
network (a) contains two communities, since the two leftmost nodes are
connected to the rest of the network but only in one direction. After a naive
graph transformation (b) it is difficult to identify any community structure.
\label{fig:dir_to_undir_naive}}
\end{figure}

\item[(b)] \textit{Deviations in clustering results:} even if one could ignore 
the ambiguities introduced in the data by the naive graph transformation
approach, these may have impact to the final outcome of a clustering algorithm.
Discarding edge directionality, valuable information is not utilized at the
clustering process, which at the end leads to deviations at the results. In
other words, clusters that exist in the initial directed network may not be
identified at the transformed one, due to the naive graph conversion process.
This mainly occurs because the existence of directed edges forms interesting
structural patterns and clusters (e.g., the flow-based clusters
that we described in Section \ref{sec:pattern-based}) that cannot be found in
undirected networks. Of course, this is something  closely related to the
definition/notion of a cluster; nevertheless, even for the same clustering
definition, the approach may lead to different results at the end. Figure
\ref{fig:dir_to_undir_naive} depicts an example of a directed graph, where a
naive transformation could distort the clustering results.
\end{itemize}

\par It becomes clear that this approach is not effective and could possibly
lead to incorrect inference about the underlying community structure. In
the following section, we will describe more meaningful graph transformation
approaches, where information about edge directionality is incorporated in the
final network and utilized properly during the clustering process.

\subsection{Transformations Maintaining Directionality}
In this section we will review the second category of approaches for the
clustering  problem in directed networks. More precisely, we
will present methods that perform  ``meaningful'' transformations of the
directed network into an undirected one, where the term meaningful is used
to denote the difference to the previously described naive transformation
method. According to this approach, the basic components of the clustering
task in directed networks can be summarized as follows:

\begin{enumerate}
 \item Transform the directed network to  undirected.
 \item Edges' direction information should be retained as much as possible
(e.g., by introducing  weights on the edges of the transformed network).
 \item Apply already proposed clustering algorithms designed for undirected
networks.
 \item The extracted communities will also correspond to the communities of the
initial directed network.
\end{enumerate}

\par
\begin{figure}[t]
\centering
  \includegraphics[width=.9\textwidth]{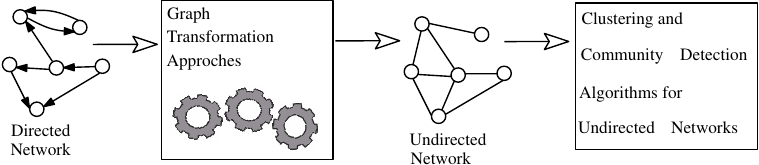}
 \caption{Schematic representation of transformations that maintain
directionality: the directed network is transformed to undirected
(weighted or unweighted, unipartite or bipartite) based on specific
methodologies. Then, any clustering  algorithm for undirected networks can be
applied. The extracted clusters   correspond to the clusters of the initial
directed network. \label{fig:graph_transf}}
\end{figure}

 More precisely, the initial graph is transformed into
an undirected one, while information and semantics about the direction of the 
edges is meaningfully incorporated in the resulting graph. For example, this can
be done by adding weights on the edges of the transformed network (or applying a
reweighting scheme in case of already weighted networks). The resulting network
can be either unipartite or bipartite. Then, algorithms that work on undirected
networks can be applied to detect the community structure; such approaches can
benefit from the plethora of diverse algorithms that have been proposed for
the community detection task in undirected networks. Schematically, a high level
description of this approach is depicted in Fig. \ref{fig:graph_transf}.

\subsubsection{Transformation to Unipartite Weighted Network} \label{sec:transf}

\par In a  commonly used transformation approach  in the related literature, the
directed network is converted into an undirected unipartite one, where
information about directionality is incorporated via weights on the edges of the
transformed network. Satuluri and Parthasarathy \cite{srini-edbd11}, investigate
how the problem of clustering directed graphs can benefit using such
symmetrization approaches. The basic insight on their approach is that, in
directed networks, a meaningful cluster can be a group of nodes that share
similar incoming and outgoing edges. In other words, a clustering approach
should not based solely on density criteria, but also the in-link and out-link
node similarity should be taken into consideration. Therefore, their approach is
able to detect groups of nodes with homogeneous in-link and out-link structure
(e.g., citation-based clusters), that do not necessarily share edges among them
(e.g., similar to the clusters in Fig. \ref{fig:pattern-based-figure} (b)). More
precisely, the authors propose a two-stage framework which is in accordance with
the above discussion: (a) transformation to undirected graph applying
symmetrization methods to the adjacency matrix and (b) clustering  the
symmetrized graph using existing algorithms. Let $G$ be the initial directed
graph with adjacency matrix $\mathbf{A}$. The authors discuss and propose
various ways to symmetrize a directed network:

\begin{itemize}
 \item \textit{$\mathbf{A} + \mathbf{A}^T$ symmetrization:} in this
approach, the produced undirected network $G_U$ will have the symmetric
adjacency matrix $\mathbf{A}_U = \mathbf{A} + \mathbf{A}^T$. The network retains
the same number of edges (i.e., every directed edge is replaced by an
undirected), but in the case of directed edges in both directions, the weight of
the new edge is the sum of the weights in the initial directed edges. However,
this simple way to symmetrize a directed network  cannot capture the notion of
node similarity based on incoming and outgoing edges; although the two
central nodes of Fig. \ref{fig:pattern-based-figure} (b) exhibit structural
similarity according to their in-links and out-links, they continue to remain
unconnected at the resulting graph, and therefore, it is very difficult to be
clustered together.

\item \textit{Symmetrization based on random walks:} according to this approach,
the normalized cut (Ref. \cite{shi-malik-pami00}) of a group of nodes in the
produced undirected network $G_U$ will be preserved with respect to
the initial directed one. In other words, the directed normalized cut of a group
of nodes will be equal to the normalized cut of the same group  in the
symmetrized undirected network. More precisely, the transformed graph will
be described by the following adjacency matrix 

\begin{equation}
 \mathbf{A}_U = \dfrac{\mathbf{\Pi P} + \mathbf{P}^T \mathbf{\Pi}}{2},
\end{equation}

\noindent where $\mathbf{P}$ is the transition matrix of the random walk and
$\mathbf{\Pi} =
\text{\texttt{diag}} (\pi_1, \pi_2, \ldots, \pi_n)$ is the diagonal matrix with
the probabilities of staying at each node in the stationary state (stationary
distribution). As noted by the authors (Ref. \cite{srini-edbd11}), this approach
makes easier the extraction of clusters that satisfy the criterion of low
normalized cuts, since this property is preserved during the symmetrization
process (compared to approaches that will be presented later at this papers and
rely on expensive spectral clustering based on the directed Laplacian matrix
(see Section \ref{sec:spectral})). However, the symmetrization method follows
the density-based clustering notion due to the dependence to the normalized
cut criterion, and therefore other types of meaningful structures like the case
of pattern-based clusters  of Fig. \ref{fig:pattern-based-figure} (b) with low
normalized cut, cannot be easily identified.

\item \textit{Bibliometric symmetrization:} both previous approaches maintain  
intact the edge set of the directed network (discarding directions); they only
reweight them according the selected symmetrization scheme. However, a natural
requirement that a symmetrization approach should meet is that at the final
graph, edges should appear between similar nodes even though in the original
network this does not happen. The prime example for this argument is the
network of Fig. \ref{fig:pattern-based-figure} (b), where there exist nodes that
do not share common edges, but both of them point to the same group of nodes
(and pointed by the same group of nodes). Therefore, these nodes share an
intuitive notion of similarity and thus a clustering algorithm should be able
to group them together.

\par The authors of Ref. \cite{srini-edbd11} propose a symmetrization approach
based on a combination of the bibliographic coupling matrix
$\mathbf{B} = \mathbf{A} \mathbf{A}^T$ and the co-citation strength matrix
$\mathbf{C} = \mathbf{A}^T \mathbf{A}$. Both these matrices are symmetric. The
former captures common outgoing edges between each pair of nodes (i.e.,
the number of common nodes that both nodes point to), while
the latter common incoming edges (i.e., the number of nodes that commonly
point to these nodes). These matrices were first introduced in the field of
bibliometrics, but later  have been  used in several applications
and domains where symmetric matrices are required (e.g., information retrieval
\cite{hits} and network analysis \cite{gkantsidis-infocom03}). Since both
incoming and outgoing edges should be of the same importance for a clustering
algorithm, the authors propose to use the sum of these matrices as a
symmetrization scheme:
 
\begin{equation}
 \mathbf{A}_U = \mathbf{A} \mathbf{A}^T + \mathbf{A}^T \mathbf{A}.
\end{equation}

\item \textit{Degree-discounted symmetrization:} one of the main properties of
real-world networks is that they follow a power-law degree distribution (e.g.,
Ref. \cite{fff}). The property states that, inside the network, there exist a
few nodes with very high degree compared to the majority of the nodes.
This observation has direct implications to the previously described 
symmetrizations, since nodes with high degree  would be shared a lot of common
edges with other nodes (and thus higher similarity). To this direction, the
authors propose a symmetrization approach where the contribution of each node to
the similarity score will be  normalized according to its degree (in-degree and
out-degree respectively). More precisely, their approach is based on two
intuitive observations:

\begin{itemize}
 \item[]Case $1$: Suppose that two nodes $i,j$ both point to a node $z$, which
has high in-degree $k^{in}_z$.

 \item[]Case $2$: Suppose that two nodes $i,j$ both point to a node $z$, with
low in-degree $k^{in}_z$ (i.e., it has incoming edges from only a
few  nodes other than $i,j$).
\end{itemize}

\noindent Based on these two points, the authors suggest that case $2$
should contribute more to the similarity between nodes $i,j$ since it is a less
frequent event and thus more informative. Hence, when two nodes $i,j$ both point
to a third one $z$, the similarity between them should be inversely
proportional to the in-degree $k^{in}_z$ of node $z$. Similarly, the
number of outgoing links of the nodes should be taken into consideration.
Thus, the out-link similarity between $i,j$ should be inversely related to the
out-degrees of nodes $i$ and $j$. Then, both the bibliographic coupling and
co-citations matrices (matrices $\mathbf{B}$ and $\textbf{C}$ respectively) are
redefined according to the degree-discounted idea as follows:

\begin{equation}
\mathbf{B} = \mathbf{D}_{out}^{-\alpha} ~ \mathbf{A} ~
\mathbf{D}_{in}^{-\beta} ~ \mathbf{A}^T ~ \mathbf{D}_{out}^{-\alpha}
\text{~~~~~~and~~~~~~}
\mathbf{C} = \mathbf{D}_{in}^{-\beta} ~ \mathbf{A}^T ~
\mathbf{D}_{out}^{-\alpha} ~ \mathbf{A} ~ \mathbf{D}_{in}^{-\beta},
\end{equation}

\noindent where $\alpha, \beta$ are the discounting parameters. Finally,
 the produced similarity matrix (and thus the adjacency matrix of the
symmetrized undirected network) will be the sum of these two matrices,
$\mathbf{A}_U = \mathbf{B} + \mathbf{C}$.
The authors report that they have empirically observed that setting
$\alpha = \beta = 0.5$ results into intuitive and meaningful
clusterings. Having now symmetrized the directed network, algorithms designed to
work on undirected graphs can be applied to extract the underlying community
structure.
\end{itemize}

\par
\begin{figure}[t]
\centering
  \includegraphics[width=.5\textwidth]{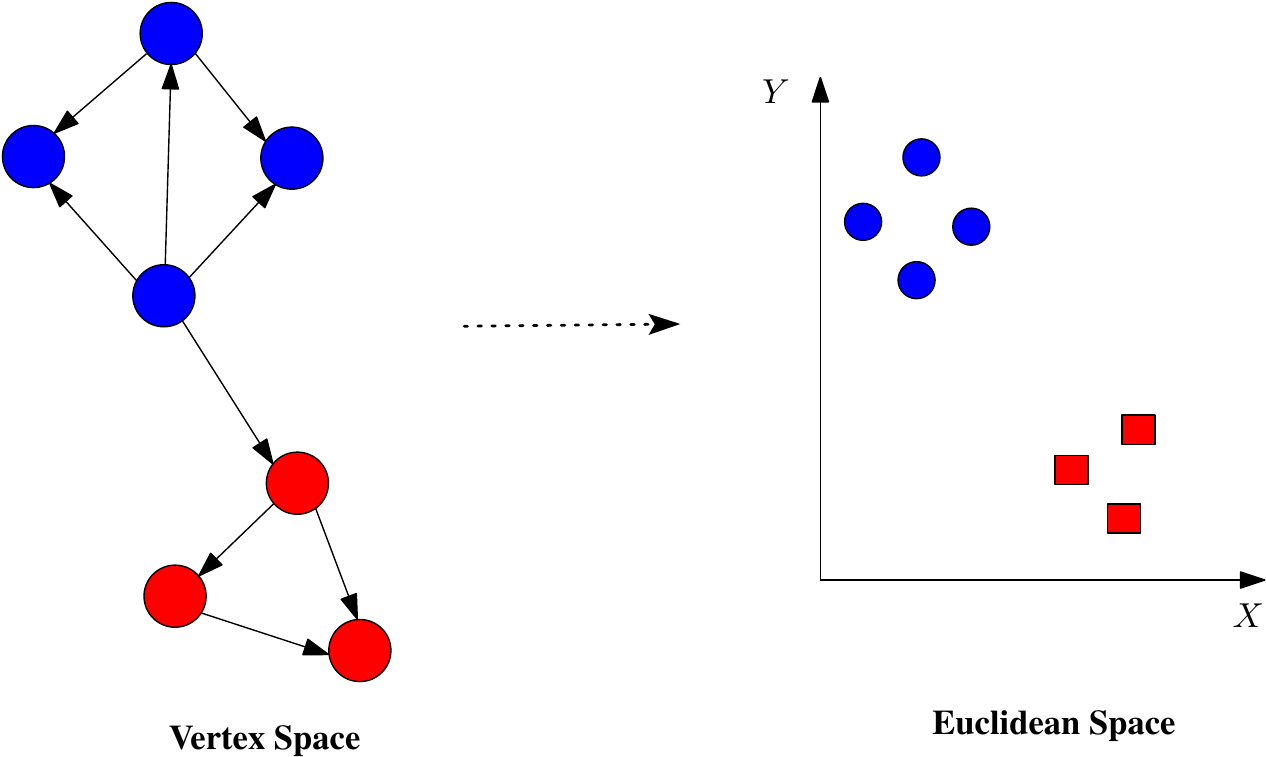}
 \caption{Visual representation of the network embedding method by Lai et
al. \cite{lai-physica10}. Each node in the directed network is treated as a
point in the Euclidean space (or similarly as a vector). The local topological
characteristics of each node should be preserved by the embedding. Figure
redesigned from Ref. \cite{lai-physica10}. \copyright 2010 Elsevier.
\label{fig:embedding}}
\end{figure}

 In a similar spirit,  Lai et al. \cite{lai-physica10}  proposed  a
symmetrization method based on network embeddings, that indirectly can be
considered as a transformation to an undirected weighted network. More
precisely, the basic idea is to embed the initial directed network into a 
vector space, preserving as
much as possible from its local topological characteristics.
According to this approach, every node in the directed network can be treated
as a point in the Euclidean space, as schematically shown in Fig. 
\ref{fig:embedding}. 

\par  This can be considered as an equivalent representation
scheme for a network, since in the general case, the adjacency matrix can be
treated equivalently as 
a representation in a Euclidean vector space and more precisely in the space  
defined by the
nodes of the network, that is $\mathbb{R}^n$. Essentially the authors rely on a
specific type of network embedding and in particular on the Laplacian
embedding for directed networks. As we described in Section
\ref{sec:background},
the Laplacian matrix is an alternative matrix representation for a network with
several interesting characteristics. Similar to the adjacency matrix, the
Laplacian  can be considered as a network representation in the Euclidean
space, where the similarities among node pairs are preserved. The authors are
based  on the extension of Laplacian matrix for directed networks proposed
by Chung \cite{chung-directed-laplacian05} (see \ref{sec:spectral} 
for more details on the Laplacian matrix) 

\begin{equation} \label{eq:lapl1}
\mathbf{L} = \mathbf{\Pi} - \dfrac{\mathbf{\Pi} \mathbf{P} + \mathbf{P}^T
\mathbf{\Pi}}{2},
\end{equation}

\noindent where  $\mathbf{P}$ is the transition matrix of the random walk
and $\mathbf{\Pi} = \text{\texttt{diag}} (\pi_1, \pi_2, \ldots, \pi_n)$  the
diagonal matrix with the probability of  staying on each node in the stationary
state. 

\par The idea is similar to the random walk symmetrization
presented earlier as again the same concept 
is utilized in order to define the Laplacian matrix (in this case, the PageRank 
random walk). However, the authors observed that the directed version of the 
Laplacian matrix in Eq. \eqref{eq:lapl1} can be expressed as $\mathbf{L} =
\mathbf{\Pi} - \mathbf{W}$, where $\mathbf{W} = \dfrac{\mathbf{\Pi} \mathbf{P} +
\mathbf{P}^T \mathbf{\Pi}}{2}$. Matrix $\mathbf{W}$ is symmetric, while
$\mathbf{\Pi}$ can be considered as the degree matrix of the network. 
Hence, $\mathbf{W}$ can be interpreted as the adjacency matrix of a new
undirected network\footnote{This is similar to the definition of Laplacian for
undirected networks $\mathbf{L}_U = \mathbf{D}_U - \mathbf{A}_U$, where
$\mathbf{D}_U$ is the diagonal degree matrix and $\mathbf{A}_U$ the adjacency
matrix of the undirected network. See also Section \ref{sec:background}.}, and
for this reason the method is considered to perform a graph transformation.
Furthermore, the authors prove that in the produced undirected
network, information about edge directionality is effectively incorporated as
weights on the edges (entries of the matrix $\mathbf{W}$). Moreover, a new
definition of modularity is presented according to this approach, which is
considered as a generalization of the
one defined on undirected networks (e.g., see Section \ref{sec:background}). The
method is able to identify pattern-based clusters and more precisely,
clusters that represent patterns of movement among the nodes of the network
(flow-based clusters, Fig. \ref{fig:pattern-based-figure} (c)). The authors
stress out that the their method has broad applicability, since it can be used
for several types of well-known directed  networks (e.g., social and biological
networks), as well as for networks where the edges represent patterns of
movement among nodes that share common properties (e.g., the web graph,
citation networks).

\par Later, the same authors presented a more extended approach
\cite{lai-jstatmech10}, where edge directionality is extracted using a PageRank
random walk (e.g., Ref. \cite{pagerank}), and introduced via weights on the
edges of the transformed undirected network. The approach is
more general regarding the types of cluster that are identified and introduces
some interesting features. In contrast to the previous method which was  based 
on the assumption that edges in the network capture only similarity between
nodes, here the information about edges' direction is utilized in order to
decide
whether an edge lies inside a community (i.e., intra-community edge, connects
two nodes that belong on the same community), or between communities (i.e.,
inter-community edge). Hence, the edges of the network are distinguished between
inter-community and intra-community
edges, very close to approaches presented for undirected
networks. In other words, the information about the directionality of
the edges, and thus the weights on the network, operate as an indicator of
how likely the associated edges will belong on the same community; while the
topological structure of the network is modified, the connectivity among nodes
is preserved and the underlying community structure becomes more clear compared
to the original network.

\par As we mentioned earlier, the method rely on the usage of random walks for
directed networks in order to determine the structure of the network and to
extract weights from edge directionality. The basic idea is the following:
assume two nodes $A$ and $B$ connected with an edge, and respective random
walks starting from these two nodes. If both $A$ and $B$ are visited mutually
during
these random walks then the edge among those nodes is more likely to be an
intra-community edge.

\par That is, according to the random walk, nodes that belong to the same
community could interact more often among each other, and thus intra-community
edges will receive higher weights than inter-community edges. The authors define
the so-called \textit{nodes' behavior vector}, where each entry represents
the expected frequency with which the specific node will be visited by the
random walk in $t$ steps\footnote{As the authors state, the length of the
random walk should not be chosen too large (i.e., no greater than $\log n$),
even if it is not very crucial for the method's performance.}. Then, the
similarity between two such vectors (i.e., how similar are the trajectories of
the random walks starting from these connected nodes) will be an indicator if
these nodes belong on the same community. The authors suggest to apply two
well-known similarity measures, namely the exponential  and the cosine
similarity. Having extracted the edge weights, the network can be
treated as undirected and thus algorithms designed for undirected networks can
be applied (the authors choose Newman's modularity optimization algorithm
\cite{newman-newman-phys-rev-e-04}). Finally, the proposed framework can
identify flow-based clusters independently of the chosen clustering algorithm.

\subsubsection{Transformation to Bipartite Network}
A somewhat different transformation approach than the ones presented earlier,
assumes that the directed network is transformed into a bipartite
undirected one and then community detection algorithms are applied on the
latter. As we described in Section \ref{sec:background}, bipartite networks
form a special class of networks with interesting properties. Treating a 
directed network as bipartite is not something new, but it has been used in the
past for other tasks in network analysis (e.g., Ref. \cite{salsa-tois}).  

\par In the context of community detection for directed networks, the authors of
Refs. \cite{module-identification-physrev07, zhou-nips05} 
utilize a transformation scheme where a bipartite graph $G_B=(V_h, V_a, E_b)$
is constructed from the original directed network $G=(V,E)$, according to the
following process:

\begin{itemize}
 \item $V_h = \{ i_h | i \in V ~ \text{and} ~ k^{out}_i > 0 \}$

 \item $V_a = \{ i_a | i \in V ~ \text{and} ~ k^{in}_i > 0 \}$

\item Each directed edge $(i,j) \in E$ between two nodes of the directed network
$G$, will be represented 
%with
by an edge $(i_h, j_a) \in E_b$ between nodes $i_h$
and $j_a$ of the produced bipartite network $G_B$.
\end{itemize}

\begin{figure}[h!t]
\centering
 \begin{tabular}{ccc}
  \includegraphics[width=.4\textwidth]{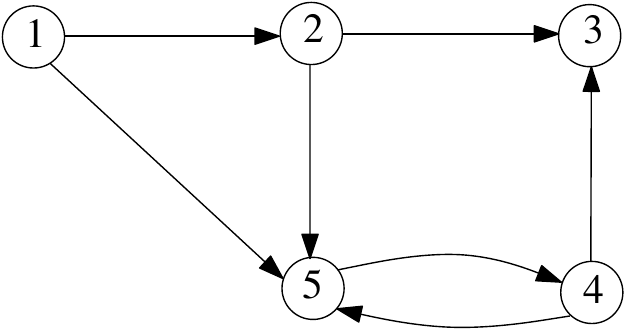} & ~~ &
  \includegraphics[width=.22\textwidth]{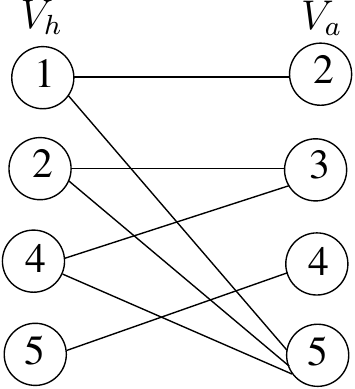} \\
   (a) Directed network & ~~ &  (b) Transformed bipartite network
 \end{tabular}
 \caption{Example of transformation to bipartite graph. Each node of the
directed network (a) is treated as a hub/authority or both,  according to its
out-degree / in-degree.
\label{fig:bipartite}}
\end{figure}

\noindent 
This implies that in the general case, the nodes of the directed network are
doubled and each one is represented by a node at each of the two partitions.
However, since some nodes may not have outgoing or incoming links (and thus
they will be isolated in the resulting bipartite network), every node is placed
in the two sets $V_h, V_a$ according to its out- and in- degree respectively.
Figure \ref{fig:bipartite} shows a construction of the bipartite network for a
given directed one.

\par The above representation scheme is inspired by Kleinberg's \textit{hub} and
\textit{authority} web model \cite{hits}, where the web pages are
distinguished in two sets: \textit{authoritative pages} and \textit{hub
pages}. The first category includes web pages relevant to a specific topic,
while the second one web pages that point to authorities of a  relevant topic.
Moreover, a web page can simultaneously belong to both sets.

\par In Ref. \cite{module-identification-physrev07}, the authors construct a
bipartite network from the original directed one (adding all nodes to both sets
of the bipartite graph), trying to identify clusters of nodes with similar
outgoing links as well as similar incoming links. They distinguish the two
partitions of the graph into the \textit{actor} partition and \textit{team}
partition ($V_h$ and $V_a$ in our description). The ultimate goal is to identify
groups of actors that are closely connected to each other through
co-participation in many teams. To this end, their approach is based on the
idea of modularity optimization. More precisely, they define a modularity
function for bipartite networks, modifying Newman's modularity measure
\cite{newman-newman-phys-rev-e-04}, and then  apply  an optimization
technique based on simulated annealing for detecting the underlying community
structure.

\par In a similar spirit, the authors of Ref. \cite{zhan-evolutionary-physrev11}
propose an approach for revealing the community structure of bipartite and
directed networks. They rely on the network transformation scheme presented
above
and suggest that every unipartite or directed network can be transformed to a
bipartite one, while the modularity is preserved\footnote{Some comments
on the performance of the method  have been proposed from
other researchers \cite{costa-physrev11}.}. Then, an adaptive genetic algorithm
called MAGA is presented, which according
to the authors is capable to effectively optimize the objective function for
the community structure detection problem (the authors select to apply the
bipartite modularity function).

\par A different approach that based on the construction of a bipartite network
is the one presented in Ref. \cite{zhou-nips05}. The authors describe a
framework for semi-supervised learning on directed networks, which can also be
applied for the task of graph clustering. The framework was introduced for the
problem of node classification in directed networks, where some nodes in the
graph bear  labels (positive or negative) and the goal is to classify unlabeled
nodes. However, in case of  absence of labeled node instances, the framework can
be used as a graph clustering tool for both directed and bipartite networks. The
main idea behind the approach is the so-called \textit{category similarity of
co-linked nodes} in directed networks; the existence of nodes with common
parents (sibling structures) and nodes with common children (co-parent
structures) should be taken into consideration at the clustering task since they
can operate as indicators regarding node similarity (the general idea is
similar with the one presented by Satuluri and Parthasarathy in Ref.
\cite{srini-edbd11}). The construction of a bipartite network from the original
directed one is also inspired by Kleinberg's hub and authority web model, where
co-linked node structures are highlighted.

\subsection{Extending Objective Functions and Methodologies to Directed
Networks}
\label{sec:extending-techniques}
In the previous section, we presented methods where the graph clustering and
community detection problem is not treated on the original directed network,
but on a new undirected network that is produced applying meaningful
transformation methods. The main advantage of these approaches is that they
can benefit from the large bulk of techniques for the undirected case of the
problem. However, the basic question behind such approaches remains the same: to
what extend the information about the directionality of the edges is retained?
To this direction, several methodologies have been proposed trying to deal with
the problem without changing the structure of the original network. That is,
instead of transforming the directed network to undirected, a different approach
would be to transform or extend existing methods making them capable to work
with directed networks.

\par Usually, two things need to be specified for the problem of clustering and
community detection in networks. The first one has to do with the quality
assessment of the produced clustering results, while the other is more general
and is related to the algorithmic framework that will be applied to extract
the community structure (according to how a ``good'' clustering should
looks like). In fact, these two things compose the basic steps of each
clustering and community detection algorithm: (i) an objective function that
quantifies the quality of a cluster and (ii) an algorithmic technique for
optimizing this function. Both these aspects-questions have been treated for the
case of undirected networks (or generally, several solutions have been
proposed). In other words, there exist several objective measures for
quantifying the quality of a clustering result, as well as algorithmic
frameworks that are trying to optimize those objective criteria in order to
identify
and extract the underlying clustering structure of undirected networks
(e.g., see Refs. \cite{fortunato, schaeffer-review}). Thus, a natural question
would be if these quality measures and general methodologies can be extended to
the case of directed networks and how this could be done.

\par In this section we present approaches for detecting clusters and
communities in directed networks that constitute extensions of the undirected
case. Note that, this category contains the most well-studied
approaches for the clustering problem in directed networks. First, we will discuss
 well-known objective criteria that have been extended to take into
consideration the directionality of the edges (e.g.,
modularity, cut-based measures). Then, we describe more general algorithmic
frameworks that were initially introduced for clustering undirected networks and how
they can be extended to the directed case. Such methodological frameworks
include spectral clustering approaches based on the Laplacian matrix, as well as
PageRank based and random-walk based methods. We must note here that some of
these approaches are not independent. For example, several modularity-based
methods apply spectral clustering in order to identify the best
clusters. Whenever necessary, we briefly review the problem for the undirected
case, trying to make this survey paper as self-contained as possible (see
Section \ref{sec:background} for more details about the background, as well as
other survey papers that focus on undirected networks).

\subsubsection{Modularity for Directed Networks} \label{sec:modularity}
One of the basic objective criteria about the quality of a particular division
into clusters for a network, is the so-called \textit{modularity} function.
Modularity was initially introduced by Newman and Girvan
\cite{newman-newman-phys-rev-e-04} for the case of undirected networks, as a
measure for assessing the strength of the partitions produced by an
hierarchical clustering algorithm (an thus indicating which partition
should be kept). The measure is based on the idea that networks with inherent
community structure usually deviate from random graphs. That is, since random
graphs are not expected to have community structure, measuring the
deviation between the concentration of edges in the original network from
that someone expect in the case of random distributed edges, would be an
indicator of the presence (or lack thereof) of community structure. Informally,
the modularity score $Q$ of each possible partition will be
\cite{newman-newman-phys-rev-e-04}:

\begin{equation}
 Q = (\text{fraction of edges within communities}) - (\text{expected fraction of
edges}).
\end{equation}

\noindent Larger positive values of modularity indicate better community
structure, since there are more edges within communities than one would expect
if edges were placed in random (the maximum value of modularity can be $1$).
The expected fraction of edges among a group of
nodes is usually based on the chosen configuration model, i.e., a random graph
with the same degree sequence of the original network. In this model, the
probability of an edge between two nodes $i, j$ with degree $k_i$ and $k_j$ is
$k_i k_j / 2m$, where $m = \frac{1}{2} \sum_{i \in V} k_i$ is the total number
of (undirected) edges in the network. Then, the modularity for undirected
networks can be expressed as

\begin{equation} \label{eq:modularity}
 Q_u = \dfrac{1}{2m} \sum_{i,j} \bigg[A_{ij} - \dfrac{k_i k_j}{2m}
\bigg] \delta(c_i, c_j),
\end{equation}

\noindent where $A_{ij}$ is the entry of the adjacency matrix which represents
the existence or not of edge between nodes $i$ and $j$,
$\delta(c_i, c_j)=1$ if $c_i=c_j$ (i.e., if nodes $i,j$ belong on the same
community) and $0$ otherwise. 

\par Generally, modularity can be used both as quality measure for a specific
network partition, as well as the basic ingredient of a framework for extracting
the community structure. The latter procedure, usually called  modularity
optimization, is one of the dominant approaches for extracting the community
structure in undirected networks (e.g., Ref. \cite{newman-modularity}). 

\par In the case of directed networks, several extensions have been proposed for
the measure of modularity. Arenas et al. \cite{arenas-modularity07} proposed
a generalization for directed networks, where their ultimate goal
was to reduce the size of the initial network (directed or undirected), while
preserving the modularity value (this is a very crucial point since optimizing
the modularity is a hard task). Their extension is based on the observation
that the existence of a directed edge $(i,j)$ between nodes $i$ and $j$, depends
on the out-degree and in-degree of nodes $i$ and $j$ respectively. Let us
consider that node $i$ has high out-degree but low in-degree, while node $j$
has high in-degree and low out-degree. Then, it is more probable to observe
the directed edge $(i,j)$ from node $i$ to node $j$, instead of observing 
edge $(j,i)$. Putting these insights together, the configuration model can be
extended to the directed case, where an edge $(i,j)$ from node $i$ to node $j$
will exist with probability $k^{out}_i k^{in}_j / m$. Then, the modularity
function for directed networks can be expressed as

\begin{equation} \label{eq:dir-modularity}
 Q_d = \dfrac{1}{m} \sum_{i,j} \bigg[A_{ij} - \dfrac{k^{out}_i
k^{in}_j}{m} \bigg] \delta(c_i, c_j),
\end{equation}

\noindent where the notation is similar to the one of Eq. \eqref{eq:modularity}.
Also, observe that there is no factor of $2$ in the denominator (the sum of
out-degrees (similarly in-degrees) is equal to $m$). Moreover, Arenas et al.
\cite{arenas-modularity07} gave the relationship between directed and
undirected modularity:

\begin{equation} \label{eq:dir-undir-modularity}
Q_d = Q_u + \dfrac{1}{4m^2} \sum_{i,j} (k^{out}_i - k^{in}_i) (k^{out}_j -
k^{in}_j) \delta(c_i, c_j).
\end{equation}

\par Leicht and Newman \cite{leicht-newman-2008} were based on the above
definition of modularity to propose an algorithm for detecting communities in
directed networks.  Their approach constitutes a generalization of the
modularity optimization method, presented by Newman and Girvan in Ref.
\cite{newman-newman-phys-rev-e-04},
where  modularity can be expressed in terms of the spectrum (eigenvalues and
eigenvectors) of a special matrix called modularity matrix. More
precisely, let us suppose that our goal is to assign the nodes of the network
into two communities, namely $\mathcal{A}$ and $\mathcal{B}$. Let $s_i, \forall
i \in V$ be an indicator variable taking value $+1$ if vertex $i$ is assigned
to community $\mathcal{A}$ and $-1$ if is assigned to community $\mathcal{B}$
and $\mathbf{s}$ be the vector whose elements are the $s_i$ values. Then,
modularity can be written as

\begin{align} 
  Q_d &= \dfrac{1}{m} \sum_{i,j} \bigg[A_{ij} - \dfrac{k^{out}_i
k^{in}_j}{m}
\bigg] \delta(c_i, c_j) \nonumber \\
&= \dfrac{1}{2m} \sum_{i,j} \bigg[A_{ij} - \dfrac{k^{out}_i
k^{in}_j}{m}
\bigg] (s_i s_j + 1) \nonumber \\
&= \dfrac{1}{2m} \sum_{i,j} B_{ij} s_i s_j \nonumber \\
&= \dfrac{1}{2m} \mathbf{s}^T \mathbf{B} \mathbf{s}, \label{eq:mod_matrix}
\end{align}

\noindent where $B_{ij} = A_{ij} - \dfrac{k^{out}_i k^{in}_j}{m}$ is the
modularity matrix. In the general case
matrix $\mathbf{B}$ is not symmetric and thus we are not able to apply a
spectral approach. However, transposing  Eq. \eqref{eq:mod_matrix},
$Q_d$ can be expressed as $Q_d = (2m)^{-1} \mathbf{s}^T \mathbf{B}^T
\mathbf{s}$. Finally, taking the average of this quantity with the one in Eq.
\eqref{eq:mod_matrix} gives

\begin{equation}
 Q_d = \dfrac{1}{4m} \mathbf{s}^T (\mathbf{B} + \mathbf{B}^T) \mathbf{s}.
\end{equation}

\noindent Matrix $\mathbf{B} + \mathbf{B}^T$ is now symmetric. Thus, 
applying well-known approaches in optimization theory, a simple clustering
algorithm can be derived from the spectrum of this matrix: compute the
eigenvector that corresponds to the largest positive eigenvalue of the matrix
$\mathbf{B} + \mathbf{B}^T$ and assign the nodes to communities $\mathcal{A}$ or
$\mathcal{B}$ according to the signs in the components of the eigenvector
(generally, node $i$ is associated with the $i$-th component of the
eigenvector). 

\par Additionally, other optimization tools can be applied (e.g., see
Section \ref{sec:background} and Ref. \cite{fortunato}). The above method can be
extended to assign the nodes is more than two communities. This can be achieved
by an iterative procedure which subdivides the produced communities until the
modularity value is not increased. The algorithm has been tested on both
synthetic and real datasets and the
results show that, considering edge directionality in the modularity
optimization process, meaningful communities can be identified (whereas ignoring
the direction of the edges this cannot be achieved).

\par However, Kim at al. \cite{linkrank-phys-rev10} observed that the directed
version of modularity in Eq. \eqref{eq:dir-modularity} exhibits two limitations:

\begin{itemize}
 \item[(i)] It cannot properly distinguish the directionality of the edges.
 \item[(ii)] It cannot be used to detect pattern-based clusters representing
patterns of movement among nodes.
\end{itemize}

\begin{figure}[ht]
\centering
 \includegraphics[width=.4\textwidth]{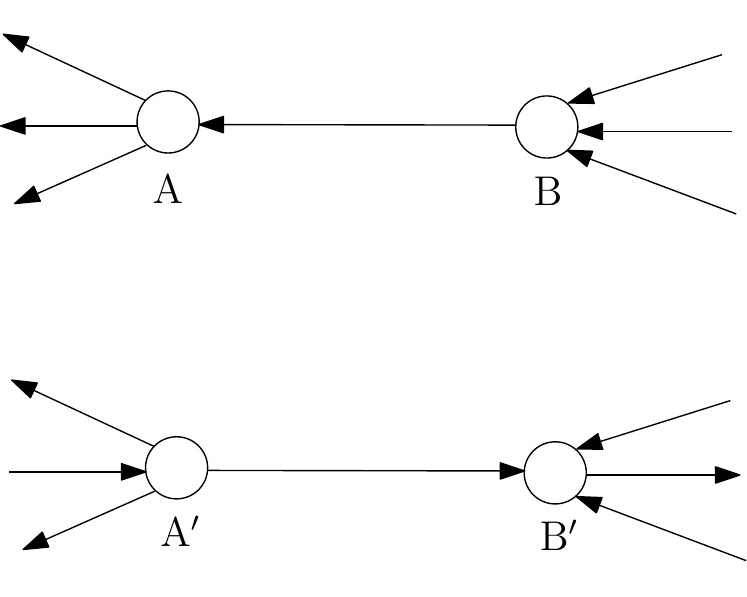}
\caption{The modularity function introduced in Ref. \cite{arenas-modularity07}
does not distinguish the directionality of the edges. Nodes $A$ and $A'$ as well
as $B$ and $B'$ have the same in- and out- degree respectively. However, in the
top figure there is a precise directed flow, while in the bottom no. Modularity
cannot distinguish these different situations (Kim et al.
\cite{linkrank-phys-rev10}). Figure redesigned from Ref.
\cite{linkrank-phys-rev10}. \copyright 2010 American Physical Society.
\label{fig:kim-modularity}}
\end{figure}

\noindent Figure \ref{fig:kim-modularity} presents an example where
the directed modularity as introduced in Ref. \cite{arenas-modularity07} and
used in the algorithm of Ref. \cite{leicht-newman-2008},  is not able to
distinguish the two different cases. According to the modularity definition,
nodes $A$ and $B$ are more likely to belong in the same community than nodes 
$A'$ and $B'$, since the edge from $B$ to $A$ is more informative than the
one from $A'$ to $B'$
(because of the fact that node $B$ has small out-degree and node $A$ small
in-degree; thus the edge from $B$ to $A$ should contribute more to the
modularity since it is a statistically surprising configuration). However, both
edges have the same contribution to the directed modularity.

\par Based on the above observations, in Ref. \cite{linkrank-phys-rev10} the
authors proposed a somewhat different formulation of modularity for directed
networks. Their approach, called LinkRank, is related to random walks and more
specifically to Google's PageRank algorithm \cite{brin-page-isdn, pagerank}.
More precisely, LinkRank indicates the importance of links (edges) in the
network (instead of nodes) as the probability that a random surfer will follow
this link in the
stationary state (similar to the definition of PageRank but now for the edges).
That is, the LinkRank value of a particular edge $(i,j)$ can be defined as

\begin{equation}
 L_{ij} = \pi_i G_{i,j},
\end{equation}

\noindent where $\pi_i$ is the $i$-th element of PageRank vector and $G_{ij}$
is the element of Google matrix $\mathbf{G}$\footnote{Similar to the transition
matrix $\mathbf{P}$ but guaranteeing the existence of a stationary vector
\cite{brin-page-isdn}.} \cite{brin-page-isdn}. Then, a generalized version of
modularity can be defined using random walk concepts as follows:

\begin{align}
 Q_{linkrank} &= (\text{fraction of time spent by random surfer while walking
within communities}) \nonumber \\
& - (\text{expected value of this fraction}).
\end{align}

\noindent According to this modularity definition (which holds for both directed
and undirected networks), a community is not just a group of nodes with more
than expected number of edges, but a group of nodes where a random surfer is
more likely to stay. Therefore, the directed modularity can be expressed as

\begin{equation}
 Q_{linkrank} = \sum_{i,j} \big[ L_{ij} - \pi_i \pi_j \big] \delta(c_i, c_j),
\end{equation}

\noindent where $\pi_i \pi_j = E(L_{ij})$ is the expected probability (in the
configuration model) that a random surfer is moving from node $i$ to $j$ (and
thus the expected value  $E(L_{ij})$ of $L_{ij}$). Furthermore, the authors
show that the proposed modularity measure $Q_{linkrank}$ is consistent with the
original modularity $Q_u$ of Eq. \eqref{eq:modularity} for undirected networks.
In other words, for undirected networks, the proposed definition of a community
as a group of nodes where a random surfer is more likely to be trapped in is
consistent with the traditional one, where a community represents a group
of, more than expected, densely connected nodes. One other important feature is
that the proposed LinkRank-based modularity function can be optimized using
already existing methods \cite{fortunato}, leading to a community discovery
algorithm for directed networks. The authors claim that their method is
able to detect mainly communities in directed networks where edges can be
considered to represent citation/reference relationships (e.g., pattern-based
clusters).

\par The same definition of modularity for directed networks was also presented
in Ref. \cite{lai-physica10} (the method was also described in Section
\ref{sec:transf}
since the original directed network is transformed into an undirected one).
The method applies PageRank random walk to define the Laplacian matrix for the
directed network, which at the end can be considered as a network embedding.
Additionally, the authors provide an alternative definition of communities as
nodes sharing common properties; nodes of the same group are more similar to
each other compared to nodes outside the group. This can be also considered as a
high level clustering definition, and is in agreement with the one we presented
in Section \ref{sec:problem-stm}.

\par In Refs. \cite{pantazis-asilomar10} and \cite{pantazis-isbi11}, Chang et
al. present an alternative formulation of modularity (compared to $Q_d$ of Eq.
\eqref{eq:dir-modularity}), that relies on a different configuration model. As
we described earlier, the configuration model used in $Q_d$
(\cite{arenas-modularity07, leicht-newman-2008}) assumes that the existence of
a directed edge from node $i$ to node $j$ is proportional to $k^{out}_i k^{in}_j
/m$. The authors present a different configuration model that rely on the idea
of conditional expected network (a similar approach has been presented for
undirected networks \cite{pantazis-isbi10}). That is, the configuration model
can be formed by conditioning on the degrees (both in- and out-) of nodes
in the original network as

\begin{equation}
E(A_{ij} | k^{in}_1, k^{in}_2, \ldots ,k^{in}_n, k^{out}_1, k^{out}_2,
\ldots, k^{out}_n) = E(A_{ij}|\mathbf{k}^{in}, \mathbf{k}^{out}).
\end{equation}

\noindent The solution of the above conditional expected model suggested by the
authors is considered for the case where the edges are distributed  according
to a Gaussian distribution and the final configuration model is the directed
Gaussian random network (DGRN). A benefit of this  model stands
from the point that one can introduce prior information to the model in the form
of the mean and covariance of the Gaussian distribution. Then, measuring the
deviation from the configuration model, the modularity can be expressed as

\begin{equation}
 Q_{dM} = \dfrac{1}{m} \sum_{i,j} \big[ A_{ij} -   E(M_{ij}|\mathbf{k}^{in},
\mathbf{k}^{out}) \big] \delta(c_i, c_j).
\end{equation}

\noindent Furthermore, using spectral techniques similar to the ones of Ref.
\cite{girvan-newman} (or generally other modularity maximization algorithms),
the modularity can be optimized obtaining a clustering assignment.

\subsubsection*{Modularity for Overlapping Communities} 
In the discussion until now, we have reviewed approaches for
the clustering and community detection problem, where each node  is assigned
into just one community with no overlaps among communities. However, a different
version of the problem is to allow nodes to be assigned in more than one
communities, leading to the concept of \textit{overlapping communities}. The
intuition
behind overlapping clustering is based on the fact that real complex networks
usually are not divided into sharp sub-networks, but typically nodes may
naturally
belong to more than one communities. For instance, in a social relationship
network, individuals usually belong to several different communities at the same
time (family's community, friendship's community, profession's community, etc).
Thus,
being able to identify the overlapping communities of directed networks, could
offer fruitful insights about network structure. 

\par To this direction, Nicosia
et al. \cite{nicosia-s-stat-mech09} extended the measure of modularity to the
more general case of directed networks with overlapping communities. The main
point of their approach is to extend the configuration model that is used in the
definition of modularity \cite{leicht-newman-2008}, allowing nodes to belong to
several communities at the same time. Typically, nodes belong to each community
with a
certain strength and each node $i \in V$ is associated with a coefficient
$\alpha_{i,c}$ that indicates how strongly this node belongs to community $c$
(i.e., each node $i$ is associated with a vector $[\alpha_{i,1},
\alpha_{i,2}, \ldots, \alpha_{i,|c|}]^T$, where $|c|$ the total number of
communities). Then, a similar coefficient can be defined for the participation
of edges to communities; for each directed edge $e=(i,j)$ the belonging
factor to community $c$ can be represented by a function of the corresponding
coefficients of nodes $i,j$, i.e., $\beta_{e,c} = \mathcal{F}(\alpha_{i,c},
\alpha_{j,c})$. Then, the $\delta(c_i, c_j)$ function can be substituted by two
different coefficients $r_{ijc}$ and $s_{ijc}$, regarding the contribution
of edge $(i,j)$ to the modularity of the network and the configuration model
respectively. Finally, the modularity can be expressed as

\begin{equation}
 Q_{ov} = \dfrac{1}{m} \sum_{\forall c} \sum_{i,j} \Bigg[ r_{ijc} A_{ij} -
s_{ijc} \dfrac{k^{out}_i k^{in}_j}{m} \Bigg].
\end{equation}

\noindent One can observe that if there is no overlap between communities, then
$r_{ijc} = s_{ijc} = \delta(c_i, c_j)$, where the edge $(i,j)$ contributes to
modularity only if $c_i = c_j$. The value $r_{ijc}$ can be thought of as the
contribution of edge $e=(i,j)$ to the modularity of community $c$ and
according to the above discussion,  $r_{ijc} = \beta_{e,c} =
\mathcal{F}(\alpha_{i,c}, \alpha_{j,c})$. For the factor $s_{ijc}$ that is
related to the configuration model, assuming that the belonging of a node to a
community is independent from the belonging of every other node on the same
community (i.e., the probability that a node $i$ belongs to  community $c$
with strength $\alpha_{i,c}$ is not related to the probability that any other
node $j$ belongs to the same community with strength $\alpha_{j,c}$), the
modularity can be defined as

\begin{equation}
 Q_{ov} = \dfrac{1}{m} \sum_{\forall c} \sum_{i,j} \Bigg[ r_{ijc} A_{ij} -
s_{ijc} \dfrac{\beta^{out}_{e,c} k^{out}_i \beta^{in}_{e,c} k^{in}_j}{m}
\Bigg],
\end{equation}
 
\noindent where

\begin{equation} \label{}
 \beta^{out}_{e,c} = \dfrac{\sum_{j \in V} \mathcal{F}(\alpha_{i,c},
\alpha{j,c})}{|V|} \text{~~~ and ~~~}
 \beta^{in}_{e,c} = \dfrac{\sum_{j \in V} \mathcal{F}(\alpha_{i,c},
\alpha{j,c})}{|V|}
\end{equation}

\noindent are the expected belonging coefficients of any edge $e=(i,j)$,
where node $i$ belong to community $c$ (i.e., the average membership for all
edges). One more thing needs to be specified for defining the modularity
measure for directed networks with overlapping communities and it concerns the
selection of function $\mathcal{F}(\alpha_{i,c}, \alpha_{j,c})$ which specifies
the belonging of an edge $(i,j)$ in a community $c$ according to the
belonging coefficients of the end nodes $i,j$. The authors suggest that the
selection of the $\mathcal{F}(\cdot)$ function should lead to a valid modularity
measure: (i) $Q_{ov}$ should equals to zero when no community structure
can be identified and all nodes belong to the same community and (ii) 
higher value of $Q_{ov}$ indicates better community structure. Every potential
function that preserves these properties can be applied to modularity.
Finally, the authors present a genetic algorithm for optimizing the
proposed modularity criterion, and therefore it can be used to identify the
underlying overlapping community structure in directed networks.

\subsubsection*{Local Definition of Modularity}
The definitions for the directed version of modularity that we have presented so
far, assume that in the configuration model (the random graph model competitor),
each node could be equally connected to any other node in the network. That
is, the probability of an edge between every pair of nodes is the same,
independent of the relative position of the nodes in the graph. In Ref.
\cite{local-modularity}, Muff et al. propose a local definition of modularity
for directed networks, where the expected number of edges within each community
$c$ is computed with respect to the subgraph consisting of the community $c$
and its neighbor  communities (and not based on the full network). That is, a
local function of modularity can be expressed as

\begin{equation}
 Q_{local} = \sum_{\forall c \in C} \Bigg[ \dfrac{L_c}{L_{cN}} -
\dfrac{k^{out}_c k^{in}_c}{L_{cN}^2}\Bigg],
\end{equation}
 
\noindent where $L_c$ is  the number of edges within community  $c$, $L_{cN}$
the number of edges  contained  in $c$'s neighbor communities and
$k^{out}_c$ ($k^{in}_c$) the total external (internal) degree of community $c$.
The authors provide experimental results where the maximization of $Q_{local}$
provides more cohesive partitions in a school network dataset (interactions
among students and their classmates) as well as in the metabolic network of
\textit{E. coli}.

\subsubsection*{Discussion}
As discussed in Section \ref{sec:background}, the modularity function
suffers from the so-called \textit{resolution
limit} \cite{modularity-resolution}, i.e., modularity optimization may fail
to identify  communities smaller than a specific size that  depends 
on the scale of the network. This limitation was initially found in the
undirected
version of modularity ($Q_u$ in Eq. \eqref{eq:modularity}), but a similar
behavior is expected for the extension in directed networks (as we mentioned
earlier, there is a close connection between $Q_u$ and $Q_d$
\cite{arenas-modularity07}). Thus, the produced partition that maximizes
modularity may correspond either to single communities or to a merging of
smaller weakly connected communities. In the literature, some possible
meta-algorithmic approaches have been proposed, that can help to overcome the
resolution limit \cite{fortunato}.

\subsubsection{Spectral Clustering and Cut-based Measures for Directed
Networks} \label{sec:spectral}
In this section, we will review methodologies for the clustering problem
in directed networks that are based on the concept of spectral clustering and
graph
cuts, and we will present the close
relationship between them; these methods mainly constitute extensions of
popular clustering approaches from the undirected case. 

\par The algorithmic framework of spectral clustering was initially considered
for the case of undirected networks, and  includes methods that partition the
nodes of a graph into clusters using 
information related to the spectrum of a matrix representation of the dataset
(e.g., Laplacian or adjacency matrix). Spectral methods can be applied not
only in networks (graph structures), but generally in every set of $N$ objects
where a pairwise similarity function between them can be defined. For a nice
tutorial about spectral graph clustering, one can refer to the
survey paper of von-Luxburg \cite{von-luxburg}. Here we will present extensions
to directed networks and more precise, we will examine the generalization of the
Laplacian matrix for directed graphs.

\par Furthermore, spectral clustering methods can also be applied in a
slightly different way for solving the graph clustering problem. This can be
achieved through their close connection with the cut-based graph clustering
method.  Broadly speaking, in the graph clustering problem the goal is to
partition the nodes of a network, in such a way that the edges between different 
groups should have low weight (or in the case of unweighted networks, the number
of edges should be small), while the edges within a group should have high
weight (note that the total weight of
every cluster is considered aggregating the weights of edges). In other words,
there are two criteria of interest when quantifying how good a community or a 
cluster is. The first one considers the number of edges between the nodes of the
cluster, while
the second the number of edges between nodes of the candidate community with the
rest of network. As noted in Section \ref{sec:background}, the objective
functions for the clustering problem can be formed according to one of these
criteria (single-criterion scores) or based on a combination of them
(multi-criterion scores) \cite{leskovec-www10}. 

\par However, the optimization of these objective cut-based criteria typically
lead to computational difficult problems, but relaxed versions of them can be
turned into spectral clustering problems. The  optimization measures can be
expressed in a matrix form and then the spectrum (eigenvectors) of this matrix
can be used to obtain the final clusters. We remind here that something similar
was presented in Section \ref{sec:modularity} for the optimization of
modularity. In that case, the  modularity was expressed in a matrix
form and applying spectral techniques the partition that maximizes 
modularity was detected. Thus, it is clear that spectral methods have a dual
use: either as clustering framework itself or as an optimization framework for
objective functions. For the latter case, first we will present how cut-based
measures can be extended to directed networks and then how spectral methods
can be applied on them as  an optimization process. In the next section, we will
discuss about the connections between cuts, spectral clustering and random
walks on graphs.

\subsubsection*{Laplacian Matrix for Directed Networks}
The Laplacian matrix of an undirected graph \cite{chung} is one of the
main tools for spectral clustering. As we discussed in Section
\ref{sec:background}, the eigenvector that corresponds to the second smallest
non-zero eigenvalue of the Laplacian matrix (the so-called Fiedler vector) can
be used to obtain a bi-partition of the nodes of the graph into two sets $S$,
$\bar{S} = V-S $ with relatively small number of edges connecting the two sets
(this can be achieved through the well-known \textit{Cheeger inequality}). That
is, the eigenvectors of the Laplacian matrix provide a solution to the
\textit{normalized cut} objective function, which captures the clustering notion
of a subset $S$ as (see also Ref. \cite{shi-malik-pami00} by Shi and Malik):

\begin{equation} \label{eq:ncut}
 \text{NCut}(S, \bar{S}) = \text{Cut}(S, \bar{S}) \bigg(\dfrac{1}{\text{Vol}(S)}
+ \dfrac{1}{\text{Vol}(\bar{S})} \bigg),
\end{equation}

\noindent where $\text{Cut}(S, \bar{S}) = |\{ (i,j) \} : i \in S, j \in
\bar{S}| = \sum_{i \in S, j \in \bar{S}} A_{ij}$ and $\text{Vol}(S) = \sum_{j
\in S, t \in V} A_{jt}$ is the total
number of edges starting from nodes in $S$. Then, the optimal bi-partition of
the graph is the one that minimizes the normalized cut value and this can be
approximated by the spectrum of the Laplacian matrix.

\par In the case of directed graphs, how the above property is generalized? The
answer was initially provided by Chung \cite{chung-directed-laplacian05}, who
proposed a version of the Laplacian matrix for directed networks, based on
a random walk process. That is, for a directed network $G$
the Laplacian matrix can be defined as

\begin{equation} \label{eq:directed-laplacian}
 \mathbf{L}_d = \mathbf{I} - \dfrac{\mathbf{\Pi}^{1/2} \mathbf{P}
\mathbf{\Pi}^{-1/2} + \mathbf{\Pi}^{-1/2} \mathbf{P}^T \mathbf{\Pi}^{1/2}}{2},
\end{equation}

\noindent where $\mathbf{P}$ is the transition matrix, i.e., $P_{ij} =
\dfrac{A_{ij}}{k^{out}_i}$ and $\mathbf{\Pi} = \texttt{diag} (\pi_1, \ldots,
\pi_n)$ the diagonal matrix with the probability of finding the random walk on
each vertex (the stationary distribution of the random walk). (One can observe
that the matrix is the same with the one  used by Ref. \cite{lai-physica10} in
Eq.
\eqref{eq:lapl1}).
Moreover, the most important point is that the Laplacian matrix of Eq.
\eqref{eq:directed-laplacian} satisfies the so-called Cheeger inequality,
making it a useful tool for the graph clustering problem. In other words, the
eigenvector of the second smallest non-zero eigenvalue of $\mathbf{L}_d$
can be used to approximate a good cut in the network. Another version of
the Laplacian matrix for directed networks (called Diplacian) with similar
interesting
properties, was recently presented by Li and Zhang in Refs.
\cite{diplacian-waw10, diplacian}.

\par Based on Chung's extension of the Laplacian matrix for directed
networks, Gleich \cite{gleich-directed} proposed an hierarchical spectral
graph clustering algorithm for directed networks. The idea utilizes
Cheeger inequality that holds for the new directed Laplacian matrix and by
recursively using the eigenvector $\mathbf{u}_1$ that corresponds to the second
smallest
non-zero eigenvalue $\lambda_1$, a partition of the graph into two clusters can
be achieved. The author suggests that this recursive process can terminate when
the resulting subgraph contains less than $p$ nodes. Moreover, a possible
extension of the algorithm to higher eigenvectors is discussed, where each
higher eigenvector (other that $\mathbf{u}_1$) offers the next best solution for
the 
normalized cut. That is, the spectrum of the directed Laplacian matrix ($k$
smallest eigenpairs) can be used to partition the network into $c$ clusters.

\par A similar solution to the problem was proposed by Zhou et al.
\cite{zhou-icml05}, who considered a normalized analogous of the directed
Laplacian matrix. More precisely, the authors define the matrix 

\begin{equation} \label{eq:dir-laplacian2}
\mathbf{\Theta} = (\mathbf{\Pi}^{1/2} \mathbf{P} \mathbf{\Pi}^{-1/2} +
\mathbf{\Pi}^{-1/2} \mathbf{P}^T \mathbf{\Pi}^{1/2}) / 2,
\end{equation}

\noindent where $\mathbf{L}_{d} = \mathbf{I} - \mathbf{\Theta}$. According to
this relation between
$\mathbf{L}_{d}$ and $\mathbf{\Theta}$, the best normalized cut will correspond
to the eigenvector of the second largest eigenvalue of $\mathbf{\Theta}$
(instead of second smallest in the case of $\mathbf{L}_{d}$). Algorithm
\ref{alg:laplacian-clustering} describes the pseudocode of the clustering
algorithm for directed networks based on the above discussion, where the graph
is partitioned into two parts (the algorithm is similar to the one presented by
Gleich in Ref. \cite{gleich-directed}).

\begin{algorithm} 
 \caption{Directed Spectral Clustering \label{alg:laplacian-clustering}}
\begin{algorithmic}[1]
 \Statex \textsc{Input:} Directed graph $G = (V, E)$
 \Statex \textsc{Output:} A partition  of the vertex set $V$ into two parts,
minimizing the normalized cut
 \Statex
 \State Define a random walk over $G$ with transition matrix $\mathbf{P}$.
 \Statex
 \State Form the matrix $\mathbf{\Theta} = (\mathbf{\Pi}^{1/2} \mathbf{P}
\mathbf{\Pi}^{-1/2} + \mathbf{\Pi}^{-1/2} \mathbf{P}^T \mathbf{\Pi}^{1/2}) / 2$,
where $\mathbf{\Pi}$ is the diagonal matrix with elements being the
stationary distribution of the random walk.
 \Statex
 \State Compute the eigenvector $\mathbf{u}_2$ of $\mathbf{\Theta}$ that
corresponds to the second  largest eigenvalue; then partition the vertex set $V$
into two parts $S
= \{ i \in V | \mathbf{u}_2(i) \ge 0 \}$ and $S' = \{ i \in V | \mathbf{u}_2(i)
< 0 \}$.
\end{algorithmic}
\end{algorithm}

\par The above algorithm can be extended in the case of a
$k$-partition (instead of a bi-partition), considering the eigenvectors that
correspond to the $k$ largest eigenvalues of $\mathbf{\Theta}$. Furthermore, in
the case
of labeled data (where each node is associated with  a label), the above
methodology can be used as a general learning (classification) framework for
directed networks. Later, the framework was extended to the case
of graphs with multiple views, where data is associated with multiple
representations \cite{zhou-icml07}. For example, in the case of the web graph,
each web page can be represented  either as a node in a directed network based
on the hyperlink structure, or using the vector-space model based on
occurrences of words in a web page. These two different views can be combined as
a directed hyperlink network, weighted according to the similarity of the web
pages. In the general case, each different view can be represented as a directed
network with the same set of nodes $V$ and the idea is to combine these
different views to improve the accuracy of the learning framework (e.g., the
graph clustering task).

\subsubsection*{Cut-based Measures for Directed Networks}
A basic point of the Laplacian-based spectral clustering algorithm that we
presented above, is that it provides a solution to the normalized cut problem.
The objective criterion that is optimized while using the
eivenvectors of the Laplacian matrix (of a directed or undirected network), is
a generalized version of the normalized cut. Additionally, other possible
cut-based objective criteria can also be applied to the clustering problem in
directed networks, as the one of weighted cuts proposed by Meil\u{a} and Pentney
\cite{meila-sdm07}. More precisely, the authors introduced
the \textit{generalized weighted cut} criterion, defined as follows

\begin{equation} \label{eq:wcut}
 \text{WCut}(S, \bar{S}) =  \dfrac{\sum_{i \in S, j \in \bar{S}}
T'_i A_{ij}}{\sum_{i \in S} T_i}
+ \dfrac{\sum_{j \in \bar{S}, i \in S} T'_j A_{ji}}{\sum_{j \in \bar{S}} T_j}.
\end{equation}

\noindent   One can observe that this criterion is similar to the one of
normalized cut (Eq. \eqref{eq:ncut}), but it is parametrized by the vectors $T$
and $T'$. That is,
the objective is to form node clusters of balanced size, where clusters' size
is parametrized by vector $T$ , while vector $T'$ plays the role of a
normalization factor for the adjacency matrix $\mathbf{A}$. An important point
is that different normalized cut-based measures can be recovered from the
definition of $\text{WCut}$, by properly setting the parameter vectors $T$ and
$T'$. This point  makes the new criterion more flexible. The authors show that
the optimization of the WCut function can be relaxed on an analogous symmetric
problem, where many
existing spectral clustering algorithms and theoretical results can be applied
for extracting the final clusters. The experimental results show that the
symmetrized version of the spectral clustering problem produced by the weighted
cut objective function, gives better results compared to the cases where the
matrix is symmetrized using simple linear algebraic transformations (e.g., some
of those presented in Section \ref{sec:transf}).

\par In the context of image processing and analysis, the authors of Ref.
\cite{Yu-shi01} present an approach for clustering directed networks,
generalizing the normalized cuts criterion. At a first
point, a new representation scheme is proposed, in which all possible pairwise
relationships are characterized according to two types of node correlations,
namely \textit{attraction} and \textit{repulsion}. According to these
relationships, the general compatibility between two pixels in the image can
be captured by two different directed networks that correspond to each of these
two relationships. The general idea behind the approach is that at the
clustering process, the attraction of
nodes that belong on the same group should be as large as possible, while the
repulsion between two different groups should be minimized. Then,
information about these two graphs is introduced in a ``dual'' clustering
criterion that extends the notion of normalized cuts. Finally, the optimization
of the produced objective functions leads to an eigendecomposition problem of a
Hermitian matrix, where the imaginary part encodes directed relationships, while
the real part encodes undirected relationships with positive numbers for
attraction and negative numbers for repulsion.

\subsubsection{PageRank and Random Walk based Methods}
\label{sec:random-walk}
The PageRank and generally random walks over graphs are closely related to
spectral clustering. That is, cut-based measures in networks (e.g., normalized
cuts) and their optimization process, can be expressed in terms of random walks
\cite{maila-shi-aistats01}. Broadly speaking, the minimization of the number of
edges that crossing a cut in a network can be described as a similar process
where the random surfer  is forced to stay more time within a cluster. In
other words, the normalized cut objective criterion presented in
Eq. \eqref{eq:ncut}, corresponds to the probability of the random walk
transitioning from the vertex set $S$ to set $\bar{S}$ in one step if it is
currently in $S$ and the random walk is started in the stationary distribution
(or vice-versa):

\begin{equation} \label{eq:ncut-randomwalk}
\text{NCut} (S, \bar{S}) = \dfrac{\Pr(S \rightarrow \bar{S})}{\Pr(S)} +
\dfrac{\Pr(\bar{S} \rightarrow S)}{\Pr(\bar{S})}.
\end{equation}

\noindent That is, if $\boldsymbol \pi$ represents the stationary distribution
of the
random walk, then the probability $\Pr(S)$  with which the
random surfer can be found in a node in $S$ can be defined as $\Pr(S) =
\sum_{i \in S} \pi_i$ (similarly for $\Pr(\bar{S})$ and also $\Pr(S) +
\Pr(\bar{A}) = 1$). Then, the probability that the random walk will move from
$S$ to
$\bar{S}$ can be defined as $\Pr(S \rightarrow \bar{S}) = \sum_{i \in S, j \in
\bar{S}} \pi_i P_{ij}$, where $\mathbf{P} = [P_{ij}]_{i,j \in V}$ is the
transition matrix.

\par Equation \eqref{eq:ncut-randomwalk} considers the general case of graphs
with the existence of directed edges and thus can be naturally applied in
directed networks (in the case of undirected networks $\Pr(S \rightarrow
\bar{S}) = \Pr(\bar{S} \rightarrow S)$, since the probability of
transition from $S$ to $\bar{S}$ is equal to the one from $\bar{S}$ to $S$). As
we described in Section \ref{sec:spectral}, this criterion can be approximated
by the eigenvalues of the Laplacian matrix for directed networks
\cite{zhou-icml05}.

\par It also holds for undirected networks that relying only on the first top
$k$ eigenvectors of the transition matrix $\mathbf{P}$ which is related to the
random walk, one is able to identify the underlying clusters
\cite{meila-multiway04}. A similar result was presented for directed networks,
which states that a clustering can be achieved by looking for piecewise constant
eigenvectors in the transition matrix $\mathbf{P}$ \cite{pentnei-meila-aaai05}.
However, as noted by Capocci et al. \cite{capocci-physica-A-2005}, one can
rely on the eigenvectors of matrix $\mathbf{P}$, if the network has a clear
modular structure. However, in practice, this is something that occurs
rarely; typically, real large networks have no clear community structure, and
the eigenvector components do not show a clear step-wise form. To deal with this
issue, Capocci et al. \cite{capocci-physica-A-2005} presented an approach where
the underlying community structure is revealed by correlations between the
same components of different eigenvectors. That is, the eigenvector components
that correspond to nodes of the same cluster, will show high correlation among
each other. In the case of directed networks, the adjacency matrix of the
network is replaced by matrix $\mathbf{A} \mathbf{A}^T$ and a similar
methodology is applied (therefore the method first transforms the directed
network to undirected by a transformation approach which introduces edges
between nodes with common neighbors).

\par In the context of community detection in the directed Web graph,
Huang et al. \cite{huang-pkdd06} proposed to extend the random-walk based
approach, using some variations of random walks that are able to identify latent
Web communities. That is, instead of only satisfying a normalized cut
criterion where two web pages (nodes) are assumed to be related  if they are
directly connected, the authors also consider the case of pattern-based clusters
where co-citation and co-reference information is taken into consideration. In
other words,
the random walk should ensure that Web pages that share a common topic or
interest should be grouped together, even if they are not directly connected
(the case of pattern-based clusters that we described in Section
\ref{sec:pattern-based}). More precisely, their first approach involves two
versions of the \textit{PageRank random walk} (or teleporting random walk)
\cite{pagerank}, one following the forward hyperlinks while the other the
backward ones. The first one can be considered as an authority-based ranking of
 nodes while the second as hub-based (e.g., see Ref. \cite{hits}). Moreover,
PageRank guarantees the convergence to a stationary distribution through  the
adoption of a damping factor (in case of absence of in/out edges for a node).
In their second approach, the authors consider a two-step random walk, in order
to reveal latent communities that  imposed by the existence of co-citation
and co-reference edges. That is, starting from a node $u$, the random surfer
first jumps one hop backward to a hub node $h$ with probability $P^-_{uh} =
A_{hu}/k^{in}_u$, and then she moves one step forward to a node $v$ (adjacent to
$h$) with probability $P^+_{hv} = A_{hv}/k^{out}_h$. Then, the two-step
transition probability between authorities $u,v$ is defined as 

\begin{equation}
 P^A_{uv} = \sum_{h} P^-_{uh} ~ P^+_{hv},
\end{equation}

\noindent where nodes are treated as authorities. Similarly, forcing the random
surfer to move firstly one step forward and then one step backward, the nodes
are treated as hubs, and the transition probability matrix $\mathbf{P}^H$ can be
similarly defined. Moreover, since both these two-step random walks require that
each node should have incoming and outgoing edges, again a teleporting
probability (damping factor) is introduced. Finally, the above two types of
random walks can be combined
through a convex combination, in order to consider both co-citation and
co-reference node similarity. In this case, the transition matrix can be
expressed as $\mathbf{P} = \beta \mathbf{P}^A + (1-\beta)\mathbf{P}^H$, where
parameter $\beta$ controls the co-citation and co-reference effects. Since the
modified transition matrix $\mathbf{P}$ has been defined, it can be applied to
the Laplacian matrix of Eq. \eqref{eq:dir-laplacian2}, and spectral methods
can be used to extract the clusters. 

\par A similar PageRank-based approach for clustering hypertext document
collections that are represented by directed networks, was introduced in Ref.
\cite{clustering-sigir08}. The proposed algorithm (called PRC) is composed by
two parts. In a first step, a set of centroid nodes are selected (according to a
node ranking criterion such as PageRank or Hits), and after that, the nodes are
assigned to clusters using a Personalized PageRank method\footnote{This is a
variant of the PageRank algorithm where a set of nodes is favored by the random 
walk. That is, the probability that the random surfer will jump to a node in a 
teleport step is not uniform for
all nodes (as in the PagerRank algorithm). In the extreme case, only one node is
favored.} (also called topic-specific or local PageRank)
\cite{topic-sensitive-pagerank}, combined with similar spectral 
optimization tools like those presented earlier. 

\par In Ref. \cite{lai-message-passing} the authors combined random walks with
the concept of affinity propagation \cite{affinity-propagation} and proposed a
message passing algorithm for community detection in both directed and
undirected networks. Affinity propagation is a mechanism that has been
previously used in the task of clustering data points, where each group is
associated with a representative point. Broadly speaking, the method of
community detection via affinity propagation can be likened to an election
process, in which nodes represent voters and the group leaders are the
representative nodes. Through message passing along the edges of the
network, the nodes are able to identify the community that they belong to; this
is determined by its community leader, examining the similarity with their
neighbors. The similarity between two nodes is computed using random walks
and a variant of the transition matrix; each node is finally represented as a
vector in $\mathbb{R}^n$ and well-known similarity metrics are applied (e.g.,
cosine similarity).

\par A somewhat different version of the problem is the one of local graph
partitioning, where instead of clustering the whole graph, the goal is to
find a ``good'' local clustering structure near a specified seed node, by
examining only  a small portion of the input graph. As we will discuss later at
this paper, this is a very interesting variant of the community detection
problem. In Ref. \cite{lang-local}, Andersen et al. propose a local clustering
algorithm
for directed networks (extending a similar method for undirected networks),
combining information from both the Personalized PageRank score of a node $v$
and the global one (i.e., the classical version of PageRank). That is, for a
specific seed node, the authors compute the Personalized PageRank score with
a single starting node (the seed node), as well as a global PageRank score with
a uniform starting distribution over all nodes. Then, it is proved that
taking the ratio of the entries in the Personalized PageRank and global PageRank
vectors and sorting the nodes of the graph according to this ratio, one
is able to identify a local set of nodes (cluster) with good clustering
properties. The quality of the obtained cluster is determined by the measure of
conductance (e.g., see Ref. \cite{leskovec-www10}) which is generalized for
directed networks.

\subsubsection{Other Extensions}

In this section we describe a few other diverse approaches for detecting
communities in directed networks, that mainly extend concepts from the
undirected case of the problem.

\subsubsection*{Clique Percolation Method for Overlapping Community Detection} 
\begin{figure}[t]
 \centering
 \includegraphics[width=.7\textwidth]{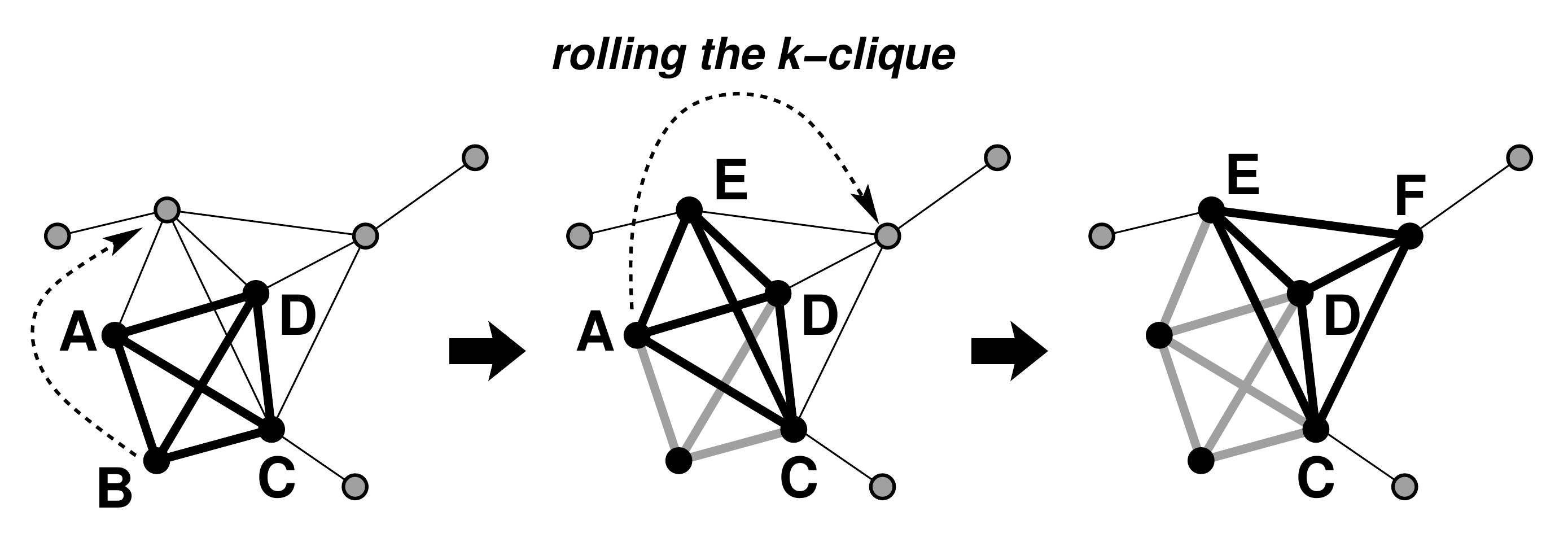}
 \caption{Illustration of the Clique Percolation Method for undirected networks
\cite{palla-directed-network-modules}. Initially, a template $k$-clique ($k=4$)
is placed on nodes A-B-C-D. The template is gradually rolled to adjacent
$k$-cliques and the final module consists of nodes A-B-C-D-E-F. The
figure is courtesy of Palla et al. \cite{palla-directed-network-modules}.
\copyright 2007 IOP Publishing. \label{fig:cpm}}
\end{figure}

Palla et al. \cite{palla-directed-network-modules} presented a technique
which extends the \textit{clique percolation method} initially proposed for
undirected networks. The goal of the method is to detect  network
modules (i.e., dense connected groups of nodes), following a local search
approach based on edge
density. Moreover, the produced modules may overlap with each other (i.e., a
node may belong to more than one communities). In the case of undirected
networks, the clique percolation method considers that the definition of 
modules is based on adjacent $k$-cliques. A $k$-clique is a complete subgraph
with $k$ nodes, while two
$k$-cliques are adjacent if they share $k-1$ nodes. A module is defined to be
the union of $k$-cliques that can be reached from each other traversing the
edges of adjacent $k$-cliques. In other words, considering a $k$-clique as a
template, the modules can be identified by rolling the template to an adjacent
$k$-clique (retaining all but one node fixed) as shown in Fig. \ref{fig:cpm}.

\par The method is extended in directed networks, defining the concept of
directed $k$-cliques as complete subgraphs of size $k$, where the nodes can be
ordered such that between any pair of nodes there is a directed edge from a
higher order node to a lower one. The ordering is obtained according to the
restricted out-degree of a node in the $k$-clique (the number of
out-neighbors in the clique). Then, the directed $k$-clique modules are defined
in a similar way as in the undirected case, by considering the union of
adjacent directed $k$-cliques. The authors discuss that the proposed definition
of $k$-cliques for directed networks is not unique, and other extensions
can be considered as well.

\subsubsection*{Local Density Clustering} 
One other clustering method that aims to detect clusters based solely on
local information, is the one presented by Schaeffer and Virtanen in Refs.
\cite{schaeffer-pakdd05, chilean-web} respectively. The basic idea of the
approach is to extend the concept of cluster density to directed networks. (The
problem is similar to the one presented earlier in Section
\ref{sec:random-walk} about local graph partitioning). More precisely, a local
search method is applied in order to find a good cluster that
contains a specified seed node (the approach can be naturally extended to
several seed nodes). That is, the internal degree of a local cluster $C \
\subseteq V$ is defined as $\text{int-deg}(C) = | \{ (u,v) \in E | u,v \in C
\}|$ (i.e., the number of edges with both  endpoints in $C$), while the
external degree is $\text{ext-deg}(C) = | \{ (u,v) \in E | u \in C, v \notin C
\}|$ (i.e., the number of directed edges $(u,v)$ that have only the start
node $u$ in $C$). The density of the directed network $G=(V,E)$ is defined as
$\delta = \dfrac{m}{n(n-1)}$, where $m=|E|$ and $n=|V|$. Similarly, the density
of a cluster $C$ (also called local density) can be defined as $\delta_\ell (C)
= \dfrac{\text{int-deg}(C)}{|C| (|C| - 1)}$ and the relative density as
$\delta_r (C) = \dfrac{\text{int-deg}(C)}{\text{int-deg}(C) +
\text{ext-deg}(C)}$. The authors combine the local and relative density and the
final quality measure is selected to be the product of them: $f(C) =
\delta_\ell(C) \cdot \delta_r(C)$. Having define the clustering quality
function, the problem of local clustering can be stated as follows: find a
subgraph $C$
with $k$ nodes (i.e., the cluster) that contains a given node $v
\in V$, maximizing $f(C)$. Since this is a computational difficult problem, the
authors propose a local search approach starting from node $v$ and
gradually expanding the subgraph around $v$.

\subsection{Alternative Approaches for Community Detection in Directed Networks}
In this section we review ``alternative'' clustering approaches for directed
networks that do not belong to one of the previous
categories. While the approaches described so far either transform the original
directed network to undirected or constitute well-known extensions from
undirected to directed networks, here we will review algorithms that follow
different and diverse methodological approaches.
We classify them in three categories, according  to the main methodology they
follow, namely (a) information-theoretic, (b) mixture models and statistical
inference, and (c) stochastic blockmodels. One additional category is devoted to
 approaches that mainly deal with variations of the clustering problem
(e.g., community detection in dynamic directed networks). We note that even some
of these approaches have been applied in the past on undirected networks, we
decide to review them independently, trying to identify and demonstrate
additional concepts that can be used for the problem. Moreover, some of them can
be applied on both directed and undirected networks (i.e., independent from edge
directionality), and we briefly discuss about this interesting feature. Although
categories (b) and (c) both refer to approaches on probabilistic models,
we decide to review them independently, since they are based on different
statistical inference techniques.

\subsubsection{Information-Theoretic Based Approaches}
\label{sec:information-theoretic}
A prominent methodology for extracting the community structure of a network is
the one that applies information-theoretic and compression principles.
Generally, the existence of communities in networks represent structural
patterns and regularities, that similar to more traditional data mining and
analysis tasks, they can be used to effectively compress the network
(data), e.g., Refs. \cite{rosvall-bergstrom-pnas07,
megalooikonomou-faloutsos-dmkd-07}. Rosvall and Bergstrom \cite{rosvall-pnas08}
proposed a method (called Isomap) to identify communities in directed networks,
by combining random walks and compression principles. That is, the modules of
the network can be recognized based on how fast information flows on them.
The authors apply the concept of random walks to describe the process of
information flow in the network and the clusters can be extracted by compressing
the description of the random walk. As we have already discussed, a community
corresponds to a group of nodes in which the random surfer is more likely be
trapped in, visiting more time nodes of the group than other nodes outside of
that. Thus, intuitively, a community would correspond to a group of nodes in
which the random walk can be compressed better and the problem can be
reformulated as a coding one: the goal is to select a partition $M$ of the $n$
nodes into $c$ communities, minimizing the description length of the random
walk.

\par At a first step, each node in the network is described by a unique
codeword based on the visiting frequency of the random walk. Using Huffman
coding, shorter codewords are assigned to more frequently visited nodes. At a
second step, the random walk trajectory on the network can be described
following a two-level description: unique names (codewords) are assigned to
the clusters of the network (coarse-grained structure), while the codewords for
the description of nodes inside a module are reused (fine-grained structure).
Thus, reporting only the codewords that have been assigned to communities,
a coarse-grained description of the network is achieved. The procedure is
similar to the one used while designing a geographic map; unique
names are assigned to cities (communities in our case), while names for the
streets (nodes in our case) of a city can be reused. Then, the clustering
problem can be expressed as finding the partition that yields the minimum
description code length. If the network has a well-defined community structure,
the above two-level description scheme will produce shorter code length:
the random walk will jump between different communities infrequently and thus
the description length will be shorter (since the codewords represent
individual nodes are shorter). The minimization of the description length can be
achieved combining greedy search and simulated annealing methods.
Regarding the clustering results, the Isomap algorithm is able to identify
pattern-based clusters and more specifically clusters of flow patterns induced
by the edges of the network.

\par
\begin{figure}[t]
\centering
 \begin{tabular}{ccc}
  \includegraphics[width=.3\textwidth]{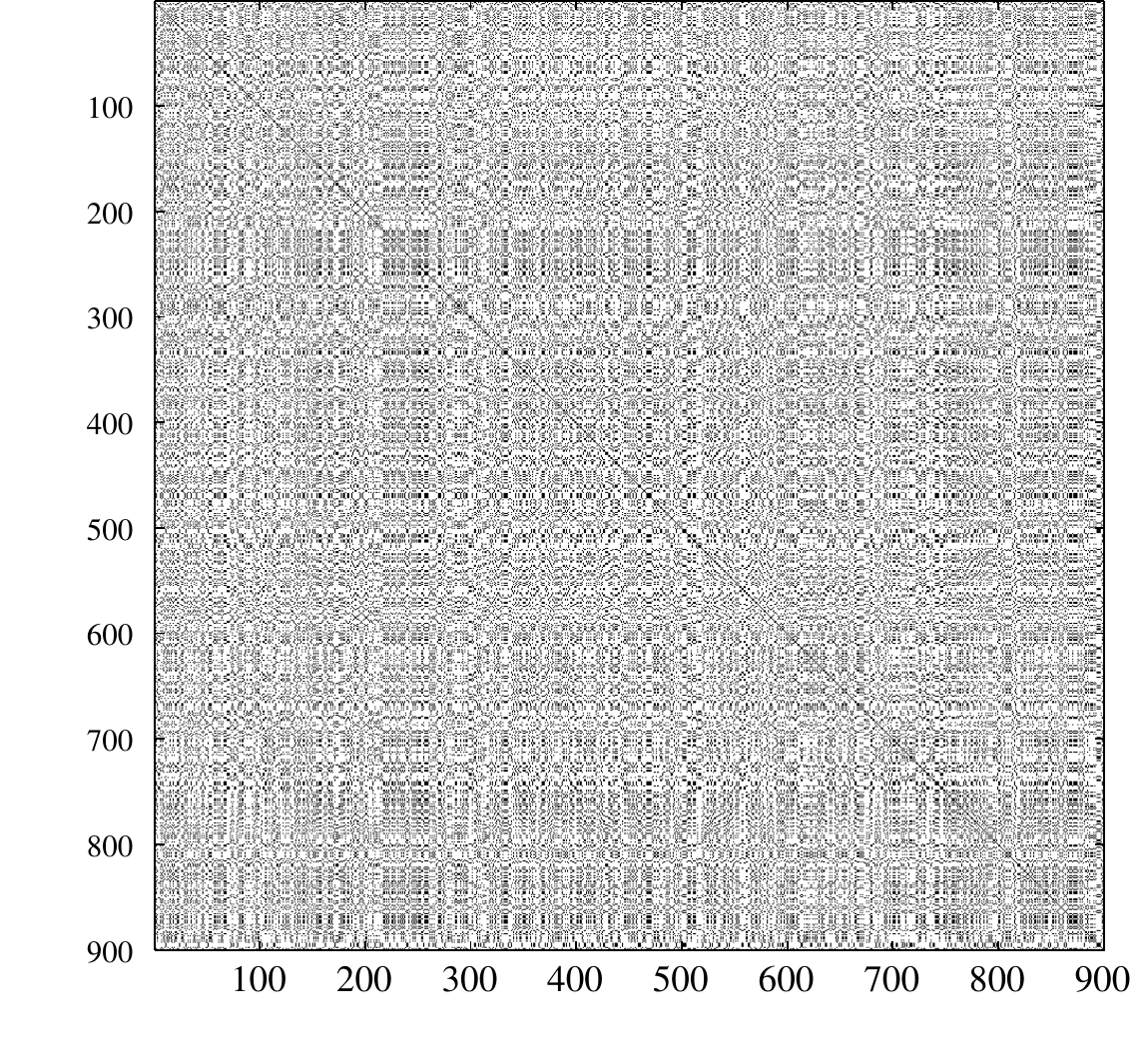} & ~~~~~~~~ &
  \includegraphics[width=.3\textwidth]{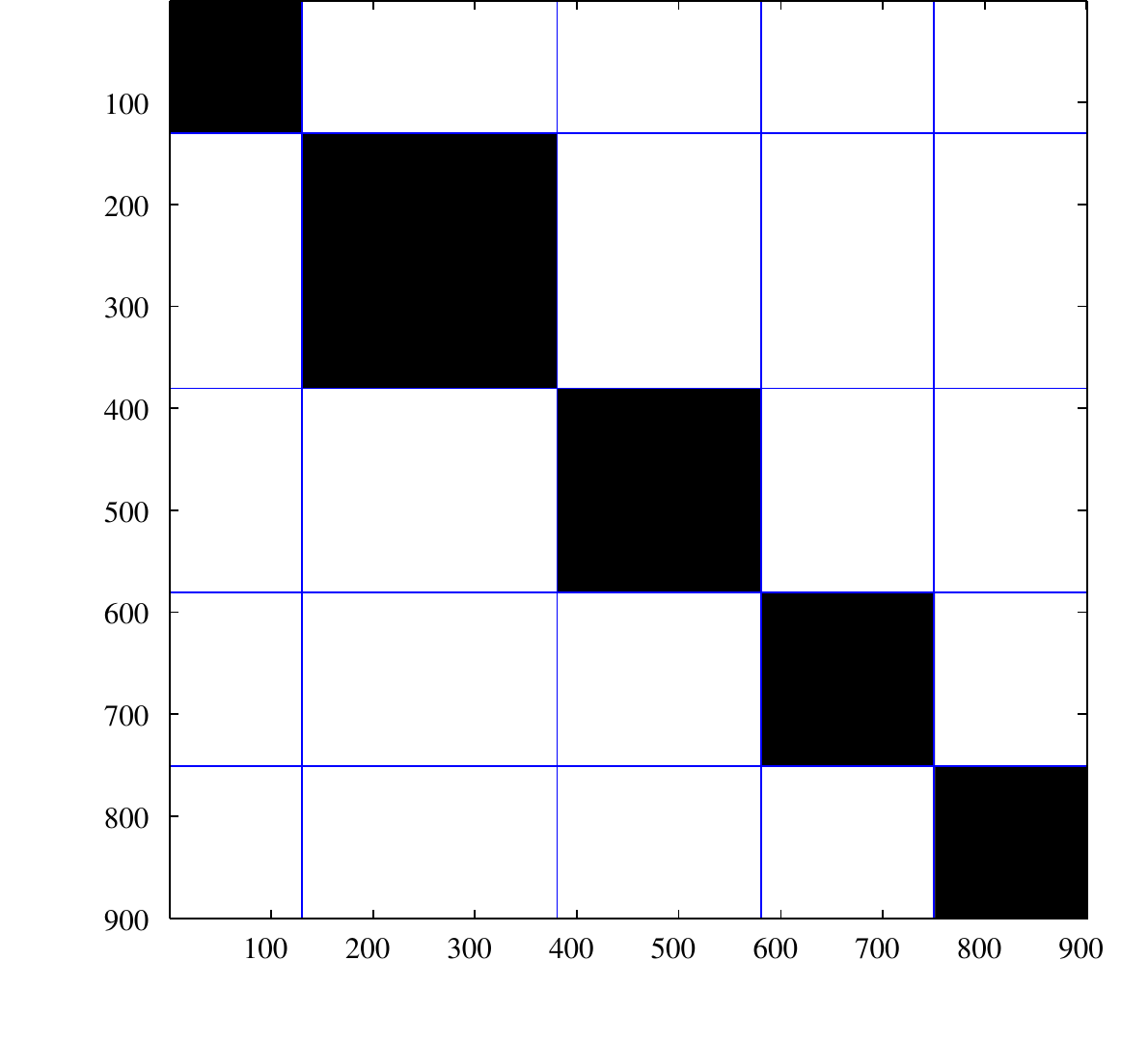} \\
   (a) Initial matrix & ~~~~~~~~ &  (b) Reordered matrix
 \end{tabular}
 \caption{An example of the community detection algorithm through
co-clustering: (a) The initial adjacency matrix. (b) The columns and rows
of the matrix are reordered forming homogeneous blocks that can be used to
better compress the matrix. At the end, these blocks correspond to co-clusters.
Figures redesigned from Ref. \cite{autopart-pkdd05}. \copyright 2004 Springer.
\label{fig:autopart}}
\end{figure}

 A somewhat different formulation of information theoretic principles in
the community detection problem has been presented by
Chakrabarti in Ref. \cite{autopart-pkdd05} (even though the algorithm is tested
on undirected networks, it seems that it can be applied in directed networks as
well). The proposed algorithm (called AutoPart) can be considered as a 
co-clustering tool for binary matrices (the adjacency matrix in our case),
where compression concepts are applied to identify the underlying clustering
structure\footnote{The term co-clustering  refers to the task of simultaneously
clustering the rows and columns of a matrix. As we will present later in this
Section, the formulation of the co-clustering problem is similar to the
blockmodeling approach.}. The goal of the algorithm is to
group the nodes of the network into clusters in such a way that the adjacency
matrix will be divided into rectangular, homogeneous blocks of high or low
density, indicating  that the certain node groups have more (or less)
connections with other groups (e.g., Fig. \ref{fig:autopart}). This can be
achieved through a reordering procedure of the adjacency matrix, where the rows
and columns of the matrix are rearranged to achieve this structure. The quality
of different possible clustering structures is evaluated in terms of
the total compression cost $T$. That is, the best compression scheme should
achieve a tradeoff between the number of produced blocks (i.e., clusters) and
how homogeneous these blocks are. In the two extreme cases, one could select
either only one block (the whole matrix) but not very homogeneous, or $n^2$
perfectly homogeneous blocks of size 1 (each cell of the matrix). This tradeoff
is achieved applying the Minimum Description Length principle (MDL) for model
selection: the best clustering (model) is the one that minimizes both the
compression cost of the data as well as the cost for the ``summary'' of the node
groups. 

\par To minimize the total compression cost $T$, a two-step iterative approach
is applied. Initially, the
graph is considered as a single cluster itself.  At each iteration, the
algorithm first finds a good node grouping for a given number of clusters, and
then,  is looking for the number of clusters $k$ to be formed by
splitting the previously created clusters with the maximum entropy per node.
This iterative procedure continues until finding the optimal number of clusters
$k$, for which the compression cost $T$ cannot further be decreased. Thus the
complexity of the method is $\mathcal{O}(Imk^2)$, where $I$ the number of
iterations to achieve convergence of the  compression cost (the author state
that in practice $I \le 20$ iterations are enough). To conclude, the main
features of the algorithm are: (i) it treats the community detection problem as
a co-clustering task where the number of  clusters is automatically determined
by the MDL principle, and (ii) it scales linearly with respect to the number of
edges.

\subsubsection{Probabilistic Models and Statistical Inference}
\label{sec:probabilistic-models}
A different formulation and solution for the community detection problem in
networks can be achieved applying \textit{statistical inference} methods.
Broadly speaking, statistical inference\footnote{Wikipedia's lemma for
\textit{statistical inference}:
\url{http://en.wikipedia.org/wiki/Statistical_inference}.} is the process of
drawing conclusions from data, subject to a set variables. Newman and Leicht
\cite{leicht-mixture07} proposed an approach for community detection in
directed networks  based on mixture models for statistical inference.
More precisely, let $c$ be the number of communities in the network and assume
that $g_i$ represents the community (group) that node $i$ belongs to. The group
memberships are initially unknown and the goal of the algorithm is to infer them
from the observed network structure. To this direction, the authors propose to
use a mixture model\footnote{Wikipedia's lemma for \textit{mixture models:}
\url{http://en.wikipedia.org/wiki/Mixture_model}.} for the underlying
communities and their properties, in which  its parameters can be adjusted to
find the best fit to the network. This point is particularly significant since
the method does not assume any prior information about the network structure.

\par Assume that $\pi_r$ is a variable that represents the fraction of nodes in
community $r$ and $\theta_{ri}$ is the probability of existence of a
directed edge from a particular node in community $r$ to a node $i$ (i.e., the
preferences of nodes in $r$ about which other nodes they link to). The following
quantities are used to define the model: the network data
$\{A_{ij}\}$, the missing data $\{g_i\}$ (i.e., community assignment), and 
model parameters $\{\pi_r\}, \{\theta_{ri}\}$. Defining a community as a set
of nodes that have similar connection patterns to each other, the task of
community detection can
be formulated as a likelihood maximization problem. In this case, the goal is
to maximize the likelihood $\Pr(\mathbf{A},g|\pi, \theta)$, i.e., the
probability that the data were generated by the given model, with respect to
model's parameters. A common approach is to maximize the log-likelihood function
instead of the likelihood itself. At the end, the expected probabilities
$q_{ir}$ that node $i$ belongs to community $r$ can be expressed in terms of
$\{\pi_i\}$ and $\{\theta_{ri}\}$ as

\begin{equation} \label{eq:em1}
 q_{ir} = \dfrac{\pi_r \prod_{j} \theta_{rj}^{A_{ij}}}{\sum_{s} \pi_{s}
\prod_{j} \theta_{sj}^{A_{ij}}}.
\end{equation}

\noindent Moreover, the authors state that the maximization of the likelihood
occurs when

\begin{equation} \label{eq:em2}
 \pi_r = \dfrac{1}{n} \sum_{i} q_{ir}, ~~ \theta_{rj} = \dfrac{\sum_{i}
A_{ij} q_{ir}}{\sum_{i} k^{out}_i q_{ir}},
\end{equation}

\noindent where $k^{out}_i$ is the out degree of node $i$. Combining Equations
\eqref{eq:em1} and \eqref{eq:em2}, an expectation-maximization (EM) algorithm
can be applied to produce the belonging probabilities $q_{ir}$ (the authors
state that the convergence of the algorithm is fast).

\par As we mentioned earlier, the major strength of this approach is that it
is independent from the underlying clustering structure of the network, making
it capable to reveal various types of  community structure. However, the
number of communities is a parameter and needs to be specified a priori, but the
authors state that it can also be inferred from the data.

\par
\begin{figure}[t]
\centering
 \begin{tabular}{ccccc}
  \includegraphics[width=.2\textwidth]{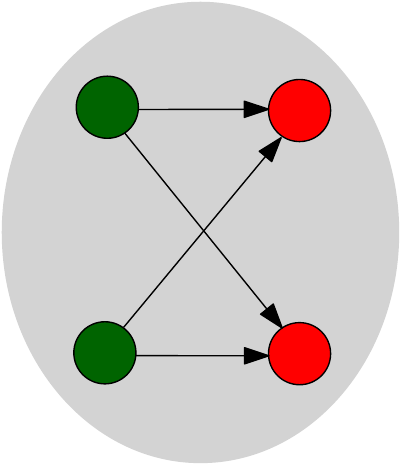} & ~~~~ &
  \includegraphics[width=.2\textwidth]{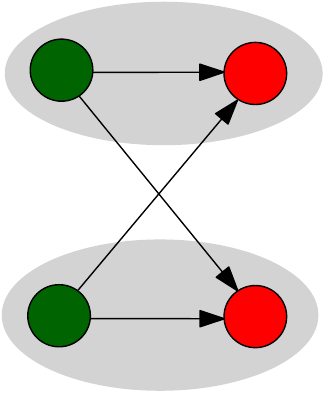} & ~~~~ &
  \includegraphics[width=.2\textwidth]{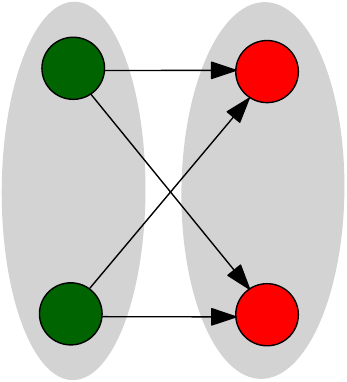} \\
   (a)  & ~~~~ &  (b)  & ~~~~ & (c)
 \end{tabular}
 \caption{A simple case where the mixture model proposed by Newman and Leicht
\cite{leicht-mixture07} has problem to assign the nodes of the network into
communities (colors indicate community membership). The possible outputs of
the method are those presented in the shadowed regions of (a) and (b), while
the natural grouping presented in (c) cannot be identified (proposed by Ramasco
and Mungan in Ref. \cite{content-based-communities-physrev08}). Figure
redesigned from Ref. \cite{content-based-communities-physrev08}. \copyright 2008
American Physical Society. \label{fig:em-community-problem}}
\end{figure}

 In a subsequent work, Ramasco and
Mungan \cite{content-based-communities-physrev08} observed that in the model
of Newman and Leicht \cite{leicht-mixture07} that presented above, the
probability $\theta_{ri}$ that a node $i$ has an incoming edge from a node in
community $r$, suggests that each community $r$ should have at least one node
with non-zero out-degree. However, this constraint may have impact at the
produced communities, as depicted in Fig. \ref{fig:em-community-problem}. In
this case, the EM algorithm cannot identify the more natural and intuitive
communities of a bipartite directed network, as shown in Fig.
\ref{fig:em-community-problem} (c). On the other hand, the
cluster assignments in Fig. \ref{fig:em-community-problem} (a) and (b) are the
possible outputs of the algorithm. 

\par To avoid this problem, the authors of
Ref. \cite{content-based-communities-physrev08} generalize the EM approach, in
such a way that the direction of edges do not restrict the possible
assignment of nodes into groups. This can be achieved replacing the edge
probabilities $\theta_{ri}$ by three new types of probabilities:
(i) $\overrightarrow{\theta_{ri}}$ representing the probability of a directed
edge from a node of community $r$ to node $i$, (ii)
$\overleftarrow{\theta_{ri}}$ for the probability of having a directed edge from
node $i$ to a node inside community $r$, and (iii)
$\overleftrightarrow{\theta_{ri}}$ for a bidirectional edge between node $i$ and
a node in community $r$. Then, the problem is formulated according to the above
new parameters and the generalized EM method is able to detect a broad range
of diverse types of communities. Moreover, the authors provide a way for
determining the number of communities in the EM formulation of the problem.

\par Another extension of the mixture models method by Newman and Leicht
\cite{leicht-mixture07} has been presented by Wang and Lai
\cite{wang-mixture-new-j-phys08}. The authors modified a subset of the
parameters of  mixture models, adding some interesting features to the
proposed APBEMA algorithm, such as independence from the degree distribution of
the network (i.e., there is no restriction about the degree of each node that
affects the produced communities) and applicability to both directed and
undirected networks without any modifications.

\subsubsection{Blockmodeling Methods} \label{sec:blockmodeling}
\begin{figure}[t]
\centering
  \includegraphics[width=.35\textwidth]{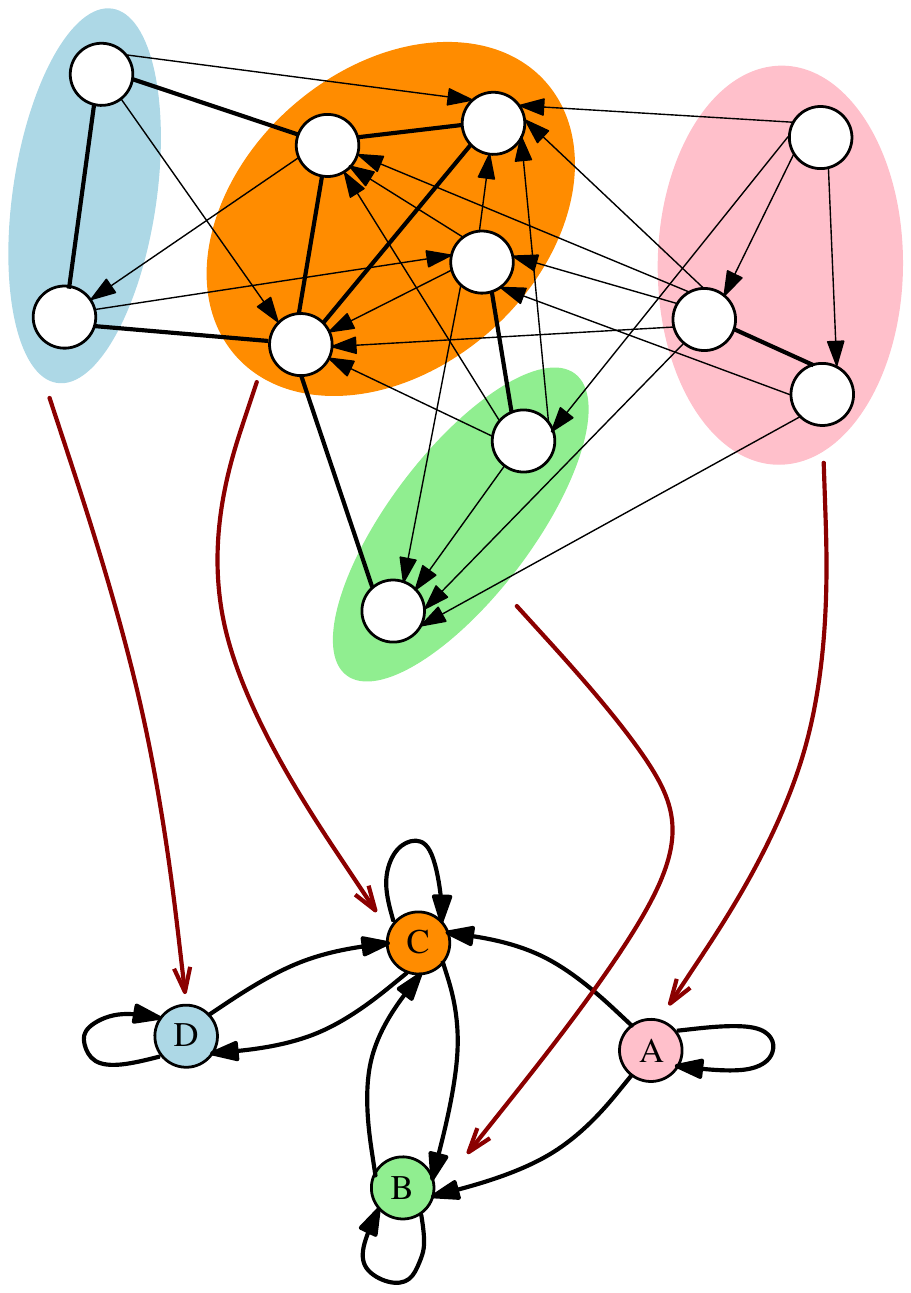}
 \caption{Schematic representation of a blockmodeling example
\cite{pajek}. The top graph corresponds to the initial directed network and the
bottom one depicts how the network is represented by a blockmodeling approach.
Figure redesigned from Ref. \cite{pajek}. \copyright 2002 Springer.
\label{fig:blockmodeling}}
\end{figure}

\textit{Blockmodeling} is an approach that has been extensively used  to analyze
and describe the structure of social networks and generally relational data
(e.g., see Refs.
\cite{Batagelj97noteson, pajek, generalized-blockmodeling}. The goal of
blockmodeling is
to represent a large and possibly incoherent network, by a smaller structure
that can be interpreted more easily. In other words, blockmodeling can be
considered as a clustering procedure, where the nodes of the network are
grouped together according to how \textit{equivalent} they are, under some
meaningful definition of equivalence. This procedure is similar to a
reordering scheme of the adjacency matrix, causing the formation of a block-wise
structure similar to the co-clustering task for community
detection  described in Section \ref{sec:information-theoretic}.
Figure \ref{fig:blockmodeling} depicts an example where a blockmodeling approach
has been applied in a directed network (top part of the figure), and how the
network is finally represented (bottom part) \cite{pajek}. That is,
the corresponding blockmodel can be described by a matrix $\mathbf{B}_{c \times
c}$, where $B_{gq}=1$ if an edge exists between  groups (communities) $g$
and $q$,
and the goal is to find the node assignment into groups and the matrix
$\mathbf{B}$ that best fits the adjacency matrix $\mathbf{A}$ of the graph.

\par Usually, two definitions of equivalence have been proposed: (i) structural
equivalence, where the nodes are equivalent if they have the same connection
patterns to the same neighbors, and (ii) regular equivalence, in which  nodes
are equivalent if they have the same or similar connection patterns to
(possibly) different neighbors. Structural equivalence can be extended to
probabilistic models, where the notion of stochastic equivalence is
introduced: the nodes of the same group are said to be stochastic equivalent if
their linking probabilities to any other node of the graph are the same. Holland
et al. \cite{holland-stochastic-blockmodels} describe the stochastic
equivalence as ``\textit{We say two nodes $a$ and $b$ are stochastically
equivalent if and only if the probability of any event about the networks is
unchanged by interchanging nodes $a$ and $b$}''. The above definition
is formed  on the basis of  the stochastic blockmodeling methods, in which every
pair of nodes that belong to the same community are stochastically equivalent.
In this case, every node belongs to a cluster and the relationships between
different nodes are related to the corresponding pair of clusters (this is in
contrast with more traditional mixture models, where nodes are assumed
to be independent given their cluster assignments). Therefore, stochastic
blockmodels can be considered as generative models for communities or blocks in
networks and in the general case they fall in the class of random graph models.
Finally, the problem is formulated as a maximum likelihood estimation. Wang and
Wong \cite{wang-blockmodels} proposed a stochastic blockmodel for directed
networks and applied it on small scale social networks. Another method for
blockmodeling that can be applied on both directed and undirected networks
(also in weighted networks), was presented by Reichardt and White
\cite{role-models-2007}.

\par Yang et al. \cite{yang-communities-sdm10} proposed a 
stochastic blockmodel for directed networks, called Popularity and Productivity
Link model (PPL), which aims to model both incoming and outgoing links 
simultaneously. In order to achieve this goal, they introduce two latent
variables, namely productivity and popularity, to explicitly capture 
outgoing and incoming edges respectively. That is, in the general case, PPL
models the joint probability $\Pr(i_{\rightarrow},j_{\leftarrow})$, i.e., the
probability that there exists a directed edge from node $i$ to node $j$, as
follows

\begin{align}
 \Pr(i_{\rightarrow},j_{\leftarrow}) &= \sum_{c}
\Pr(i_{\rightarrow}|c) \Pr(j_{\leftarrow}|c) \Pr(c) \nonumber \\
&= \sum_{c} \bigg( \dfrac{\gamma_{ik} \alpha_i}{\sum_{i'} \gamma_{i'c}
\alpha_{i'}} \dfrac{\gamma_{jk} \beta_j}{\sum_{i'} \gamma_{i'c}
\beta_{i'}} \sum_{i'} \gamma_{i'c} w_{i'} \bigg), \label{eq:ppl}
\end{align}

\noindent where $\gamma_{ic}$ represents the probability of node $i$ to belong
to community $c$, $\alpha_{i}$ the productivity of node $i$ (i.e., how likely
an edge starts from $i$), $\beta_{j}$ the popularity of node $j$ (i.e., how
likely an edge is received by $j$) and $w_{i}$ the weight of node $i$ for
deciding the $\Pr(c)$ that  belongs to community $c$. A generative process can
be defined for Eq. \eqref{eq:ppl} and finally, through an EM algorithm the MLE
solution can be derived (the complexity per iteration for the EM algorithm will
be linear).

\par A limitation of the stochastic blockmodels is that each node  belongs
only to one community. However, this may not hold for several types of network
data; in many cases nodes belong to more than one communities. Airoldi et al.
\cite{airoldi-jmlr08} proposed the \textit{mixed membership} model, an
extension of the stochastic blockmodel, where each node belongs to any possible
communities via a membership probability. That is, allowing multiple membership
of nodes into communities, one is able to capture  different underlying
roles that nodes may exhibit in the network (similar to the concept of
overlapping communities). More precisely, each node $i$ is
associated with a $K$-dimensional vector $\vec{\pi}_i$, where
$\pi_{i,g}$ denotes the probability that node $i$ belongs to group (community)
$g$ and $K$ is the number of groups. Moreover, for each pair of nodes $i, j$,
the indicator vector $\vec{z}_{i \rightarrow j}$ denotes the
membership of node $i$ regarding its interaction with node $j$, and
$\vec{z}_{j \rightarrow i}$ the group membership of node $j$
regarding node $i$. Then, the mixed membership stochastic blockmodel for
a graph $G=(V,E)$ (directed) is drawn according to the following procedure:

\begin{itemize}
 \item For each node $i \in V$:
      \begin{itemize}
       \item Draw a $K$-dimensional  mixed membership vector
$\vec{\pi}_i \sim \text{Dirichlet} (\vec{\alpha})$
      \end{itemize}

\item For each node pair $(i,j) \in V \times V$
      \begin{itemize}
       \item Draw membership indicator for  $\vec{z}_{i \rightarrow j} \sim
\text{Multinomial}(\vec{\pi}_i)$
       \item Draw membership indicator for  $\vec{z}_{j \rightarrow i} \sim
\text{Multinomial}(\vec{\pi}_j)$
       \item Sample the value of their interaction $E(i,j) \sim
\text{Bernoulli}(\vec{z}^{~T}_{i \rightarrow j}~ \textbf{B}~ \vec{z}_{j
\rightarrow i} )$
      \end{itemize}
\end{itemize}

\noindent where matrix $\mathbf{B}_{K \times K}$ represents the probabilities of
interactions between different communities. The authors discuss how one can
compute the parameters of the model, and they provide several experiments on
real data (e.g., social networks, protein interaction data).

\par A different kind of blockmodel was recently presented by Rohe and
Yu \cite{rohe-coclustering-2012}, and is based on the notion of
co-clustering (i.e., the task in which  both  rows and columns of the
adjacency matrix are clustered simultaneously). The blockmodel is also
accompanied by a new spectral clustering algorithm for directed networks. More
precisely, at a first step a new co-clustering algorithm
for directed networks is introduced, based on the decomposition of a graph's
Laplacian defined by the authors. The idea behind this approach is based on the
fact that the
co-clustering task may be more meaningful for the case of directed networks: two
rows will belong to the same co-cluster if they have common endpoints, while two
columns will be in the same co-cluster if they receive edges from several common
nodes. The authors extend the spectral clustering algorithm presented
in Section \label{sec:spectral} (either for directed or undirected networks) to
the co-clustering task, where the eigendecomposition is replaced by the singular
value decomposition for dealing with the asymmetric matrix. To better
understand the properties of the algorithm, the authors present a stochastic
blockmodel for directed networks, based on the concept of co-clustering. That
is, the notion of stochastic equivalence is relaxed into two types, in order to
capture the two different roles of  nodes (senders and receivers):

\begin{itemize}
 \item Two  nodes $i$ and $j$ are stochastically equivalent senders if and only
if $\Pr(i \rightarrow v) = \Pr(j \rightarrow v), ~ \forall v \in V$.
 \item Two nodes $i$ and $j$ are stochastically equivalent receivers if and
only if $\Pr(v \rightarrow i) = \Pr(v \rightarrow j), ~ \forall v \in V$.
\end{itemize}

\noindent In the case of blockmodels for the traditional clustering task, both
these conditions for stochastic equivalence should occur. On the other
hand, considering these concrete cases of stochastic equivalence, a
blockmodeling based notion of co-clusters can be defined. The authors define
formally the \textit{stochastic co-blockmodel} for directed networks and study
thoroughly the performance of their spectral algorithm under this model.

\subsubsection{Other Approaches} \label{sec:other-techniques}
In this section we describe  diverse approaches that  can generally be
applied in the task of community detection and exploration in directed networks. 
In most cases, these methods adopt completely different methodological
approaches and they typically deal with variations of the community
detection problem (e.g., community detection for time evolving networks or
community exploration methods). Some of these topics will also be discussed
in Section \ref{sec:future}, since they constitute interesting extensions and
future research directions for the problem.

\subsubsection*{Community Structure Exploration and Evaluation Methods}
\begin{figure}[t]
\centering
  \includegraphics[width=.8\textwidth]{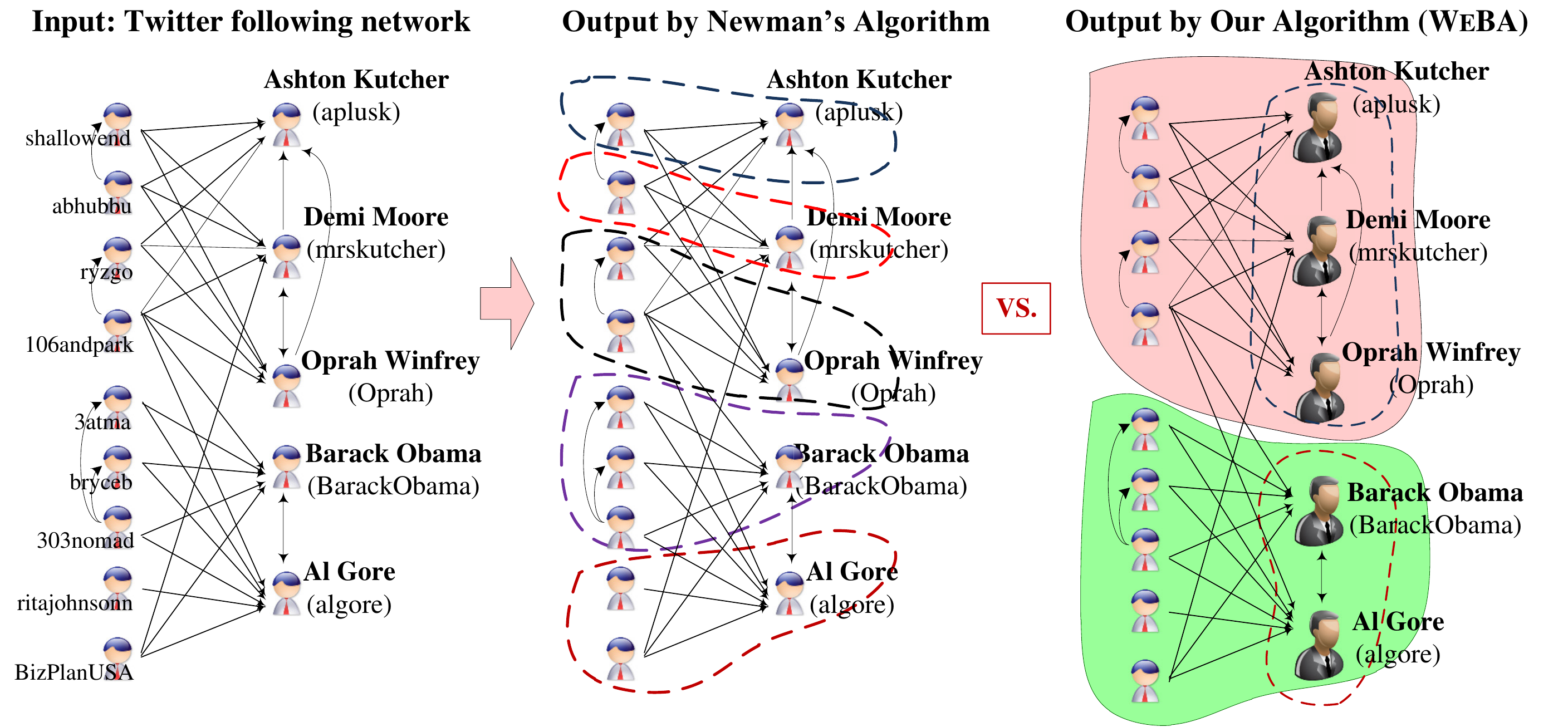}
 \caption{An example of community kernel detection (rightmost figure)  in the
Twitter network (leftmost figure) and how the outcome is differentiated from a
traditional community detection method based on modularity (central figure)
\cite{wang-community-kernels-icdm11}. The figure is courtesy of Wang et al.
\cite{wang-community-kernels-icdm11}. \copyright 2011 IEEE. \label{fig:weba}}
\end{figure}

\paragraph{\normalfont \textbf{Community Kernels}}
Most of the approaches presented so far are based on the assumption that
communities correspond to subgraphs with dense internal connections and sparse
external connections, while there is no special treatment of the most
influential nodes of the network. However, in many cases (usually in social
networks), there exist some influential nodes (e.g., important/popular users in
online social networking applications such as twitter) whose community structure
is quite different from that of the other nodes. To deal with these cases, Wang
et al. \cite{wang-community-kernels-icdm11} proposed the notion of
\textit{community kernels} and  studied the problem of community kernel
detection in social networks (both directed and undirected), as a way for
exploring the community structure of large networks. Usually, social networks
(e.g., the Twitter's who-follows-whom network) form a near bipartite structure,
where one partition corresponds to a few influential nodes (e.g.,
celebrities or politicians) while the other to the rest of the nodes; the
partitions are typically connected via a large number of edges targeting to
influential nodes, as shown in the leftmost part of Fig. \ref{fig:weba}. Most of
the well known community detection algorithms (which are based on density-based
measures) cannot identify this underlying structure, partitioning the influential
users into different communities and placing them  in the same
communities with
their followers (Fig. \ref{fig:weba} (center)). However, one would expect that
the influential nodes should be placed in the same communities according to 
common interests (e.g., politicians, entertainers) forming the community
kernels, while for each kernel there should exist a corresponding auxiliary
community that is associated with that kernel (rightmost part of Fig.
\ref{fig:weba}).

\par Wang et al. \cite{wang-community-kernels-icdm11}, at a first step
defined the notion of community kernel; each member of the kernel has more
connections with  members of the same kernel, than outside of it.
Moreover, each member (node) of an auxiliary community has more
connections with the associated kernel than to any other kernel. The notion of
community kernels can also be applied in several settings. For example, in a
co-authorship network, a kernel may correspond to a group of senior researchers
or professors in a specific area, while the auxiliary community to a group of
students or junior researchers in this area. Two algorithms are proposed to
extract the community kernels of large scale social networks, a greedy one and
\textsc{WeBA} which provides approximation guarantees (since the problem of
identifying the best community kernel is computationally difficult). Both of
them scale
linearly with respect to the size of the network.

\begin{figure}[t]
\centering
  \includegraphics[width=.7\textwidth]{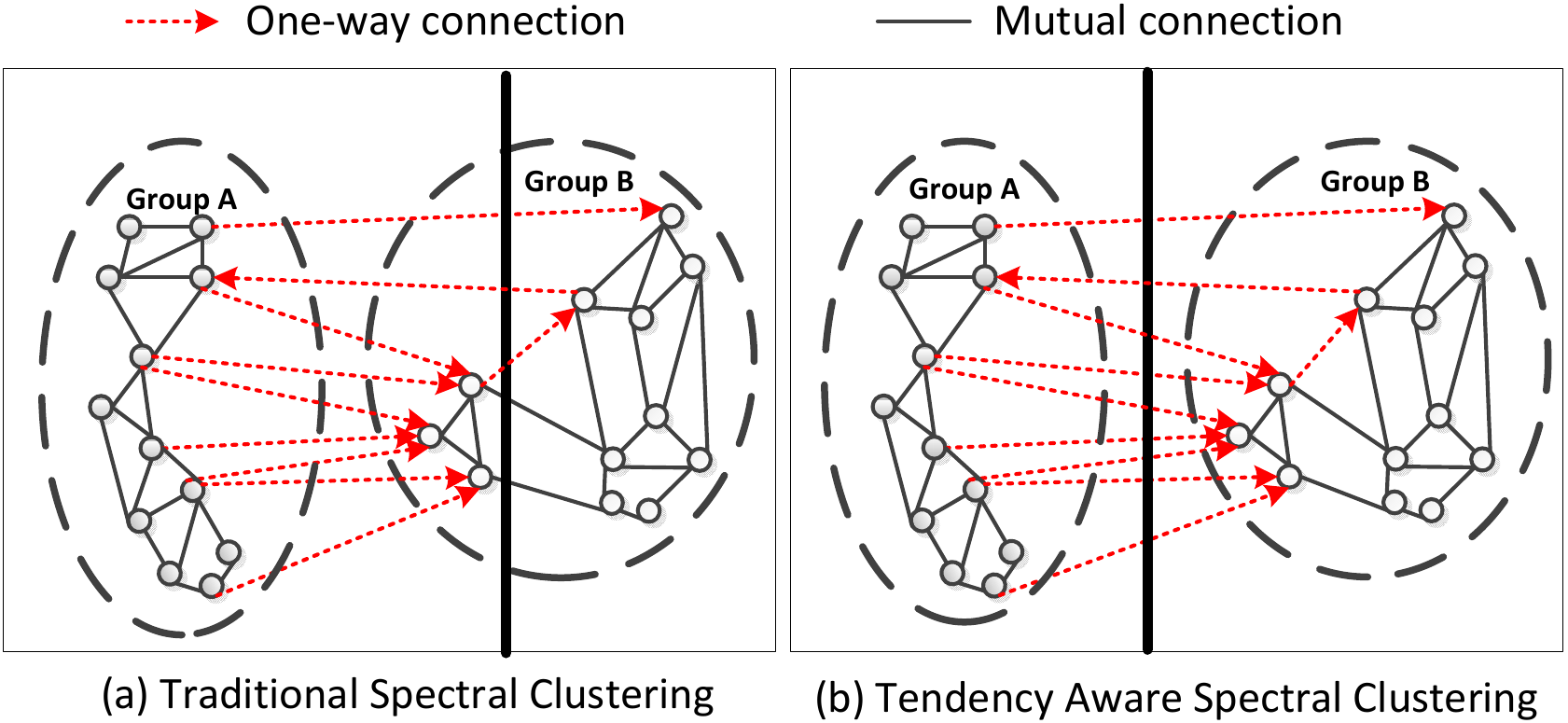}
 \caption{An example of the difference between traditional spectral graph
clustering (a) and tendency aware spectral clustering (b) defined by Li et al.
\cite{mutual-love-waw12}. The dashed line edges represent one way edges while
 normal lines show mutual connections. The figure is courtesy of Li et al.
\cite{mutual-love-waw12}. \copyright 2012 Springer. \label{fig:mutuality}}
\end{figure}

\paragraph{\normalfont \textbf{Mutuality-Tendency Aware Community Detection}}
Most of the
approaches presented so far do not explicitly distinguish the existence of
mutual (i.e., two nodes $u,v$ are mutually connected if both directed edges
$(u,v), (v,u)$ exist in the network) or one-way connections between the nodes of
a directed network in the graph clustering task. In
other words, by simply minimizing the number of inter-cluster edges, clustering
methods do not capture the existence of possible tendencies
between node pairs to be mutually connected. This point is of particular
interest since the existence of mutual connections in a cluster may be an
indicator of cluster's stability. Towards this direction, Li et
al. \cite{mutual-love-waw12} developed a  spectral clustering algorithm for
directed networks, which takes into account the tendencies of node pairs to
form reciprocal (mutual) connections. More precisely, the mutuality tendency
among a pair of nodes  is quantified using graph theoretic
concepts and according to this, a mutuality tendency aware criterion for the
clustering algorithm can be defined by maximizing the intra-cluster mutuality
tendency and minimizing the
inter-cluster mutuality tendency. Figure \ref{fig:mutuality} depicts an example
of a social network where nodes of the same group tend to have reciprocal
connections, while nodes across different groups are connected by one way
edges. When applying traditional spectral clustering methods (Fig.
\ref{fig:mutuality} (a)), the nodes of the second group (Group B) are
partitioned into two clusters, while a tendency aware clustering approach (Fig.
\ref{fig:mutuality} (b)) is able to utilize edges' mutuality information and
thus the majority of mutual connections are placed within the same clusters
\cite{mutual-love-waw12}.

\paragraph{\normalfont \textbf{Connected Components based Method}}
A simpler explanation
about  the notion of communities is given in Ref. \cite{cason11}. The authors
argue that a specific type of connected components in the network, can be used to
represent and explore the community structure. Generally, the definition of
connected components in directed networks is more complicated
than the undirected case, and the main types are the following:
weakly connected component (WCC), connected component (or unilaterally connected
component - UCC) and strongly connected component (SCC) (see Section
\ref{sec:background} for details). The authors of Ref. \cite{cason11} applied a
more strict definition and more precisely the one of strongly $p$-connected
components ($p$-SCC). A $p$-SCC corresponds to a subgraph $G' = (V', E')$,
in which $\forall u,v \in V'$, there is a directed path of length at most $p$
between $u$ and $v$  and one between $v$ and $u$, with $p \ge 2$. In other
words, the notion of $p$-SCC represents a SCC with an additional constraint on
the path length between two nodes. The authors combined the $p$-SCC concept
 with a merging routine that adjusts the size of the produced communities, to
detect clusters in directed networks.

\paragraph{\normalfont \textbf{Core-based Community Exploration}}
The notion of community structure is also closely related to the one of
collaboration between the nodes of a network. A natural mechanism for
the formation of a community in networks is related to the notion of cohesion,
which actually quantifies the collaboration nature among its members. In other
words, quantifying the degree of cohesion of a community, one can estimate the
collaboration among its elements. To the direction, Giatsidis et al.
\cite{d-cores} introduced novel metrics for evaluating the cohesion of
directed networks, extending the $k$-core concept from the undirected
setting to the one of $D$-cores for directed networks. Broadly speaking, a core
can be defined as a maximum size subgraph that is dense enough, i.e., for each
node in the subgraph, there exist at least $k$ incident edges that are adjacent
to nodes of the same subgraph. The concept is extended to directed networks,
where the $(k, \ell)$-$D$-core is defined, which corresponds to a maximal size
subgraph where for every node $i$ in the core, $k^{in}_i \ge k$ and $k^{out}_i
\ge \ell$ (the in- and out- degree respectively), leading to a degeneracy-based
community exploration and evaluation approach.

\paragraph{\normalfont \textbf{Game-Theoretic Approaches}}
A different formulation of the
clustering and community detection task in directed networks can be achieved
based on game-theoretic notions. Torsello et al. \cite{game-theoretic06}
presented a framework for clustering in directed networks within the context of
object grouping in computer vision and pattern recognition. The grouping
process is expressed as a non-cooperative game of the competition between the
hypotheses of group membership, where groups (clusters) correspond to
evolutionary stable strategies.

\subsubsection*{Extracting the Best Clusters on Large Scale Directed Networks}
An important characteristic of real networks which recently has gained an
increased interest is their size (scale). As networks grow in size, the
complexity of the analysis tasks applied to them (including the
clustering/community detection task) increases, and therefore the
feature of scalability should be taken into consideration.
Moreover, regarding the graph clustering task, many applications require only a
subset of the ``best'' clusters and not all the possible clusters  produced
by an algorithm. In other words, depending on the application domain, not all
possible clusters of the entire graph are useful, but the most strongly
connected ones are typically needed. For example, in a social networking
application, only
the most strongly connected groups of individuals may be of interest.
Methods that first identify all the clusters of a network, rank them and
keep only the top clusters tend to be inefficient, both in time and space
requirements. However, it is possible to reduce the searching space of an
algorithm by a pruning process, examining only those clusters with the highest
scores (regarding their quality).

\par Macropol and Singh \cite{singh-vldb10} proposed the TopGC (Top Graph
Clusters) algorithm, for finding the best connected, clique-like clusters in
large networks. The algorithm works on both directed and undirected networks
finding variable size clusters, while its running time is linear with
respect to the size of the network. The basic idea behind the algorithm is based
on the observation that nodes with similar neighbors in a graph, generally
should belong on the same cluster. In other words, the overlap between the
neighbor sets for two nodes is an indicator for whether or not these nodes
should be clustered together (e.g., the neighbor sets of the nodes of a clique
(except from the node itself) match exactly). The TopGC algorithm exploits the
fast similarity search based on \textit{Locality Sensitive Hashing} (LSH) in
order to find nodes with similar neighborhoods. In addition, the LSH
method is modified to achieve reduced memory consumption, since only the best
clusters of the graphs need to be extracted. The authors also state that the
algorithm is highly parallelizable, making its implementation possible in the
MapReduce framework \cite{mapreduce}.

\subsubsection*{Dynamic Networks}
One important aspect of several real-world networks is their dynamic nature,
i.e., they are not static but typically evolve over time with the
addition/deletion of nodes/edges, forming graph streams. A graph stream
$\mathcal{G}$ is defined as a sequence of graphs $G^{(t)}$, i.e., $\mathcal{G}
:= \{G^{(1)}, G^{(2)}, \ldots, G^{(t)}, \ldots\}$ \cite{sun-graphscope07}.
Dynamic networks have recently gained much attention from the research
community due to their interesting structural properties (e.g., Refs.
\cite{leskovec-tkdd, aggarwal}). Regarding the graph clustering and community
detection problem, the
works presented so far in this paper mostly concern with the problem applied
on static directed networks. In other words, we are interested to extract the
community structure of the network at a specific time point $t$ of its
evolution, working on the  snapshot $G^{(t)}$ of the
network at timestamp $t$. However, an interesting question is how the problem
of community detection can be extended and tackled in the case of dynamic networks. In this
case, we need approaches that will be able to incrementally find communities 
on networks, as well as to monitor and detect changes in the community
structure over time. Thus, two sub-problems need to be addressed:

\begin{enumerate}
 \item[(a)] \textit{Community discovery:} Node assignment into  clusters
(communities), following any of the definitions presented in Section
\ref{sec:problem-stm}.

 \item[(b)] \textit{Change point detection:} How to quantify and detect the
change of the community structure over time?
\end{enumerate}

\noindent More precisely, the community detection problem in dynamic networks
can be treated as a two-step incremental procedure: (a) the community discovery
subtask refers to static snapshots of the time-evolving graph, while (b), in
the change detection subtask, a measure of similarity between different
partitions over time needs to be determined in order to detect change
points in the graph stream. These change points correspond to time points where
a significant change in the already identified community structure has occurred.
The above problem has been studied for both undirected \cite{sun-graphscope07}
and directed networks. In the latter case, the authors of Ref.
\cite{dynamic-networks-cnikm09} present an approach
for detecting the community structure and the change points in dynamic weighted
directed networks. The first subtask is achieved by applying the method of
Random Walks with Restart \cite{rws-icdm06} (for computing the relevance scores
of the nodes), combined with a local version of modularity for examining the
quality of the produced partition. For the change detection subtask, the
authors propose a similarity metric between partitions, in order to detect
change points over time, accompanied by an algorithm for updating the partition
of a graph segment when a new graph is added on the stream.

\section{Definition-Based Classification of Clustering Methods}
\label{sec:definition-based}
In Section \ref{sec:edge-dir} we presented our main classification for the
clustering approaches in directed networks, targeting methodologies and
algorithms that have been presented so far for the
problem. In this section, we consider a different but also important (and in
some cases supplementary) classification scheme, where the methods are
grouped according to the clustering notion they adopt. 

\par In Section \ref{sec:problem-stm} we discussed the two main
clustering notions in directed networks,  otherwise the two
main approaches to characterize a subset of nodes as a cluster. The first
and most common one, namely \textit{density-based clusters}, is based on the
concept of intra-cluster and inter-cluster edge density (see
Section \ref{sec:density-based} for more details). That is, a cluster is
considered to be a subgraph with high internal edge density and low external
one. Depending on the objective measure that is used to quantify  edge
density, some variations of the general intra- and inter- density measures can
be considered (e.g., considering only the intra-cluster edge density of a
subgraph). Nevertheless, in all these cases, the feature that dominates on the
characterization of a cluster is expressed as a function of edge density. 

\par The second clustering notion in directed networks is more broad and includes cases 
where nodes are clustered together based on criteria beyond the classical one
of edge density -- see Section \ref{sec:pattern-based} for a more detailed
description. That is, in several cases, the presence of (directed) edges create
interesting structures that
deviate from the well-known density rule and can be naturally considered as
clusters. Examples of pattern-based clusters are the ones presented in Fig.
\ref{fig:pattern-based-figure}. It is clear that these clusters represent a
variety of interesting patterns (e.g., citation-based clusters or the case where
the cluster represent a subgraph that imposes ``strong'' information
flow). Furthermore, as we will present shortly, in most  cases the
notion of pattern-based clusters co-exist in a network with the one of
density-based. That is, a large number of methods is able to group
the nodes of a network following a density rule combined by other more
sophisticated types of structural patterns, that the ``pure'' density-based
approaches are not able to distinguish. To conclude, one can say
that both types of clusters represent diverse and interesting structures
and patterns, that induced by the edges of the networks (link-density for the
former case and link-pattern for the latter one
\cite{ljubljana-communities-ejpb12}).

\par The goal of this section is to provide a categorization of the approaches
reviewed in Section \ref{sec:edge-dir} according to their clustering
type-notion. A summary of this classification with the most representative
algorithms is also presented in Table \ref{tbl:summary}. Since density-based and
pattern-based clusters may co-exist on a directed network, we also provide a
discussion on these cases.

\subsection{Density-based Clusters}
In this category of ``pure'' density-based clustering, usually
belong approaches that become well-known extensions from the undirected case. 
That is,  spectral clustering methods based on the directed Laplacian matrix
\cite{gleich-directed, zhou-icml05, zhou-icml07, rohe-coclustering-2012} are
examples of approaches that identify density-based clusters. The graph
clustering method based on the weighted cuts criterion (proposed by  Meil\u{a}
and Pentney \cite{meila-sdm07}), also leads to similar clustering results. The
attraction and repulsion approach of Ref.
\cite{Yu-shi01} proposed in the field of image analysis and computer
vision, also belongs to  this category, as it constitutes a generalization of
the normalized cuts criterion for directed networks.
Furthermore, this category includes the local partitioning \cite{lang-local}
and local density methods \cite{schaeffer-pakdd05, chilean-web}. Other
approaches that based solely on the density-based notion of clusters are the
co-clustering algorithm (AutoPart) presented in Ref. \cite{autopart-pkdd05}, the
directed clique percolation method \cite{palla-directed-network-modules} and
techniques based on the  blockmodeling concept for statistical inference
\cite{wang-blockmodels, yang-communities-sdm10, role-models-2007,
airoldi-jmlr08}.

\subsection{Pattern-based Clusters}
The second category of methods are those based on the concept of pattern-based
clusters, i.e., clusters beyond the edge density notion. One interesting thing
is that most of these methods  are able to identify a mixed type of 
density-based and pattern-based clusters. In other words, they still identify
clusters based on the density concept, however they enhance these clusters with
other significant patterns. For example, using the idea of graph symmetrization
\cite{srini-edbd11}, one can transform the directed network into an
undirected one using measures that capture the incoming and outgoing edge
similarity, leading to the concept of citation-based clusters (see
Fig. \ref{fig:pattern-based-figure} (a), (b)). In other words, the graph is
transformed to an undirected one using nodes' incoming and outgoing edge
similarity, and therefore at the clustering process it is possible  two nodes
to belong on the same cluster even if they are not directly connected in the
original directed network. This constitutes an important feature, especially for
the case of networks where such underlying information exists (e.g., citation
networks).

\par The directed version of modularity presented by Arenas et al.
\cite{arenas-modularity07} and Leicht and Newman \cite{leicht-newman-2008},  has
formed the basis for several community detection approaches in directed
networks. It is able to extract density-based clusters, but it also has the
ability to recognize significant patterns imposed by edge directionality: it can
classify the nodes of a network into clusters, in such a way that directed edges
link from the one cluster to the other. Furthermore, similar behavior can be
observed even in networks with no underlying community structure; this is an
additional evidence that directed edges lead to various interesting underlying
patterns. However, as  we presented in Section \ref{sec:modularity}, the above
version of modularity has the drawback that it cannot distinguish properly the
direction of edges (see Fig. \ref{fig:kim-modularity}).

\par Another interesting type of pattern-based clusters is the one presented by
Rosvall and Bergstrom \cite{rosvall-pnas08}
and is based on the concept of patterns of movement  among the nodes of a
directed network. The community detection method presented in that work
(Infomap) is based on random walks and the main intuition is that a community
can be defined as a group of nodes where the random surfer is more likely to be
trapped in. This concept can be treated as an increased flow circulation pattern
between the nodes of the same community, as presented in Fig.
\ref{fig:pattern-based-figure} (c). Several community detection approaches
for directed networks have also been built upon this flow-based
pattern. Lai et al. \cite{lai-physica10} presented a Laplacian network
embedding algorithm which apart from density-based clusters, it is able to
detect flow patterns among the nodes. Their subsequent approaches based on
random walk similarity \cite{lai-jstatmech10} and affinity propagation
\cite{lai-message-passing}, are moving in a similar axis. The LinkRank method
introduced by Kim et al. \cite{linkrank-phys-rev10} is also able to extract
flow-based patterns. Moreover, as discussed in the paper, their
generalized version of modularity can distinguish in a proper way the
direction of the edges, compared to the one of Leicht and Newman
\cite{leicht-newman-2008}. Additionally, the approaches that
utilize the Directed Gaussian Random Network (DGRN) as null model
\cite{pantazis-asilomar10, pantazis-isbi11}, are also able to cluster the nodes
of directed networks based on information flow patterns.

\par In the approach of Guimer\`{a} et
al. \cite{module-identification-physrev07}, the directed network is converted
into a bipartite one, and a bipartite version of modularity is applied to
extract the community structure. While the method mainly considers density
features, it is also able to detect communities based on common incoming and
outgoing edges; it relies on the idea of actors co-participation in a team
(for bipartite networks). Additionally, other methods that consider
the directed network as a bipartite one, are also able to detect both
density-based  as well as  citation-based clusters \cite{zhou-nips05,
zhan-evolutionary-physrev11}. A similar behavior is presented in the two-step
random walks method by Huang et al. \cite{huang-pkdd06}; the two-step
random walk model is able to capture important connectivity patterns by
exploiting the existence of co-citation and co-reference relationships.

\par The method of Newman and Leicht \cite{leicht-mixture07} based on
mixture models can detect diverse types of clusters, including
assortative\footnote{Assortativity is the property where the nodes of a network
tend to link to other nodes that are similar in some way (see Wikipedia's
lemma for \textit{Assortativity}:
\url{http://en.wikipedia.org/wiki/Assortativity}).} and dissassortative
structures. However, as discussed in Section \ref{sec:probabilistic-models},
the method cannot identify communities that do not have at least one
node with non-zero out-degree (see Fig. \ref{fig:em-community-problem}). The
approach of Ramasco and Mungan \cite{content-based-communities-physrev08}
and  Wang and Lai \cite{wang-mixture-new-j-phys08} overcomes this problem,
recovering  clusters that do not necessarily follow the density-based
notion.

\subsection{Empirical Comparison of Clustering Approaches in Directed Networks}
\label{sec:comparison}
Having reviewed the methods proposed so far for the clustering problem in
directed networks, we will now proceed with a brief empirical comparison of
them. Table \ref{tbl:summary} presents a summary of the major approaches along
with their basic features.

\par As we discussed earlier, the proposed approaches follow diverse
methodologies and in many cases they are built upon different notions of
clusters/communities in directed networks. In the first case which follows
naturally from the the problem in undirected networks, only density features are
considered to characterize a cluster. Typically, the edges between
nodes of the network represent pairwise relationships, which operate as 
similarity measure among different entities (nodes). However, as several
research works propose, due to the existence of directed
edges in a network, it is possible to exist other interesting type of
clusters. In many cases, the proposed approaches are able to identify
density-based clusters combined with interesting structures beyond
the density pattern. For example, applying symmetrization schemes in the
directed network, it is possible to group nodes in the same cluster, even
if they do not share an edge in the initial network \cite{srini-edbd11}.
This constitutes an important characteristic for a set of networks where
co-citation and co-reference relationships may be of interest (e.g., citation
networks). Similarly, another interesting clustering definition is the one that
based on random walks and the concept of information flow and movement among the
nodes of a network (e.g., Ref. \cite{rosvall-pnas08}).

\par A natural question that arises from the above discussion is which method
should a researcher or a practitioner use. The answer is not clear but mainly
depends on the network under consideration and on the application domain (for
the latter we discuss in Section \ref{sec:applications}). In case
of networks where  edges represent pairwise relationships, it may be more
useful to apply density-based methods or methods that, at least, are able
to identify density-based clusters (e.g., modularity optimization, spectral
clustering, etc.). On the other hand, when edges represent patterns of movement
among nodes or generally some kind of information flow, methods that are able
to recover flow-based clusters may be preferable.

\par Although the clustering notion-definition is an important feature for
selecting a community detection method for directed networks, it is not the
only one. As we discussed in previous sections, there exist a plethora of
algorithms that seems to follow the same clustering definition. For these
approaches, additional features should be compared in order to select the most
suitable one for a specific application or for a specific graph dataset that
needs to be analyzed. Some important features are the ones presented in Table
\ref{tbl:summary}. For example, one may select an appropriate algorithm
examining the objective function that is used to characterize the quality of a
community (e.g., modularity, normalized cuts). Some methods pose additional
characteristics, such that their ability to identify overlapping
communities, that may be significant for specific applications. The time
complexity of an algorithm is also a crucial factor, especially for large scale
networks. Since in most cases the clustering problem is expressed as an
optimization one, the complexity depends heavily on the selected optimization
method \cite{fortunato}. 

\par Thus, it becomes clear that selecting the proper clustering
approach depends on multiple criteria. In Section \ref{sec:applications} we will
see which of these approaches have been used in the related literature for
specific applications in several domains. This may be useful for practitioners
without the required know-how in the field.

\begin{small}
\begin{landscape}
 \begin{table}[h!t]
\centering
 \begin{tabular}{lcllll} \toprule
  Method & Category & Clustering Type & Objective Function &
Additional Features & Complexity  \\ \midrule \\

Symmetrization \cite{srini-edbd11} & $1^{\text{st}}$ & All &  & 
& $\mathcal{O} (\sum_{i} \mathbf{D}(i))$  \\

Network Embedding \cite{lai-physica10}  & $1^{\text{st}}$, $2^{\text{nd}}$ &
All & Directed modularity & Laplacian matrix & Selected algorithm   \\

Random Walk Sim \cite{lai-jstatmech10} & $1^{\text{st}}$ & All &  &
 & Selected algorithm  \\

Bipartite Modularity \cite{module-identification-physrev07} & $1^{\text{st}}$,
$2^{\text{nd}}$ & All & Bipartite modularity & Directed to Bipartite &
Modularity Opt. \\

MAGA method \cite{zhan-evolutionary-physrev11} & $1^{\text{st}}$,
$2^{\text{nd}}$ & All & Bipartite modularity & Genetic algorithm (GA) &
Complexity of GA \\

Semi-supervised Learning \cite{zhou-nips05} & $1^{\text{st}}$, $2^{\text{nd}}$ &
All & Modified modularity &  &  Spectral Opt. \\

Directed Modularity \cite{leicht-newman-2008, arenas-modularity07} &
$1^{\text{st}}$, $2^{\text{nd}}$ & All & Directed modularity &
 &  Spectral Opt.  \\

LinkRank \cite{linkrank-phys-rev10} &$2^{\text{nd}}$ & All & Directed
modularity &  & Modularity Opt. \\

DGRN \cite{pantazis-asilomar10, pantazis-isbi11} &$2^{\text{nd}}$ & All &
Directed modularity &  &  Spectral Opt. \\

Overlapping Modularity \cite{nicosia-s-stat-mech09} &$2^{\text{nd}}$ &
Density-based & Dir.-Overl. modularity & Genetic Algorithm &
$\mathcal{O}(|C|n^2)$ \\

Local Modularity \cite{local-modularity} &$2^{\text{nd}}$ & Local cohesive
groups & Local modularity &  & Modularity Opt.   \\

Directed Laplacian \cite{gleich-directed} &$2^{\text{nd}}$ & Density-based &
Normalized Cuts &  & Spectral Opt. \\

Directed Normalized Laplacian \cite{zhou-icml05} &$2^{\text{nd}}$ &
Density-based & Normalized Cuts & & Spectral Opt. \\

Multiple Views Spectral Clus. \cite{zhou-icml07} &$2^{\text{nd}}$ &
Density-based & Multiple Cuts & Multiple Views & Spectral Opt. \\

Weighted Cuts \cite{meila-sdm07} &$2^{\text{nd}}$ & Density-based & Weighted
cuts &  & Spectral Opt. \\

Attraction and Repulsion \cite{Yu-shi01} &$1^{\text{st}}$, $2^{\text{nd}}$ &
Density-based & Normalized cuts &  & Spectral Opt. \\

Two-step Random Walks \cite{huang-pkdd06} & $2^{\text{nd}}$ & All & Normalized
cuts &  & Spectral Opt. \\

Local Partitioning \cite{lang-local} & $2^{\text{nd}}$ & Density-based &
Conductance & Local Method &  Spectral Opt. \\

Message Passing  \cite{lai-message-passing} & $2^{\text{nd}}$ & All &  &
Affinity propagation & $\mathcal{O}(In |C| + 2n^2)$ \\

Directed Clique Percolation \cite{palla-directed-network-modules} &
$2^{\text{nd}}$ & Density-based & $k$-clique & & $\mathcal{O}(\exp(n))$\\

Local Density \cite{schaeffer-pakdd05,chilean-web} & $2^{\text{nd}}$ &
Density-based & Density & Local method & \\

Infomap \cite{rosvall-pnas08} & $3^{\text{rd}}$ & Pattern-based & Code length &
Compression & Optimization Method \\

AutoPart \cite{autopart-pkdd05} & $3^{\text{rd}}$ & Density-based & Code length
& Compression & $\mathcal{O}(m |C|^2)$ \\

Mixture Models \cite{leicht-mixture07, content-based-communities-physrev08,
wang-mixture-new-j-phys08} & $3^{\text{rd}}$ & All &  & Mixture Models, EM &
Convergence of EM \\

Blockmodels \cite{wang-blockmodels, yang-communities-sdm10, role-models-2007,
airoldi-jmlr08} & $3^{\text{rd}}$ & Density-based &  & & Parameter Estimation\\

Co-Blockmodel \cite{rohe-coclustering-2012} & $3^{\text{rd}}$ &
Density-based &  & Spectral & Spectral decomposition\\

Community Kernels \cite{wang-community-kernels-icdm11} & $3^{\text{rd}}$ &
All &  & Kernel's Score & Linear approximation\\

&  &  & &  & \\
 \bottomrule

\end{tabular}

\caption{Summary of the basic graph clustering approaches for directed networks.
The field of Category represents one of the three main categories presented
in Section
\ref{sec:edge-dir} (excluding the naive approach): $1^{\text{st}}$:
transformations maintaining directionality, $2^{\text{nd}}$: extending objective
functions and methodologies to directed networks, $3^{\text{rd}}$: alternative
approaches. The type (notion) of clusters that identified by the methods is also
depicted. The fourth column shows the objective function applied by each method
(in case where an objective function is clearly described), while the next
column presents some additional features of the methods. The last column
describes the time complexity of each method (more generally the factors that
determine the complexity). Typically, the complexity of spectral optimization
(Spectral Opt.) relies on the complexity of eigenvalue decomposition (spectral
decomposition).
\label{tbl:summary}}
\end{table}
\end{landscape}
\end{small}

\section{Evaluation Metrics and Benchmarking} \label{sec:evaluation}
In this section we discuss on the task of assessing the results of
a graph clustering algorithm. Generally, a network can be 
divided into several meaningful partitions and one should decide
which of them is the most appropriate as a clustering result. 
Typically, quality measures are used for this task, but in some cases, their
accuracy may not be a good indicator
\cite{fortunato}. Moreover, since a wealth of diverse algorithms has been
proposed for both directed and undirected networks, one wants
to decide which algorithm results in better performance -- in
terms of clustering quality; of course, the performance of an algorithm in terms
of time complexity is also a crucial factor. In other words, an important
problem in the area  is the one of evaluating the performance of clustering
algorithms by comparing their results.  In the case where  real datasets
with known community structure (ground truth) are available, this can been done
by comparing the results with the a priori known node assignments to
communities/clusters. Moreover, a widely applied approach is to evaluate the
performance of a community detection algorithm on benchmark graphs with an
inherent community structure. However, in the case of directed networks both
techniques are still premature. On the one hand, it is very difficult to find
directed graph datasets with known community structure and sufficient size,
while only a few benchmark graphs exist.

\par The directed clustering evaluation task is closely related to the
respective one in the undirected case and a more thorough discussion is
presented in Refs. \cite{fortunato, schaeffer-review}. In the case of quality
measures, typically the directed version of modularity
\cite{arenas-modularity07, leicht-newman-2008} is applied to quantify the
significance of a partition, while for benchmarking purposes some very recent
benchmarks are reviewed.

\subsection{Evaluating Partitions}
The problem of evaluating the quality of communities produced by an algorithm is
rather broad and several approaches have been proposed and applied in the
undirected case of the problem. Most of them are applied directly to the
clustering  problem in directed networks, since the algorithms and the 
results for both problems can be treated similarly. An approach is to
examine quality indices for partitions in directed networks, and some of them
were presented in Section \ref{sec:edge-dir} (e.g., modularity). As noted by
Schaeffer \cite{schaeffer-review}, although the optimization
process of these measures is a difficult issue, their evaluation for a given
partition of the network is a more lightweight operation. However, the
evaluation process may be biased  regarding the characteristics of the
quality measure. Moreover, while there are several comparative studies on
quality measures for undirected graph clustering \cite{brandes-esa2003,
leskovec-www10}, similar studies are missing for the directed version of the
problem.

\par An alternative approach to evaluate the produced clusters is related to the
stability of the results under perturbations of the input graph. The motivating
idea behind this technique is the following: if a cluster is significant, then
after some modifications at the original network, its significance will be
retained and the cluster itself will be still identified by the algorithm. 
Typically,  the stability of an algorithm is examined by measuring the 
number of operations needed to transform the original clustering results to the
ones produced after the perturbation of the graph. In the related literature
different perturbation approaches have been proposed for undirected networks
\cite{gfeller-physreve05, karrer-physreve07, raghavan-stability}, but their
applicability to directed ones is not straightforward. (For a more detailed
presentation one can refer to Ref. \cite{fortunato}.)

\subsection{Comparing Algorithms}
Instead of examining  the quality of the clustering results
produced by an algorithm, it may be preferable to compare the results
produced by several algorithms, towards selecting the most accurate one. For
this task, one has to define a similarity criterion between results produced by
different algorithms. In the case where a true assignment of nodes into clusters
is known a priori (also known as ground truth clustering), these criteria can be
used to evaluate the performance of a clustering algorithm. More precisely,
suppose that $C_A(v), \forall v \in V$ represents the cluster assignment of node
$v$ using algorithm $A$. Then, a similarity measure for two algorithms $A, B$ 
with node cluster assignments $C_A(1), C_A(2), \ldots, C_A(n)$ and $C_B(1),
C_B(2), \ldots, C_B(n)$, can be defined as

\begin{equation}
 S(A,B) = \dfrac{1}{n} \sum_{v \in V} \dfrac{|C_A(v) \cap
C_B(v)|}{|C_A(v) \cup C_B(v)|},
\end{equation}

\noindent where a score value close to one indicates similarity for
the clustering results. However, this measure does not behave
well if the results of the one algorithm have been produced by a merging process
of two or more clusters of the other algorithm \cite{schaeffer-review}. 

\par In a similar spirit, one can use the measures of \textit{precision} and
\textit{recall}, with respect to the ground truth clustering assignment. Suppose
that $\mathcal{C} = \{C_1,C_2, \ldots, C_K\}$ is the output of a clustering
algorithm, where $K$ represents the number of clusters and $C_j$ is the ground
truth clustering. Then, for any output cluster $C_i$ the precision and recall of
this cluster can be defined as

\begin{equation}
 \text{Prec}(C_i, C_j) = \dfrac{|C_i \cap C_j|}{|C_i|} \text{~~~~~~and~~~~~~}
\text{Rec}(C_i, C_j) = \dfrac{|C_i \cap C_j|}{|C_j|}.
\end{equation}

\noindent A measure that integrates precision and recall is the so-called
\textit{F-measure} $\text{F}(C_i, C_j)$  defined as the harmonic mean of
precision and recall\footnote{Wikipedia's lemma for precision, recall and
F-measure:
\url{http://en.wikipedia.org/wiki/Precision_and_recall}.}: 

\begin{equation}
 \text{F}(C_i, C_j) =  \dfrac{2 \cdot \text{Prec}(C_i,
C_j) \cdot \text{Rec}(C_i, C_j)}{\text{Prec}(C_i, C_j) + \text{Rec}(C_i,
C_j)}.
\end{equation}

\noindent Each produced cluster $C_i$ is matched with its corresponding ground
truth cluster $C_j$ for which the F-measure is maximized, $\text{F}(C_i) =
\max_j \text{F}(C_i, C_j)$. Then, the average F-measure of the produced
clustering is defined as the average F-measure over the set of clusters as

\begin{equation}
 \text{F}(\mathcal{C}) = \dfrac{\sum_i |C_i| \cdot \text{F}(C_i)}{\sum_{i}
|C_i|}.
\end{equation}

\par Another important category of similarity measures for assessing the
results of clustering algorithms, originates from the field of information
theory. Danon et al. \cite{danon-jstat05} used the measure of \textit{Normalized
Mutual Information} (NMI), which considers information-theoretic concepts and is
based
on the fact that if two clusters are similar to each other, then only a small 
amount of additional information is needed  to infer one clustering
assignment from the other. The definition is based on the concept of
confusion matrix $\mathbf{N}$ (the rows correspond to the ground truth
clusters while the columns to the clusters identified by the algorithm), and it
can be expressed as follows

\begin{equation} \label{eq:nmi}
 \text{NMI}(A,B) = \dfrac{-2 \sum_{i=1}^{C_A} \sum_{j=1}^{C_B} N_{ij}
\log(\frac{N_{ij} N}{N_{i;}N_{;j}})}{\sum_{i=1}^{C_A} N_{i;}
\log(\frac{N_{i;}}{N}) + \sum_{j=1}^{C_B} N_{;j} \log(\frac{N_{;j}}{N})},
\end{equation}

\noindent where $|C_A|, |C_B|$ represent the number of ground truth clusters
and the number of produced clusters respectively. The element $N_{ij}$
corresponds to the number of nodes in real cluster $i$ that appear in the
produced cluster $j$, while $N_{i;}$ is the sum over row $i$ and $N_{;j}$ the
sum over column $j$ of the confusion matrix $\mathbf{N}$. In the case where the
produced results are identical with the ground truth, the $\text{NMI}(A,B)$
measure takes its maximum value one, while in the case where the two clusterings
totally disagree, the $\text{NMI}(A,B)$ score is zero. Using
information-theoretic concepts, the numerator of Eq. \eqref{eq:nmi} corresponds
to the mutual information between the two clustering results, while the
denominator represents the sum of the corresponding entropies (actually the
mutual information and entropies of the random variables that represent the
cluster assignments). We also stress here that these measures can also be
applied
to compare two clustering assignments and not necessarily a comparison to ground
truth data. Other information-theoretic criteria for comparing different
clustering results have been presented by Meil\u{a}
\cite{Meila-comparing-clustering} (see the related paper by Fortunato
\cite{fortunato} for more details).

\subsection{Testing Algorithms}
The step that follows the design of a new community detection algorithm,
involves the testing process. Usually, in this task, the algorithm is applied 
to a network with specific community structure and the results are compared
to the known structure. To do this, the algorithm should be applied to a
network with well defined community structure, in order to extract
meaningful conclusions about its function. 
For this reason, the use of benchmark graph datasets is involved. 
In the case of undirected networks, there exist a few real graphs with known
community structure that are commonly used for testing community detection
algorithms. The most known of them is Zachary's social  network (see e.g., Ref.
\cite{newman-modularity}), which represents friendship relationships between the
members of a karate club. However, in most
cases, these graph datasets are of small scale and they are not adequate for
testing the performance (in terms of accuracy) of an algorithm at larger scale. 
Moreover, for the directed graph clustering problem, there does not exist
any such commonly used set of benchmark graphs that can be used to assess the
accuracy of algorithms.

\par A similar way to test a graph clustering algorithm is to examine
its performance on synthetic benchmark graphs, i.e., computer generated graphs
with built-in community structure. These graphs are artificial and typically
produced by a mechanism which is controlled by some parameters. Several
benchmark graph datasets have been proposed for the undirected graph clustering
problem, such as the \textit{planted $\ell$-partition model} and the
\textit{LFR} benchmark (see Ref. \cite{fortunato} for more details). For the
case of directed networks which is the focus of this paper, the problem of
generating realistic benchmarks  has received relatively little attention from
the research community. In several research works where a new  algorithm is
proposed, some specific benchmark directed graph datasets are also presented and
used for testing the algorithm. For example, Rosvall and  Bergstrom
\cite{rosvall-pnas08} proposed the directed network presented in Fig.
\ref{fig:pattern-based-figure}(c) as a benchmark for clustering algorithms that
consider flow-based types of clusters (similar synthetic benchmarks
have been also presented in other research works).

\par Recently, Lancichinetti and Fortunato  \cite{fortunato-benchmark09}
presented an algorithm for generating directed graph benchmarks for testing
purposes (the generator has also the ability to produce weighted graphs with
overlapping communities). The benchmark graphs constitute an extension of the
LFR undirected graph generator and consider that the in-degree $\{y_i\}$ and
out-degree $\{z_i\}$ sequences follow some specific distributions. Moreover, the
size of the produced communities follow a power-law distribution. The generation
mechanism is as follows: initially, we sample the in-degree sequence $\{y_i\}$
from a power-law distribution, and the out-degree sequence $\{z_i\}$ from a
$\delta$ distribution (by drawing $N$ random numbers for each of them). Each
node in the graph shares some of its edges with other nodes inside its community
and the rest of the neighbors are outside the community depending on its degree
(in- and out-). For this reason, two topological mixing parameters are
introduced for each node, to define the proportion of incoming and outgoing
edges that will fall inside and outside node's community. According to these
parameters, nodes  of the same community (stubs) are randomly connected
(preserving both in-degree and out-degree distributions) and some extra random
edges are placed  between them and nodes of different communities.

\par The LFR benchmark graphs are constructed based on a density rule, which
places internal and external edges between nodes of the same and
of different communities respectively. However, as we have already seen
throughout this survey paper, the existence of directed edges may reflect other
interesting structural patterns, such as the flow-based pattern. From a first
view it seems that the LFR generator is not suitable to generate graphs with
patterns of flow between nodes. However, as the authors discuss, this can be
achieved  by constraining the number of incoming and outgoing edges of
communities.

\section{Real-world Applications on Directed Networks} \label{sec:applications}
The task of graph clustering and community detection in general, and the one of
directed graph clustering  in particular, lies at the heart of many applications
and research agendas. A large number of research works in several scientific
disciplines have been devoted to
increase our understanding of real-world complex systems, applying
graph clustering approaches.  In this section we review some important
applications of the clustering problem in directed networks, in several fields,
such as social and information sciences, biology, and neuroscience. Since the
clustering task constitutes one of the most common and prominent analysis tool
in networks, the following list of possible applications can be extended in
every other field where directed networks appear.

\subsection{Social, Information and Technological Networks}
Social networks are used to represent the interactions among
individuals/entities, under a wide range of settings. The study of social
networks has its
roots in the field of sociology and constitutes a prominent research area for
decades (see e.g., Ref. \cite{scott-book-2000}). More recently,  the advances in
communication and information technologies along with the widespread penetration
of the Internet and the World Wide Web (WWW), have led to the explosure of 
available social networking data. Characteristic examples are the online social
networking applications, such as Facebook (\url{www.facebook.com}), Twitter
(\url{twitter.com}) and Google+ (\url{plus.google.com}). Of similar importance
are several information and technological networks that are part of our everyday
life, including the Internet, the World Wide Web network (the hyperlink
structure between webpages), and mobile phone communication networks.
In many cases, the relationships between entities on such  networks are
not reciprocal, forming directed edges. Cluster analysis in  social,
information and technological networks has been proved to be a useful task that
can be used to shed light on the structure of these complex systems.

\par
\begin{figure}[t]
\centering
 \includegraphics[width=.5\textwidth]{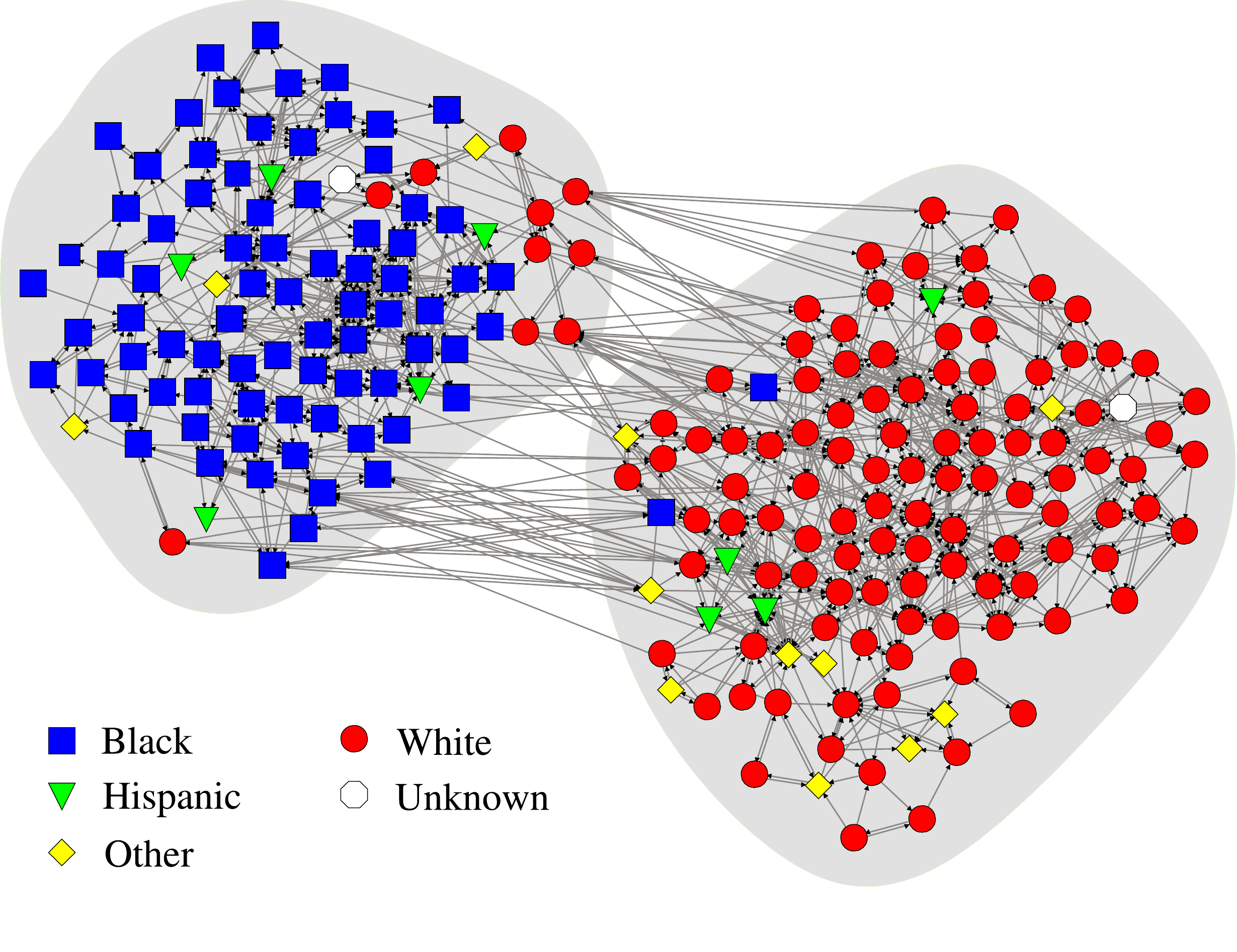} 
 \caption{A directed social network representing  friendship connections
between high school students. The shadowed regions represent the two
clusters extracted by the algorithm of Newman and Leicht
\cite{leicht-mixture07}. The vertex shapes show the ethnicity of the students.
The figure is courtesy of Newman and Leicht \cite{leicht-mixture07}. \copyright
2007 National Academy of Sciences, U.S.A.. \label{fig:friendship}}
\end{figure}

 Wang and Wong \cite{wang-blockmodels} applied a graph clustering
approach based on stochastic blockmodels for directed networks, in order to
analyze the strength of ties between students from different socioeconomic
backgrounds. According to their study, a set of $27$ students ($13$ male and
$14$ female) from a single classroom were asked to indicate liking for the other
students using one of the following facial expressions: (a) big smile,
(b) moderate smile and (c) no smile. The case of big smile indicates an
increased liking among  students and this information is represented as a
directed edge in the network. A more extended study of a directed friendship
network of high school students (U.S. National Longitudinal Study of
Adolescent Health) has been performed in Refs. \cite{leicht-mixture07,
lai-physica10}. Figure \ref{fig:friendship} presents the clustering results
on this dataset, produced by the method of Newman and Leicht
\cite{leicht-mixture07}.

\par Similar cluster detection studies have been performed for
several other social networks of largest scale, that typically arise in the
context of social networking applications. For example,
the authors of Refs. \cite{mutual-love-waw12, wang-community-kernels-icdm11,
cason11} performed cluster analysis in large scale directed
social networks with thousands of nodes and edges, such as Slashdot
(\url{slashdot.org}), Epinions (\url{www.epinions.com}), Twitter and e-mail
exchange networks.

\par Clustering  algorithms can also be applied in the
case of directed information networks. The most prominent example here is the
World Wide Web (WWW). The nodes of the Web network represent webpages, while the
edges  hyperlinks between different webpages. Communities in the hyperlink
structure of the Web may represent webpages that belong on the same thematic
category, and
therefore, identifying communities can be helpful in several
practical applications such as recommender systems (e.g., Ref.
\cite{huang-pkdd06}). In a similar way, other hyperlink structures that
correspond to directed information networks can be benefited by clustering
methods. For example, the well-known electronic encyclopedia of Wikipedia
(\url{http://www.wikipedia.org/}) can be naturally represented as a directed
network, where each node corresponds to a lemma and the edges to hyperlinks
between different lemmas. Applying clustering methods to the directed
network of Wikipedia, one can identify meaningful categories of lemmas
\cite{srini-edbd11}.

\par
\begin{figure}[t]
\centering
 \includegraphics[width=.6\textwidth]{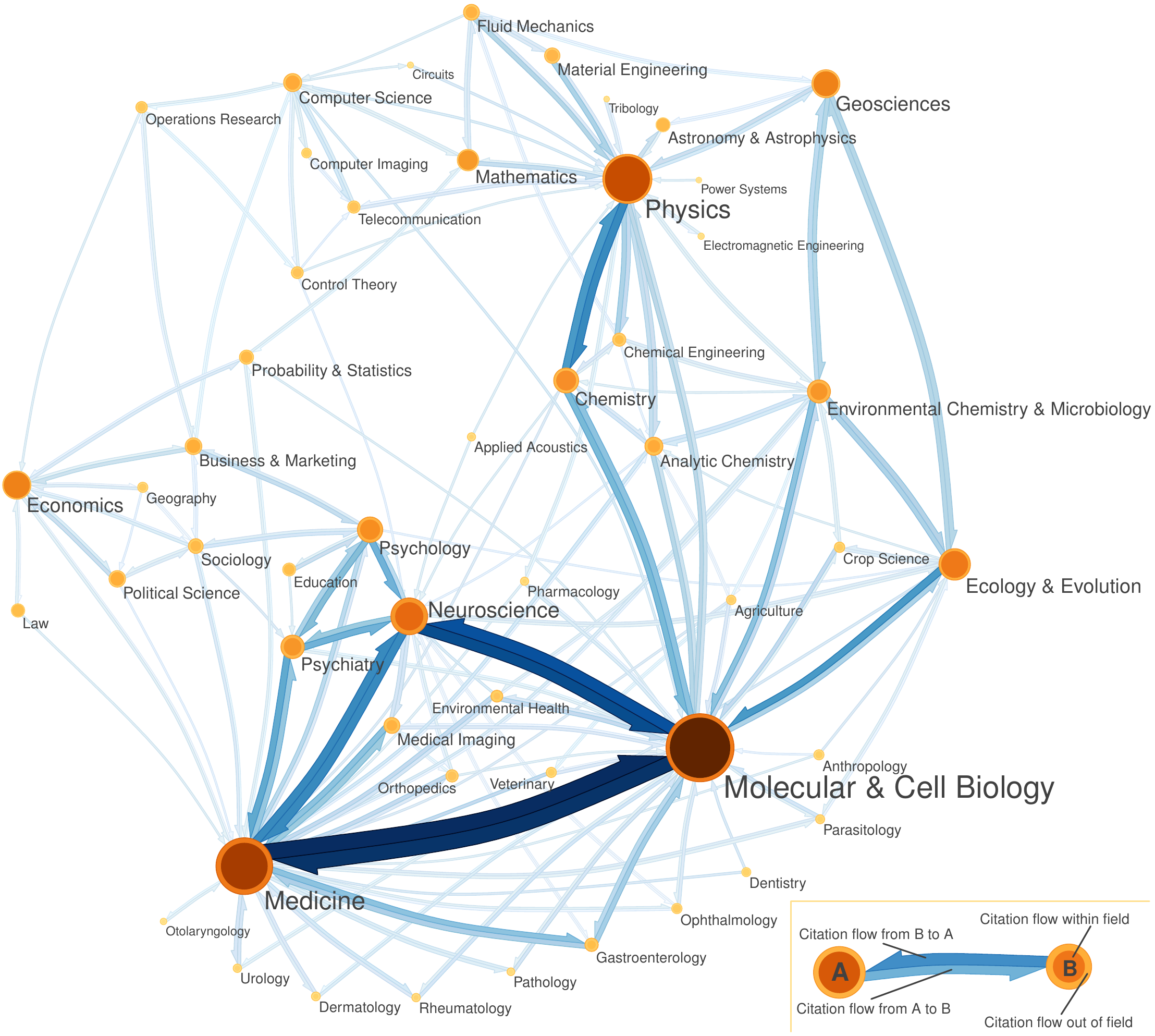} 
 \caption{A map of scientific disciplines based on citation patterns of about
6.5 million citations. The figure is courtesy of Rosvall and Bergstrom
\cite{rosvall-pnas08}. \copyright 2008 National Academy of Sciences, U.S.A..
\label{fig:map}}
\end{figure}

 Another important application  on information
networks concerns the case of citation networks, i.e., directed networks where
the nodes represent documents and the edges capture citation relationships
(see e.g., Refs. \cite{newman, citation-networks12}).
Clustering methods have been applied in the past to citation networks, in
order to understand the connection patterns between scientific papers and more
generally, to comprehend the connections between different scientific
disciplines. Rosvall and Bergstrom \cite{rosvall-pnas08} applied their 
flow-based graph clustering method (see Section \ref{sec:information-theoretic})
to a citation network of scientific journal papers, creating a map of scientific
disciplines as shown in Fig. \ref{fig:map}. The general observation is that the
structure of sciences
can be represented as the letter \textbf{U}, with the social sciences on the one
side, engineering on the other, joined by medicine, molecular biology,
chemistry and physics. Chen and  Redner \cite{informetrics10} studied the
community structure of the citation network of the Physical Review journals
family between 1893 and 2007 using modularity optimization techniques, observing
interesting properties. Similarly, the authors of Ref. \cite{metroeconomica10}
applied community detection methods to a patent citation network, in order to
understand the patterns of knowledge transfer between technology
fields.

\par Clustering algorithms have been also applied to partition  software
systems into smaller units. A software system can be represented by a call
graph, i.e., a directed network in which  nodes represent the programs of
the system and the edges depict calls from one program to another. For
example, Bisseling et al. \cite{call-graph} applied graph clustering to
partition Java and Cobol programs into smaller modules.

\subsection{Biological Networks}
A large number of biological systems can be represented as directed networks.
Such an example is the network of metabolic pathways, i.e., series of chemical
reactions occurring within a cell. These reactions are connected by their
intermediates: the products of the one reaction are the substrates for
subsequent reactions\footnote{Wikipedia's lemma for \textit{Metabolic pathway}:
\url{http://en.wikipedia.org/wiki/Metabolic_pathway}.}. Moreover, metabolic 
pathways are usually considered in one direction. Numerous distinct
pathways co-exist within a cell, forming the so-called metabolic network. A
major challenge in biology is to understand the structure and evolution of these
networks \cite{metabolic-plos05} and graph clustering approaches have been
applied towards this goal \cite{pathways-glycosylation-plos09}.

\par Another important type of biological directed networks are the gene
regulatory networks (GRNs). GRNs show the regulatory relationships among genes
in a cellular system and are involved  in the production of proteins. Broadly
speaking, the nodes of a GRN are proteins, mRNAs and protein-protein complexes,
while the edges  represent individual reactions (protein-protein and
protein-mRNA interactions)\footnote{Wikipedia's lemma for \textit{Gene
regulatory network}:
\url{http://en.wikipedia.org/wiki/Gene_regulatory_network}.}. Clustering
tools can be applied to reveal the structural properties of these
networks, since densely connected groups may have an important biological
interpretation (e.g., Ref. \cite{nemo-bmc10}). 

\par In the context of  lateral gene transfer and prokaryote genome sequence
data, the donor-recipient relations between genomes are modeled by directed
networks called Lateral Gene Transfer networks (LGT). Applying graph
clustering methods on these networks, we are able to test hypotheses
regarding LGT patterns and mechanisms operating in nature
\cite{genome-research}.

\par The list of applications in the biological domain continues with the case
of food web networks \cite{food-web-pnas02, newman}, which represent trophic
relationships in ecosystems. Typically, the nodes of the network correspond to
species of an ecosystem and the directed edges capture pray-predator
relationships. Regarding the community structure of food webs, the basic
question is whether the networks are organized into compartments (the term
used to describe the communities-clusters in food webs), where species within
the same compartment interact frequently among each other, but show fewer
interactions between species of different compartments (see e.g., Refs.
\cite{food-webs-nature03, compartmentalization-ecology10,
compartments-ecology-letters09}). The existence of community structure in food
webs is an important property, due it its relationship to the robustness of the
network under perturbations.

\subsection{Neuroscience}
Neuroscience is the scientific discipline that studies the nervous system and
the brain\footnote{Wikipedia's lemma for \textit{Neuroscience}:
\url{http://en.wikipedia.org/wiki/Neuroscience}.}. With the advances in brain
mapping and neuroimaging techniques\footnote{Wikipedia's lemma for
\textit{Neuroimaging}: \url{http://en.wikipedia.org/wiki/Neuroimaging}.} (i.e.,
techniques used to image the
structure and function of the brain), the brain can be modeled by graph
structures known as complex brain networks. In recent years, there has
been several studies concerning  graph theoretical analysis of human brain
networks for a wide range of mapping techniques, such as MRI, fMRI and EEG/MEG
(see the following review articles in the area \cite{brain-networks09,
brain-connectivity10}). Depending on the mapping approach, the nodes of the
networks can be defined as the electroencephalography electrodes or
multielectrode-array electrodes, or as specific anatomically defined regions of
the brain (e.g., regions of MRI or diffusion tensor imaging data). Then, a
measure of association between  nodes should be selected. Some examples are the
measures of spectral coherence or Granger causality between two MEG sensors or
the connection probability between two regions of diffusion tensor imaging
data. Finally, the adjacency matrix of the graph is typically formed by
the pairwise association between nodes (see Ref. \cite{brain-networks09} for
more details). In many cases, the association is not symmetric, forming directed
brain networks.

\par Cluster analysis in human brain networks is an important task, which can
help neuroscientists to extract functional subdivisions of the brain. Thus, in
the case of directed brain networks, the clustering  methods presented
throughout this survey paper can be proved useful in the field of neuroscience.
Liao et al. \cite{human-brain-neuroimage11} studied the directed
influence brain network of resting-state fMRI recordings. Among other things,
they observed that the network has a modular structure, applying the modularity
maximization technique of  Leicht and Newman (see
Section \ref{sec:modularity} and Ref. \cite{leicht-newman-2008}). The authors
also discuss some specific properties that are shared among  nodes of the
same module. In Ref. \cite{neuro}, the authors examined the directed networks of
spontaneous activity correlation or Resting State Networks (RSN), i.e., brain
networks that capture spontaneous activity (not stimulus or task driven),
applying also the directed version of modularity \cite{leicht-newman-2008}.
Vertes and Duke \cite{spikes-hfsp10} studied a mechanism of neuronal encoding,
investigating the effects of network topology based on spatiotemporal patterns
of spikes. The produced network of neurons is directed and the authors,
among other things, applied clustering methods on the analysis. In Ref.
\cite{cortex-plos-one11}, directed cortex networks were analyzed,
focusing on structural properties such as node betweenness centrality, average
shortest path length and community structure. As a last application
example from the area of neuroscience, we refer to the work of Pan et al.
\cite{nervous-plos-one10}, who studied the nervous system  of the nematode
\textit{Caenorhabditis elegans}, focusing on its organization into modules.

\subsection{Other Applications}
Directed graph data arises in several other application domains, broadening the
scope of  clustering methods presented throughout this paper. In the case
of financial networks, the authors of Ref. \cite{communities-corporate10}
analyzed the community structure of an Italian corporate ownership network,
where nodes represent firms (companies), while a directed edge between two
firms $i,j$ captures stock ownership relationship (between the shareholder $i$
and the owned company $j$). The authors applied modularity maximization
techniques \cite{leicht-newman-2008} on the connected part of this directed
network (consisting of $141$ nodes), observing strong community structure.

\par Clustering methods in directed networks can also be used in the field
of computer vision and image processing, and more precisely in the task of image
segmentation.  An image can be modeled as a graph, where pixels are represented
as nodes in the graph, and  edges capture similarity between different pixels
(as computed by a pairwise similarity function on a set of image attributes
such as color and intensity). The similarity (or affinity) measure can be
either symmetric or asymmetric. In the
case of asymmetric affinity measures, the produced graph is directed. Since the
initial image has been transformed into a graph, the image segmentation task
becomes a graph clustering problem and several new clustering approaches have
been proposed for this grouping problem with asymmetric affinities (e.g., Refs.
\cite{Yu-shi01, game-theoretic06}).

\par The range of  applications for the clustering problem in directed networks
can become very broad, since a large number of non-graph data can be represented
as directed graphs using appropriate transformation techniques. Given
a set of data points $x_1, x_2, \ldots, x_n$, where each $x_i \in \mathbb{R}^N$,
and a similarity measure $s_{ij} \ge 0 $ for all pairs of points $(x_i,
x_j)$, the data can be
represented by a similarity graph $G=(V,E)$. In this graph, each node $v_i
\in V$ represents a data point $x_i$, while two nodes $v_i, v_j$ 
are connected via edge if the similarity $s_{ij}$ between data points $x_i$ and
$x_j$ is
larger than a specific threshold (the edge can be weighted by the term
$s_{ij}$). Then, the data clustering problem can be considered as a graph
clustering task; the goal is to partition the similarity graph in such a
way that edges across different clusters should have low weight, while 
high weight edges should be placed within the  same cluster. Depending on the
method
applied to construct the similarity graph, the latter can be either directed or
undirected \cite{von-luxburg}. For example, in the approach of
$k$-nearest neighbor graph, node $v_i$ is connected to $v_j$ if $v_j$ is
among the $k$-nearest neighbors of $v_i$. This forms a directed graph, since the
neighborhood relationship is not symmetric. Hence, in the general case on non
symmetric similarity functions, i.e., $s_{ij} \neq s_{ji}$, the corresponding
similarity graph can be directed.

\section{Open Problems and Future Research Directions} \label{sec:future}
In this section we discuss  interesting open problems and future research
directions for the graph clustering and community detection task in directed
networks. Most of the topics that will be presented shortly also suit to the
undirected case of the problem (e.g., see Ref. \cite{fortunato}). However, it
seems that the graph clustering problem in directed networks is  more
challenging compared to the undirected version. Moreover, since the directed
case is a generalization of the undirected one (each undirected graph can be
also represented as a directed, considering edges to both directions), effective
methods for the former case can be used for the latter one as well.

\subsubsection*{Formal Definition of the Problem}
The major point that should be addressed in the area is  a formal and precise
definition of the graph clustering and community detection problem in directed
networks. We have observed  that most of the
proposed methods follow two high-level clustering notions-definitions (or a
combination of them), as described in Section \ref{sec:problem-stm}. The most
prevalent one, the density-based clustering notion, can be considered as a
direct generalization from the undirected case, and according to this viewpoint
some well-known approaches have been proposed. Although in many of these methods
(e.g., modularity optimization) the direction of the edges is taken into
consideration in the clustering task, there is no a clear way of how this can be
properly done. For example, in the
directed version of modularity (e.g., Refs. \cite{leicht-newman-2008,
arenas-modularity07}),  the existence of a directed edge $(i,j)$ between nodes
$i,j$, depends on the out-degree of node $i$ and in-degree of node $j$. Then,
the configuration (null) model is suitably adjusted to meet this
fact. However, in many cases, different approaches set up diverse
requirements for the problem and therefore, many of the proposed methods are not
consistent with each other (this has been pointed out for the
undirected graph clustering problem as well \cite{fortunato, schaeffer-review}).

\par As we mentioned in Section \ref{sec:introduction} regarding the challenges
posed by the problem, even some direct generalizations from the undirected
graph clustering problem are not straightforward. We have discussed that graph
concepts applied to characterize and evaluate the community structure in
undirected networks (e.g., density) cannot be directly extended to the directed
case, making the theoretical foundations of the  problem not yet fully explored.
To conclude, we consider that the foremost task regarding the problem
is to establish the theoretical tools towards a formal definition of how  good
clusters or communities in directed networks should look like. Of course, one
should not expect that a single definition would fit to all needs, since as
noted by Schaeffer \cite{schaeffer-review}, the problem is highly
application-oriented (this fact holds for both directed and undirected
networks; an evidence for the former case is the plethora of diverse
applications presented in Section \ref{sec:applications}). Later at this
section, we will discuss possible data-driven and application-driven methods, as
an interesting future direction.

\subsubsection*{Algorithm Design and Evaluation}
Several interesting points typically arise in the context of designing and
evaluating algorithms for the clustering problem in directed networks. Having
define the basic functionalities of the algorithm as well as the type-notion of
clusters that we are looking for, three other aspects are of particular
significance: 

\begin{itemize}
 \item[(a)] \textit{Algorithm's parameters:} issues related to the parameters
of the algorithm, such as their number, how the user is going to set input
values for the parameters, as well as the sensitivity of the algorithm on the
parameters' selection.

 \item[(b)] \textit{Algorithm's scalability:} issues related to the ability of
the algorithm to perform well, while the size of the input (i.e., the size of
the graph in our case) increases.

\item[(c)] \textit{Algorithm's evaluation:} issues related to the evaluation of
the algorithm, both in terms of computational efficiency and effectiveness.
\end{itemize}

\noindent Any graph clustering algorithm should contain as few
parameters as possible, with the ideal case to be the design of parameter-free
algorithms. As has been noted in several research articles, this point is
extremely crucial while designing data mining algorithms
(e.g., see Ref. \cite{keogh-parameter-free}) for several reasons.  First of all,
a non-expert user should be able to use the algorithm without any technical
difficulties related to the selection of appropriate values for the
parameters. Moreover, in the clustering task, incorrect selection of the
parameters may lead the algorithm to extract incorrect patterns from the data.
Finally, the output of the algorithm (the extracted clusters/communities)
should not be highly sensitive to the settings of parameters' values. Some
of the community detection algorithms presented in this paper do not require
any input parameters (e.g., the methods based on modularity optimization),
while others require to select the number of clusters.

\par The scalability is always considered as an important performance issue in
the design and evaluation of graph clustering algorithms, and over the last
years it has received great attention from the research community due to the
enormous available graph data. Usually, the objective measures
used to identify the clustering structure, lead to computational
difficult problems; then, either approximation techniques  or other novel
approaches (or heuristics) are applied to cope with the computational complexity
constraints. For example, in the case of undirected networks, Satuluri and
Parthasarathy \cite{satuluri-stochastic-flow-kdd09} presented a clustering
technique based on stochastic flows, that improves in terms of scalability, the
well-known Markov Clustering algorithm (MCL) proposed by van Dongen
\cite{vanDongen2000} (the authors also note that their method can be easily
extended to directed networks). Recently, distributed computing techniques based
on the MapReduce framework \cite{mapreduce} have become the standard approaches
for processing massive data. Several methods have been proposed for
analyzing graph data in this framework (e.g., the \textit{Pegasus}
system \cite{pegasus-icdm09}), including spectral graph
analysis \cite{kang-spectral-analysis-pakdd11} and centrality estimation
\cite{kang-centrality-sdm11}. To this direction, it  would be interesting to
extend already developed algorithms for the clustering problem in directed
networks or to design new ones in the MapReduce framework, letting us
to study the community  structure of very large networks (e.g., billion-node
networks).

\par One other significant point related to the clustering problem in directed
networks has to do with the evaluation of methods, i.e., how to decide
if the results of an algorithm are ``good'' or which of several possible
clustering results is the best one. As we discussed in Section
\ref{sec:evaluation}, in case
of datasets with a priori knowledge of the community structure, this information
can be used to evaluate the performance of the algorithm. However, ground truth
data is not always available; then, the goal is to define reliable
benchmark graphs that can be used to test and evaluate algorithms. The work
of Lancichinetti and Fortunato \cite{fortunato-benchmark09} presents such a
generator for benchmark directed networks but we consider that more effort
should be put on this topic by the research community due to its importance. 

\par Furthermore, in order to become more clear which clustering algorithm may
be better or at least has better performance on specific directed networks,
experimental and comparative  studies should be done. In the case of undirected
networks, several experimental comparisons have been performed for different
scale and different types of networks (e.g., social networks, information
networks. See Ref. \cite{leskovec-www10, danon-jstat05, coscia-review}
for more details). We consider that similar studies in directed networks will
shed more light on how to select a clustering algorithm.

\subsubsection*{Towards Data-driven and Application-driven Approaches}
It becomes clear that the problem of clustering  in
directed networks is very challenging and several diverse methods have been
proposed to deal with it. We consider that the problem is
from its nature application-oriented and thus, there should always be
space for new possible solutions depending on the characteristics of the network
data and the application domain.

\par A possible direction is the design of data-driven algorithms for
the clustering task in directed networks. According to this approach, the
goal is to study the structure of the networks we are interested in, and then
take into account possible structural observations that may affect the
community detection task; even better would be  to exploit possible patterns or
interesting structures that may be contained in the data, in the design phase of
an algorithm. Such studies and approaches have
already been performed in the case of undirected networks. Some of them study
the quality of communities in real networks. Leskovec et al.
\cite{leskovec-communities-www08, leskovec-IM} observed that in large scale
social and information networks, ``good'' communities exist only at very
small scales (of about $100$ nodes), while at larger scales the communities
gradually blend in with the rest of the network. Similar observations have been
presented by Malliaros et al. \cite{sdm-robustness}, studying the
robustness of large social graphs. Another example is the community detection
algorithm presented by Prakash et al. \cite{eigenspokes}, which exploits the
\textit{eigenspokes} pattern observed in large scale
sparse real graphs (i.e., pattern related to the eigenvectors
of the adjacency matrix of the graph). Thus, it would be useful to perform
exploratory analysis regarding the structural properties of directed networks
and utilize possible interesting findings to the algorithm design process.

\par A different possible direction is to follow application-driven approaches,
i.e., design domain-specific and application-specific clustering algorithms for
directed networks. As we have already mentioned, the graph clustering  task can
be applied in a wide range of applications, from social network analysis to
biological networks and from  economic networks to the domain of neuroscience.
All these diverse applications demonstrate different features and therefore it
should be more appropriate to follow different methodological approaches with
respect to the application under consideration.

\subsubsection*{Other Research Directions}
In this section we present other, diverse interesting topics for future
research regarding the clustering/community detection problem in directed
networks (generally, most of the research directions in the case of undirected
networks (e.g., Refs. \cite{fortunato, schaeffer-review}) can also be considered
as important aspects for directed networks).

\par A possible direction is to examine local (versus global) definitions
of clusters in directed networks and therefore local algorithms
for detecting the community structure. In other words,
instead of partitioning the full graph, it would be interesting to define
measures and design algorithms for evaluating a subgraph in terms of community
structure. Such approaches can also operate as \textit{community structure
exploration} tools and may be useful either for large scale networks or
for networks with no clear community structure (e.g., Ref. \cite{leskovec-IM}).

\par Generally, in the discussion until now, the directed networks represent
mainly connectivity information between entities of the same type; the edges may
or may not contain weights, which quantify the significance of the ties and
typically are interpreted as deterministic values with positive meaning.
However, new domains and applications in the context of network analysis, impose
new kind of information that should be also taken into consideration in the
clustering task. Three well known representative examples are the so-called
signed networks, probabilistic (or uncertain) networks and the case of
heterogeneous networks. The \textit{signed networks} (e.g., Refs.
\cite{leskovec-signed-chi10, kunegis-clustering-signed-sdm10}) are trying to
capture the notion of positive and negative interactions among the nodes of a
network. A positive edge denotes similarity (proximity), while negative edges
represent dissimilarity (distance). Signed networks can be both directed and
undirected.  Characteristic examples of directed signed networks are the trust
networks between users in product review websites, like Epinions
(\url{www.epinions.com}). The \textit{probabilistic graphs}
(e.g., Refs. \cite{potamias-knn-uncertain-vldb10,
probabilistic-freq-subgraphs-kdd10, kollios-clustering-probabilistic-tkde11})
capture uncertainty that is introduced under several conditions
(such as for privacy-preserving reasons), and every edge in the network is
associated with its probability of existence. Since probabilistic graphs is the
natural extension of deterministic ones to represent uncertainty factors, they
can be both directed and undirected. For the undirected case, Kollios et al.
\cite{kollios-clustering-probabilistic-tkde11} recently presented an algorithm
for the graph clustering task. Finally, \textit{heterogeneous networks} (in
contrast to homogeneous ones) \cite{sun-heterogeneous-book-2012}, directed or
undirected,  are used to represent multi-typed networks, i.e., networks that
contain multiple objects and link types (e.g., in a bibliographic network the
nodes may correspond to authors, publications and venues, while the edges
link these different types of nodes). An interesting research question is
how can we extend existing techniques or design new ones for the clustering task
in directed networks under the settings presented above.

\par Finally,  an important research direction which has been already discussed,
is the case of dynamic directed networks, i.e., networks that evolve over time
with the addition/deletion of nodes/edges. In Section \ref{sec:other-techniques}
we described a method for time-evolving directed networks, where the goal was
both the extraction of the community structure and the detection of a change
point (regarding community structure) over time.

\section{Conclusions} \label{sec:conclusions}
In this survey we have reviewed thoroughly the problem of clustering and
community detection in directed networks and to the best of our knowledge, this
is the first comprehensive review fully devoted to the problem. The main goal
was to organize and present in a unified manner the work  conducted so
far for the problem. In a first step, we have presented a classification of the
approaches in four main categories, according to the methodology they follow:
(i) naive graph transformation approaches, (ii) transformations maintaining
directionality, (iii) approaches that extend objective functions and
methodologies to directed networks, and (iv) alternative approaches. Since a
large portion of the methods constitute extensions from the undirected version
of the problem, we have followed an incremental presentation, describing the
basic features of the undirected problem and how they can be extended to the
directed case. Furthermore, we have presented a second orthogonal classification
of the methods, based on the clustering notion (or type of clusters) they
follow. This classification scheme is supplementary to the previous one and it
may be useful for practitioners that need to select an appropriate algorithm for
a specific application. We have also presented methods and tools for evaluating 
and testing the results of a graph clustering algorithm in directed
networks. Moreover, since the problem is highly application
oriented, we have demonstrated interesting application domains. To
conclude, we consider that more effort should be put on the problem by the
research community  due to its high importance, and towards this direction we 
have provided interesting topics for future research.

\section*{Acknowledgments}
The authors would like to thank the anonymous reviewer for the valuable and
constructive comments, as well as the authors that kindly offered visual
content of their articles. The authors are partially supported by the DIGITEO
Chair grant LEVETONE in France. Fragkiskos D. Malliaros is a recipient of the
Google Europe Fellowship in Graph Mining, and this research is supported in part
by this Google Fellowship.

\bibliographystyle{elsarticle-num}
\bibliography{my_references.bib}

\end{document}